{}
{}

\documentclass[a4paper,12pt]{article}

\newif\ifpublic\publictrue
\newif\ifjournal\journalfalse

\usepackage{amsmath}            
\usepackage{amssymb}            
\usepackage{graphicx}           
\usepackage[nosort]{cite}
\usepackage{fixmath}            
\ifpublic\else\usepackage{showkeys}\fi
\usepackage[bookmarks=true,hyperfigures=true]{hyperref}
\usepackage[bulletsep]{collref}

\expandafter\def\expandafter\bfseries\expandafter{\bfseries\ifmmode\else\boldmath\fi}
\expandafter\def\expandafter\mdseries\expandafter{\mdseries\ifmmode\else\unboldmath\fi}
\expandafter\def\expandafter\normalfont\expandafter{\normalfont\ifmmode\else\unboldmath\fi}

\newcommand{\remark}[2][.]{\ignorespaces}
\ifpublic\else
\RequirePackage{color}
\renewcommand{\remark}[2][.]{{\color{red}\renewcommand{\bfdefault}{b}\rmfamily\if.#1\else\textbf{#1:} \fi#2}}

\fi

\allowdisplaybreaks[3]

\numberwithin{equation}{section}

\usepackage[font=small,labelfont=bf,width=0.85\textwidth]{caption}

\RequirePackage{amsmath}
\makeatletter
\newcommand{\brk@ord}{\bBigg@{0}}
\newcommand{\brk@ordl}{\mathopen\brk@ord}
\newcommand{\brk@ordr}{\mathclose\brk@ord}
\newcommand{\brk@ordm}{\mathrel\brk@ord}
\newcommand{\brk@var}{\brk@ord}
\newcommand{\brk@varl}{\left}
\newcommand{\brk@varr}{\right}
\newcommand{\brk@varm}{\mathrel\brk@var}
\newcommand{\brk@altname}[3]{\expandafter\def\csname#2\expandafter\@gobble\string#1\endcsname{#1[#3]}}
\newcommand{\brk@usearg}[3]{%
  \def\brk@star{*}\def\brk@blank{}\def\brk@arg{#1}%
  \ifx\brk@arg\brk@blank\def\brk@arg{brk@ord}\fi%
  \ifx\brk@arg\brk@star\def\brk@arg{brk@var}\fi%
  \csname\brk@arg #2\endcsname#3}

\newcommand{\DeclareMathBrackets}[3]{
  \newcommand{#1}[2][]{\brk@usearg{##1}{l}{#2}##2\brk@usearg{##1}{r}{#3}}
  \brk@altname{#1}{big}{big}\brk@altname{#1}{lr}{*}}
\newcommand{\DeclareMathBiBrackets}[4]{
  \newcommand{#1}[3][]{\brk@usearg{##1}{l}{#2}##2#3##3\brk@usearg{##1}{r}{#4}}
  \brk@altname{#1}{big}{big}\brk@altname{#1}{lr}{*}}
\newcommand{\DeclareMathBiMBracketsStar}[4]{
  \newcommand{#1}[3][]{\brk@usearg{##1}{l}{#2}##2\brk@usearg{##1}{m}{#3}##3\brk@usearg{##1}{r}{#4}}
  \brk@altname{#1}{bi}{big}}
\newcommand{\DeclareMathBiBracketsStar}[4]{
  \newcommand{#1}[3][]{\brk@usearg{##1}{l}{#2}##2\brk@usearg{##1}{}{#3}##3\brk@usearg{##1}{r}{#4}}
  \brk@altname{#1}{big}{big}}

\makeatother

\DeclareMathBrackets{\brk}{(}{)}
\DeclareMathBrackets{\vev}{\langle}{\rangle}
\DeclareMathBrackets{\bra}{\langle}{|}
\DeclareMathBrackets{\ket}{|}{\rangle}
\DeclareMathBrackets{\sbra}{\langle}{|}
\DeclareMathBrackets{\sket}{|}{\rangle}
\DeclareMathBrackets{\cbra}{[}{|}
\DeclareMathBrackets{\cket}{|}{]}
\DeclareMathBrackets{\sbrk}{[}{]}
\DeclareMathBrackets{\set}{\{}{\}}
\DeclareMathBrackets{\abs}{|}{|}
\DeclareMathBrackets{\eval}{.}{|}
\DeclareMathBiBrackets{\comm}{[}{,}{]}
\DeclareMathBiBrackets{\acomm}{\{}{,}{\}}
\DeclareMathBiBrackets{\gcomm}{[}{,}{\}}
\DeclareMathBiMBracketsStar{\braket}{[}{|}{\}}

\def\showkeysrefformat#1{{\normalfont\tiny\ttfamily#1}}
\makeatletter
\def\SK@@ref#1>#2\SK@{%
 {\@inlabelfalse\leavevmode\vbox to\z@{%
 \vss\SK@refcolor\rlap{\vrule\raise .75em%
  \hbox{\showkeysrefformat{#2}}}}}}
\makeatother

\usepackage[a4paper,text={160mm,247mm},centering]{geometry}

\newcommand{\order}{\mathcal{O}}                            

\newcommand{\alg}[1]{\mathfrak{#1}}
\newcommand{\grp}[1]{\mathrm{#1}}

\def\eqn{\eqref}

\DeclareMathOperator{\tr}{tr}

\DeclareMathOperator{\str}{str}

\newcommand{\superN}{\mathcal{N}}
\newcommand{\trans}{{\scriptscriptstyle\mathsf{T}}}

\newcommand{\sfrac}[2]{{\textstyle\frac{#1}{#2}}}
\newcommand{\half}{\sfrac{1}{2}}

\DeclareMathOperator{\pathord}{P}

\newcommand{\gen}[1]{\mathrm{#1}}
\newcommand{\genY}[1]{\widehat{\mathrm{#1}}}
\newcommand{\gengauge}[1]{\gen{#1}_*}

\newcommand{\genfield}[1]{\mathbb{#1}}
\newcommand{\genYfield}[1]{\widehat{\mathbb{#1}}}
\newcommand{\genYfieldnl}[1]{\widehat{\mathbb{#1}}_\text{bi}}
\newcommand{\genpath}[1]{\mathbf{#1}}
\newcommand{\genmatr}[1]{\mathcal{#1}}
\newcommand{\genYpath}[1]{\widehat{\genpath{#1}}}
\newcommand{\gaugegen}{\mathbb{G}}
\newcommand{\kappagen}{\delta}
\newcommand{\kappagenfull}{\kappagen_*}
\newcommand{\genfull}[1]{\genfield{#1}_*}
\newcommand{\genYfull}[1]{\genYfield{#1}_*}
\newcommand{\genYfullnl}[1]{\genYfield{#1}_{*,\text{bi}}}

\newcommand{\der}{\mathrm{d}}
\newcommand{\cder}{\mathcal{D}}
\newcommand{\cdel}{\mathcal{D}}
\newcommand{\del}{\partial}
\newcommand{\sdel}{D}
\newcommand{\scdel}{\mathcal{D}^{\text{cov}}}
\newcommand{\scdelb}{\bar{\mathcal{D}}^{\text{cov}}{}^{\hspace{0.25pt}}}
\newcommand{\copro}{\mathrm{\Delta}}

\newcommand{\gauge}{A}
\newcommand{\sgauge}{\mathcal{A}}

\newcommand{\scalar}{\phi}
\newcommand{\sscalar}{\Phi}

\newcommand{\sfstr}{\mathcal{F}}
\newcommand{\sconfrem}{\mathcal{G}}

\newcommand{\swilson}{\mathcal{W}}
\newcommand{\storsion}{T}
\newcommand{\sviel}{\mathcal{E}}
\newcommand{\svielB}{e}
\newcommand{\svielF}{\der\theta}
\newcommand{\svielC}{\der\bar\theta}

\newcommand{\idxall}[1]{{\mathcal{#1}}}
\newcounter{numidx}
\newcommand{\idxconf}[1]{{\setcounter{numidx}{\number`#1}\addtocounter{numidx}{-97}%
  \ifcase\arabic{numidx}\delta\or\rho\or\kappa\or\sigma\or\upsilon\else\omega\fi}}
\newcommand{\Real}{\mathbb{R}}

\newcommand{\nln}{\nonumber\\}

\def\[{\begin{equation}}
\def\]{\end{equation}}

\providecommand{\href}[2]{#2}
\newcommand{\arxivlink}[1]{\href{http://arxiv.org/abs/#1}{arxiv:#1}}

\makeatletter
\def\mr@ignsp#1 {\ifx\:#1\@empty\else #1\expandafter\mr@ignsp\fi}%
\newcommand{\multiref}[1]{\begingroup
\xdef\mr@no@sparg{\expandafter\mr@ignsp#1 \: }%
\def\mr@comma{}%
\@for\mr@refs:=\mr@no@sparg\do{\mr@comma\def\mr@comma{,}\ref{\mr@refs}}%
\endgroup}
\renewcommand{\eqref}[1]{(\multiref{#1})}
\makeatother

\makeatletter
\newcommand{\namedref}[2]{\hyperref[#2]{#1~\ref*{#2}}}
\newcommand{\secref}{\@ifstar{\namedref{Sec.}}{\namedref{sec.}}}
\newcommand{\appref}{\@ifstar{\namedref{App.}}{\namedref{app.}}}
\newcommand{\tabref}{\@ifstar{\namedref{Tab.}}{\namedref{tab.}}}
\newcommand{\figref}{\@ifstar{\namedref{Fig.}}{\namedref{fig.}}}
\makeatother

\let\oldbib=\thebibliography
\def\thebibliography{\phantomsection\addcontentsline{toc}{section}{\refname}\oldbib}

\let\oldtoc=\tableofcontents
\def\tableofcontents{\phantomsection\addcontentsline{toc}{section}{\contentsname}\oldtoc}

\RequirePackage{verbatim}

\makeatletter
\newwrite\mpi@out
\def\mpi@write#1{\immediate\write\mpi@out{#1}}
\immediate\openout\mpi@out\jobname.mp
\mpi@write{\@percentchar generated from `\jobname'}%
\AtEndDocument{\mpostdone}

\def\mpostdone{
  \immediate\closeout\mpi@out%
  \ifpublic\else%
    \immediate\write18{mpost -tex=latex \jobname.mp}
  \fi%
  \gdef\mpostdone{}
}

\newcommand{\mpi@putlineno}{%
  \mpi@write{\@percentchar---------------------------------------}%
  \mpi@write{\@percentchar l.\the\inputlineno}%
}

\newcommand{\mpi@verbatim}{
  \@bsphack
  \let\do\@makeother\dospecials
  \catcode`\^^M\active
  \def\verbatim@processline{\mpi@write{\the\verbatim@line}}%
  \verbatim@start
}

\newenvironment{mpostcmd}{%
  \mpi@putlineno%
  \mpi@verbatim%
}%
{\mpi@write{}\@esphack}

\newenvironment{mpostfile}[1]{%
  \mpi@putlineno%
  \mpi@write{filenametemplate "#1";}%
  \mpi@write{beginfig(0)}%
  \mpi@verbatim%
}%
{\mpi@write{endfig;}\@esphack}

\newcommand{\includegraphicsex}[2][]{%
  \xdef\mpi@tmp{#2}%
  \IfFileExists{\mpi@tmp}%
    {\includegraphics[#1]{\mpi@tmp}}%
    {\textbf{??}\typeout{file \mpi@tmp{} missing}}%
}

\makeatother

\makeatletter
\newsavebox{\apb@box}\newlength{\apb@width}
\newcommand{\autoparbox}[2][c]{\sbox{\apb@box}{#2}%
 \settowidth{\apb@width}{\usebox{\apb@box}}%
 \parbox[#1]{\apb@width}{\usebox{\apb@box}}}
\newcommand{\includegraphicsbox}[2][]{\autoparbox{\includegraphicsex[#1]{#2}}}
\makeatother

\providecommand{\hypersetup}[1]{}
\providecommand{\texorpdfstring}[2]{#1}

\hypersetup{plainpages=false}
\hypersetup{pdfpagemode=UseNone}
\hypersetup{bookmarksnumbered=true}
\hypersetup{pdfstartview=FitH}
\hypersetup{colorlinks=false}
\hypersetup{citebordercolor={.5 1 .5}}
\hypersetup{urlbordercolor={.5 1 1}}
\hypersetup{linkbordercolor={1 .7 .7}}

\makeatletter
\let\@keywords\@empty
\let\@subject\@empty
\providecommand{\keywords}[1]{\gdef\@keywords{#1}}
\providecommand{\subject}[1]{\gdef\@subject{#1}}
\def\thetitle{\@title}
\def\theauthor{\@author}
\def\thesubject{\@subject}
\def\thedate{\@date}
\def\thekeywords{\@keywords}
\makeatother
\AtBeginDocument{
\hypersetup{pdftitle={\thetitle}}%
\hypersetup{pdfauthor={\theauthor}}%
\hypersetup{pdfsubject={\thesubject}}%
\hypersetup{pdfkeywords={\thekeywords}}%
}

\begin{mpostcmd}
prologues:=2;

verbatimtex
\documentclass{article}
\usepackage{amsfonts,amssymb,latexsym}
\begin{document}
etex
\end{mpostcmd}

\begin{mpostcmd}
picture copyrightline,copyleftline;
copyrightline := btex \copyright\ \textsf{2015 Niklas Beisert} etex;
copyleftline := btex $\circledast$ etex;
def putcopyspace =
label.bot(btex \vphantom{gA} etex scaled 0.1, lrcorner(currentpicture));
enddef;
def putcopy =
label.ulft(copyrightline scaled 0.1, lrcorner(currentpicture)) withcolor 0.9white;
label.urt(copyleftline scaled 0.1, llcorner(currentpicture)) withcolor 0.9white;
currentpicture:=currentpicture shifted (10.5cm,14cm);
enddef;
\end{mpostcmd}

\begin{mpostcmd}
def pensize(expr s)=withpen pencircle scaled s enddef;
pair pos[];
path paths[];
def midarrow (expr p, t) =
fill arrowhead subpath(0,arctime(arclength(subpath (0,t) of p)+0.5ahlength) of p) of p;
enddef;
\end{mpostcmd}

\begin{mpostfile}{FigDomainReg.mps}
paths[3]:=(0,0.3){dir 50}..(0.4,0.5){dir 20}..{dir 40}(0.8,1)..{dir 50}(1,1.3);
paths[1]:=subpath(0,2) of paths[3];
paths[2]:=(subpath(3,2) of paths[3]) rotated -45 yscaled -1 rotated 45 shifted (-1,0);
pos[1]:=point 0 of paths[1];
pos[2]:=point 2 of paths[1];
pos[3]:=point 0 of paths[2];
pos[4]:=point 1 of paths[2];
pos[5]:=(0.2,0.2);
paths[3]:=pos[4]--(paths[1] intersectionpoint (((0,0)--(1,0)) shifted pos[4]));
paths[4]:=pos[5]--(paths[1] intersectionpoint (((0,0)--(0,1)) shifted pos[5]));
pos[6]:=(0.7,0.7);
paths[5]:=pos[6]--(paths[1] intersectionpoint (((0,0)--(-1,0)) shifted pos[6]));
xu:=3cm;
fill buildcycle(paths[1],reverse(paths[3]),(0,1)--(0,0)) scaled 1xu withcolor 0.9white;
fill buildcycle(paths[1],(1,1)--(0,1),paths[2],paths[3]) scaled 1xu withcolor 0.85white;
fill (paths[2]--(0,1)--cycle) scaled 1xu withcolor (0.8white+0.2red);
fill (paths[1]--(1,1)--(0,0)--cycle) scaled 1xu withcolor (0.8white+0.2blue);
draw ((0,0)--(1,1)--(0,1)) scaled 1xu dashed evenly scaled 1.5 pensize(0.5pt);
draw paths[1] scaled 1xu dashed evenly scaled 1 pensize(0.5pt);
draw paths[2] scaled 1xu dashed evenly scaled 1 pensize(0.5pt);
draw paths[3] scaled 1xu dashed withdots scaled 0.5 pensize(1pt);
drawarrow ((0,-0.05)--(0,1.2)) scaled 1xu pensize(1pt);
drawarrow ((-0.05,0)--(1.2,0)) scaled 1xu pensize(1pt);
draw ((-0.05,1)--(+0.05,1)) scaled 1xu;
draw ((1,-0.05)--(1,+0.05)) scaled 1xu;
draw paths[4] scaled 1xu pensize(1pt);
draw paths[5] scaled 1xu pensize(1pt);
label.bot(btex $0$ etex, (0,-0.05) scaled 1xu);
label.bot(btex $1$ etex, (1,-0.05) scaled 1xu);
label.lft(btex $1$ etex, (-0.05,1) scaled 1xu);
label.lft(btex $0$ etex, (-0.05,0) scaled 1xu);
label.llft(btex $\tau_1$ etex, (1.2,0) scaled 1xu);
label.llft(btex $\tau_2$ etex, (0,1.2) scaled 1xu);
label.lft(btex $\varepsilon$ etex, (0.5*pos[1]+0.5*(0,0)) scaled 1xu);
label.lft(btex $\bar\varepsilon$ etex, (0.5*pos[4]+0.5*(0,1)) scaled 1xu);
label.top(btex $\varepsilon$ etex, (0.5*pos[3]+0.5*(0,1)) scaled 1xu);
label.top(btex $\bar\varepsilon$ etex, (0.5*pos[2]+0.5*(1,1)) scaled 1xu);
label.lrt(btex $\varepsilon(\tau_1)$ etex, point 0 of paths[4] scaled 1xu);
label.lrt(btex $\bar\varepsilon(\tau_2)$ etex, point 0 of paths[5] scaled 1xu);
\end{mpostfile}

\begin{mpostfile}{FigDomain2Reg.mps}
paths[3]:=(0,0.3){dir 50}..(0.4,0.5){dir 30}..{dir 40}(0.8,1)..{dir 50}(1,1.3);
paths[1]:=subpath(0,2) of paths[3];
paths[2]:=(subpath(3,2) of paths[3]) rotated -45 yscaled -1 rotated 45 shifted (-1,0);
paths[3]:=paths[1] rotated -45 yscaled -1 rotated 45;
paths[4]:=paths[2] rotated -45 yscaled -1 rotated 45;
pos[1]:=point 0 of paths[1];
pos[2]:=point 2 of paths[1];
pos[3]:=point 0 of paths[2];
pos[4]:=point 1 of paths[2];
xu:=3cm;
fill (paths[1]--paths[2]--cycle) scaled 1xu withcolor 0.9white;
fill (paths[2]--(0,1)--cycle) scaled 1xu withcolor (0.8white+0.2red);
fill (paths[1]--(1,1)--(0,0)--cycle) scaled 1xu withcolor (0.8white+0.2blue);
fill (paths[3]--paths[4]--cycle) scaled 1xu withcolor (0.9white+0.1green);
fill (paths[3]--(1,1)--(0,0)--cycle) scaled 1xu withcolor (0.8white+0.2red);
fill (paths[4]--(1,0)--cycle) scaled 1xu withcolor (0.8white+0.2blue);
draw ((0,0)--(1,1)--(0,1)) scaled 1xu dashed evenly scaled 1.5 pensize(0.5pt);
draw ((1,1)--(1,0)) scaled 1xu dashed evenly scaled 1.5 pensize(0.5pt);
draw paths[1] scaled 1xu dashed evenly scaled 1 pensize(0.5pt);
draw paths[2] scaled 1xu dashed evenly scaled 1 pensize(0.5pt);
draw paths[3] scaled 1xu dashed evenly scaled 1 pensize(0.5pt);
draw paths[4] scaled 1xu dashed evenly scaled 1 pensize(0.5pt);
drawarrow ((0,-0.05)--(0,1.2)) scaled 1xu pensize(1pt);
drawarrow ((-0.05,0)--(1.2,0)) scaled 1xu pensize(1pt);
draw ((-0.05,1)--(+0.05,1)) scaled 1xu;
draw ((1,-0.05)--(1,+0.05)) scaled 1xu;
label.bot(btex $0$ etex, (0,-0.05) scaled 1xu);
label.bot(btex $1$ etex, (1,-0.05) scaled 1xu);
label.lft(btex $1$ etex, (-0.05,1) scaled 1xu);
label.lft(btex $0$ etex, (-0.05,0) scaled 1xu);
label.llft(btex $\tau_1$ etex, (1.2,0) scaled 1xu);
label.llft(btex $\tau_2$ etex, (0,1.2) scaled 1xu);
label(btex $+$ etex scaled 1.5, (0.25,0.75) scaled 1xu);
label(btex $-$ etex scaled 1.5, (0.75,0.25) scaled 1xu);
\end{mpostfile}

\begin{mpostcmd}
verbatimtex
\end{document}
etex

end;
\end{mpostcmd}

\mpostdone

\title{Integrability of Smooth Wilson Loops\texorpdfstring{\\}{ }in \texorpdfstring{$\superN=4$}{N=4} Superspace
}
\author{Niklas Beisert, Dennis M\"uller, Jan Plefka, Cristian Vergu}

\begin{document}
\pdfbookmark[1]{Title Page}{title}
\thispagestyle{empty}

\begingroup\raggedleft\footnotesize\ttfamily
HU-EP-15/33\\
\arxivlink{1509.05403}\par
\endgroup
\vspace{10mm}

\ifpublic\else\vspace{-\baselineskip}\centerline{\smash{\raisebox{5mm}{\LARGE\bfseries DRAFT}}}\fi

\begin{center}
{\Large\bfseries\thetitle\par}%
\vspace{15mm}

\begingroup\scshape\large
Niklas Beisert${}^{1}$, Dennis M\"uller${}^{2}$, Jan Plefka${}^{1,2}$%
\\
and Cristian Vergu${}^{1,3}$
\par
\endgroup
\vspace{5mm}

\textit{${}^{1}$ Institut f\"ur Theoretische Physik,\\ Eidgen\"ossische
  Technische Hochschule
  Z\"urich,\\ Wolfgang-Pauli-Strasse 27, 8093 Z\"urich, Switzerland}\\[0.1cm]
{\small\verb+nbeisert@itp.phys.ethz.ch+} \vspace{5mm}

\textit{${}^{2}$ Institut f\"ur Physik und IRIS Adlershof,
\\ Humboldt-Universit\"at zu Berlin, \phantom{$^\S$}\\
  Zum Gro{\ss}en Windkanal 6, D-12489 Berlin, Germany} \\[0.1cm]
{\small\verb+{dmueller,plefka}@physik.hu-berlin.de+} \vspace{5mm}

\textit{${}^{3}$ Department of Mathematics, King's College London \\
The Strand, WC2R 2LS, London, UK} \\[0.1cm]
{\small\verb+c.vergu@gmail.com+} \vspace{8mm}


\textbf{Abstract}\vspace{5mm}\par
\begin{minipage}{14.7cm}
We perform a detailed study of the Yangian symmetry of smooth 
supersymmetric Maldacena--Wilson loops in planar $\superN=4$ super Yang--Mills theory. 
This hidden symmetry extends the global superconformal symmetry present for these observables. 
A gauge-covariant action of the Yangian generators on the Wilson line is established 
that generalizes previous constructions built upon 
path variations. 
Employing these generators the Yangian symmetry is proven 
for general paths in non-chiral $\superN=4$ superspace 
at the first perturbative order.
The bi-local piece of the level-one generators requires the use of a regulator 
due to divergences in the coincidence limit. 
We perform regularization by point splitting in detail, 
thereby constructing additional local and boundary contributions
as regularization for all level-one Yangian generators.
Moreover, the Yangian algebra at level one is checked and
compatibility with local kappa-symmetry is established. 
Finally, the consistency of the Yangian symmetry is shown to depend on two properties:
The vanishing of the dual Coxeter number of the underlying superconformal algebra and the
existence of a novel superspace ``G-identity'' for the gauge field theory. This tightly
constrains  the conformal gauge theories to which integrability can possibly apply.
\end{minipage}\par
\end{center}
\newpage

 \renewcommand{\thefootnote}{\arabic{footnote}} \setcounter{footnote}{0}

\ifpublic
 \setcounter{tocdepth}{2} \hrule height 0.75pt
 \tableofcontents
 \vspace{0.8cm} \hrule height 0.75pt \vspace{1cm}
\newpage
\fi


\section{Introduction}
\label{sec:introduction}

The maximally supersymmetric gauge field theory in four dimensions  
\cite{Brink:1976bc,Gliozzi:1976qd} --
$\superN=4$ super Yang--Mills (SYM) -- with  \(\grp{SU}(N)\) gauge group
has become something like the drosophila of gauge field theories. 
This is due to its remarkable symmetry structure and the 
observed integrability \cite{Beisert:2010jr} in the 
planar limit of the theory. 
Integrability arises in particular in the spectral problem of scaling dimensions 
of local gauge-invariant operators \cite{Minahan:2002ve,Beisert:2003tq} 
and (tree-level) scattering amplitudes \cite{Drummond:2009fd}. 
Wilson loop operators \cite{Wilson:1974sk} constitute 
another prominent class of gauge-invariant observables 
for quantum gauge field theories.
In ordinary gauge theories, they serve as order parameters for confinement. 
For conformal and thus non-confining $\superN=4$ SYM, 
they are of relevance because they may, in principle, provide
an ideal testing ground for integrability.
According to the AdS/CFT correspondence \cite{Maldacena:1997re} the $\superN=4$ SYM model is dual
to the type IIB superstring on an $AdS_{5}\times S^{5}$ space-time background. Within this
setting there is a natural Wilson loop for $\superN=4$ SYM that extends the standard definition,
where one integrates the gauge field $A_{\mu}$ along a closed path in $\Real^{1,3}$.
The presence of six real adjoint scalar fields in the theory allows for a coupling to an extended
path $\tilde C$ on $\Real^{1,3}\times S^{5}$, to wit \cite{Maldacena:1998im,Rey:1998ik}
\[
\frac{1}{N} \tr W= \frac{1}{N} \tr  \pathord \exp{ \left(\oint_{\tilde{C}} \mathrm{d} \tau
\left(  \dot{x}^{\mu} A_{\mu}(x) +  n^i \sqrt{\dot{x}^{\mu}\dot{x}_{\mu}} \, \phi_i (x)\right) \right)}
\qquad\text{with } ~
(n^{i})^2=1 \, . 
\label{eq:MWbosdef}
\]
From a ten-dimensional perspective the coupling to the loop $\tilde C$ is such 
that the path is light-like: $\dot x^{2}- (n^{i})^{2}\, \dot x^{2}=0 $
(using mostly minus signature). 
This so-called Maldacena--Wilson loop operator is UV-finite and invariant under conformal transformations. 
Its dual description at strong coupling
is given by a minimal (bosonic) string surface in $AdS_{5}\times S^{5}$ space 
ending on the boundary at the curve $\tilde C$. 

However, it is clear that the operator $W$ is merely the leading 
component of a supersymmetric object generalizing the path $\tilde C$ into superspace. This is
necessary in order to have a Wilson loop which respects the ordinary $\genfield{Q}$ and
conformal $\genfield{S}$ supersymmetries of $\superN=4$ SYM. 
Such a super Maldacena--Wilson loop was first considered in \cite{Ooguri:2000ps} and
further studied in \cite{Muller:2013rta}. The complete definition and a detailed study 
of its local symmetries was recently performed by us \cite{Beisert:2015jxa}. 
It is defined as 
\begin{align}
\label{eq:ourloop}
\frac{1}{N} \tr {\mathcal{W}}= \frac{1}{N} \tr  \pathord \exp \biggl( \oint_{\tilde{Z}} \mathrm{d} \tau
\, \Bigr ( & p^{\mu}  \sgauge^{\text{cov}}_{\mu}(x,\theta,\bar\theta)  +
q^{i}\, \Phi_{i}(x,\theta,\bar\theta)\, \nonumber \\
&+
\dot\theta^{a\alpha} \sgauge^{\text{cov}}_{\alpha a}(x,\theta,\bar\theta) +
\dot{\bar\theta}^{\dot\alpha}{}_{a} \, \sgauge^{\text{cov} \, a}{}_{\dot\alpha}(x,\theta,\bar\theta)
  \Bigr) \, \biggr) \, ,
\end{align}
where $\tilde{Z}=\{x,\theta,\bar\theta;q\}$ defines a path
in a non-chiral $\mathcal{N}=4$ superspace extended by coordinates $q^i$
on an auxiliary space $\Real^6$.
The combination $p^{\mu}=\dot x^{\mu}+ \theta \sigma^{\mu} \dot{\bar\theta} -
\dot{\theta} \sigma^{\mu} \bar{\theta}$ 
is a covariant generalization of $\dot x^\mu$,
and the constraint $q^{i}q^{i}=p^{\mu}p_{\mu}$ restricts the 
path on the internal space to a sphere $S^5\subset\Real^6$
in analogy to $(n^i)^2=1$.
Furthermore, $\sgauge^{\text{cov}}_{\mu}$, $\sgauge^{\text{cov}}_{\alpha a}$, 
$\sgauge^{\text{cov} \, a}{}_{\dot\alpha}$ and $\Phi_{i}$ are
fields on superspace.%
\footnote{Note that in our previous work \cite{Beisert:2015jxa} we
did not use the superscript ``cov'' for the connection superfield. Here we do so in order to
distinguish from a torsion-free situation. Similarly, $\scdel$ here corresponds to 
$\cdel$ in \cite{Beisert:2015jxa}.}
This Wilson loop operator enjoys superconformal 
$\alg{psu}(2,2|4)$ and local kappa-symmetry
along the path \cite{Beisert:2015jxa}, the latter being a generalization of the local $1/2$ BPS
property of the bosonic operator \eqref{eq:MWbosdef}. The expectation value of this operator 
for closed non-self intersecting smooth paths $\tilde Z$ is UV-finite at least at the 
leading non-trivial order in perturbation theory. 
\unskip\footnote{In principle our analysis may be extended to more general UV-divergent cases, 
see the \hyperref[sec:conclusions]{conclusions} for a brief discussion of this.}

In \cite{Muller:2013rta} the integrability of the above super Maldacena--Wilson loop operator at
leading order in the Gra{\ss}mann coordinates $\theta$ and $\bar\theta$ was demonstrated: The one-loop
vacuum expectation value $\vev{\tr\mathcal{W}}$ 
is invariant under an infinite-dimensional Yangian algebra $Y[\alg{psu}(2,2|4)]$ 
represented by variational operators acting on the path $\tilde Z$. 
In two-dimensional quantum field theory a Yangian symmetry is an expression of integrability, i.e.\
its Casimir operators imply the existence of an infinite number of conserved local charges
which constrain the dynamics substantially.

Recently, the question of Yangian symmetry of the super Maldacena--Wilson loop has been investigated
also at strong coupling. Already in the initial paper \cite{Muller:2013rta} the level-one
Yangian symmetry of the bosonic string minimal surface problem was shown. This was generalized
to the full Green--Schwarz superstring minimal surface problem in \cite{Munkler:2015gja}, where
the full $Y[\alg{psu}(2,2|4)]$ Yangian symmetry could be established and the general solution
to the equations of motion ending on the boundary path $\tilde Z$ was found in the
vicinity of the boundary. In \cite{Munkler:2015xqa} the extension of this algebra 
to the level-one hypercharge bonus Yangian symmetry was determined, 
mirroring the structure observed for scattering amplitudes \cite{Beisert:2011pn}.


In this work we complete the analysis of smooth super Wilson loops initiated in \cite{Beisert:2015jxa} 
and address the Yangian
invariance of the super Maldacena--Wilson loop \eqref{eq:ourloop}. 
In fact there have been previous integrability studies of bosonic Wilson loops 
in refs.~\cite{Alday:2010ku, Alday:2010vh, Drukker:2012de, Basso:2013vsa, Basso:2014koa, Basso:2014nra}. 
However, these references studied only contours which were piecewise light-like, $\dot x^{2}=0$,
for which the coupling to the scalars in \eqref{eq:MWbosdef} disappears. These light-like polygonal-shaped 
Wilson loops are dual to maximally-helicity-violating scattering amplitudes in $\superN=4$ SYM 
\cite{Alday2007a,Brandhuber2008,Drummond2008a}. 
In order to achieve superconformal symmetry for light-like polygonal Wilson loops
and to provide a dual to next-to-maximally helicity violating scattering amplitudes, 
supersymmetric Wilson loops have been considered in chiral \cite{Mason2010, Caron-Huot2011} 
and in full non-chiral \cite{Beisert:2012gb,Beisert2012} $\superN=4$ superspace.  

However, such Wilson loops suffer from UV-divergences arising through the
light-like polygonal contours and their cusps \cite{Drummond2010b}. They parallel the
IR-divergences of the massless scattering amplitudes.
The standard and very convenient procedure to deal with these divergences
is dimensional reduction. 
Other regularization schemes one may choose for Wilson loops are framing and boxing. 
The first prevents the divergences by forcing the propagators to end on different contours, 
while the second defines a finite ratio of Wilson loops \cite{Alday:2010ku}.
Further regularizations for scattering amplitudes are mass regularization, 
proposed in ref.~\cite{Alday:2009zm} and spectral regularization proposed in ref.~\cite{Ferro:2013dga}, 
but they do not appear to have a simple correspondent for Wilson loops.
Unfortunately, so far no regulator is known that preserves all the superconformal and Yangian
symmetries of the tree-level amplitudes.
In contrast, for the smooth non-null Wilson loops of \eqref{eq:ourloop} 
discussed in \cite{Ooguri:2000ps,Muller:2013rta,Beisert:2015jxa}, 
we expect the superconformal and non-local Yangian symmetry to be non-anomalous as the 
correlator is finite for smooth paths.
We therefore consider them highly natural observables to study within the context of
AdS/CFT integrability.

\medskip

Our paper is organized as follows. In \namedref{section}{sec:action} the action of the superconformal and
level-one Yangian generators on the Wilson line operator is constructed. 
The level-one generators contain
a local and a  bi-local piece, where the latter is given by the path-ordered
product of two level-zero generators. It is shown that gauge covariance necessitates 
the departure from a variational derivative representation of the level-zero generators 
acting on the coordinates of the path 
to a gauge-covariant representation which acts on the fields and is importantly 
augmented by a compensating gauge transformation. Moreover, we 
show that the cyclicity of the
Wilson loop is compatible with the level-one Yangian
generators provided that the dual Coxeter number of the level-zero algebra vanishes
(which it does for $\alg{psu}(2,2|4)$) and a novel so-called
G-identity holds for a particular contraction of the field strength two-form with the
conformal Killing vectors. 
In \namedref{section}{sec:Yalgebra} the level-one Yangian algebra for our gauge-covariant
generators is established and the compatibility with kappa-symmetry is proven. 
In \namedref{section}{sec:superspace} we start out with a brief review of the superspace formalism 
and the results for the two-point functions of our previous work \cite{Beisert:2015jxa}. 
Then the representation of the superconformal algebra 
in our $\superN=4$ non-chiral extended superspace $\{x,\theta,\bar\theta;q\}$ is given
and the above mentioned G-identity is proven. 
We show in \namedref{section}{sec:oneloop} the Yangian invariance of the
vacuum expectation value of an arbitrary smooth super Wilson loop at first perturbative order
to all orders in anti-commuting variables. In order to do so it is mandatory to regulate the
bi-local part via point splitting in the coincidence limit. This procedure induces the form
of the local part of the level-one generators which we construct for all algebra elements.
Finally, in \namedref{section}{sec:conclusions} we conclude.

\section{Yangian Action on Wilson Lines}
\label{sec:action}

In the following we discuss the action of Yangian generators
on Wilson line operators in the classical theory.
We will take a generic gauge theory rather than $\superN=4$ super Yang--Mills
as the starting point for two reasons: 
On the one hand, it will streamline the discussion, and let us focus on 
the features of the Yangian algebra. 
On the other hand, it allows us to see more clearly
which of the many exceptional features of $\superN=4$ super Yang--Mills theory
are responsible for its integrability.

The Wilson line is defined by
\[
\swilson = \pathord \exp \int_\gamma \sgauge,
\qquad
\der \swilson = \swilson \sgauge(1) - \sgauge(0) \swilson,
\]
where $\gamma:\tau\mapsto X(\tau)$ is an (open) path between $\tau=0$ and $\tau=1$ (for concreteness)
and $\sgauge$ is the gauge connection one-form (pulled back to the path).
We consider a general spacetime with coordinates $X^{\idxall{A}}$
on which a conformal algebra can act.
The gauge connection can be expanded as $\sgauge = \der X^\idxall{A} \sgauge_\idxall{A}$
in terms of the flat vielbein $\der X^\idxall{A}$.
For the sake of simplicity
we shall assume the coordinates to be bosonic throughout this chapter
in order to avoid a proliferation of sign factors.
The latter can be reinstated for a graded spacetime according to 
simple rules, see \secref{sec:superspace}.

\subsection{Conformal Action on the Wilson Line}
\label{sec:insertions}

We start by defining the action of the
conformal generators, which constitute the zeroth level of the Yangian algebra,
on the Wilson line.
In fact, there are several options for their definition
which, ultimately, differ in boundary contributions.
Unfortunately, such boundary terms easily spoil Yangian symmetry,
by a finite or even an infinite amount.
Therefore, a precise definition of the action of
the conformal and Yangian generators is crucial.

\paragraph{Action on the path.}

A finite element $G$ of the conformal group
acts on the Wilson loop by transforming the path
\[\label{eq:geoact}
G\swilson_{\gamma} = \swilson_{G\gamma}.
\]
For the conformal algebra we consider
an infinitesimal group element $G=1+\epsilon \gen{J}+\order\brk{\epsilon^2}$
with an algebra generator $\gen{J}$ and the expansion parameter $\epsilon$.
Supposing that the generator $\gen{J}$ acts on the path $X(\tau)$ by the variational operator $\genpath{J}$
we can write the action as
\[\label{eq:actbypath}
\genpath{J}=\int \der\tau\, \genpath{J}(\tau),
\qquad
\genpath{J}(\tau) = \gen{J}X^{\idxall{A}}(X(\tau))\, \frac{\delta}{\delta X^{\idxall{A}}(\tau)}\,.
\]
With abuse of notation, $\gen{J}X^{\idxall{A}}(X)$ 
denotes the infinitesimal conformal displacement 
of the point $X^{\idxall{A}}$. In the conformal action $\genpath{J}(\tau)$
the latter is evaluated on the path $X(\tau)$.
This definition and an analogous one for the Yangian algebra
was used in \cite{Muller:2013rta} to discuss Yangian invariance of smooth Wilson loops.

Next we act with this generator on a Wilson line.
The generator $\genpath{J}(\tau)$ acts on 
the gauge connection $\sgauge(\sigma)$ via the path element 
$\der X^\idxall{A}=\der\sigma\, \partial_\sigma X^\idxall{A}$
as well as 
the evaluation point $X$ of the gauge field $\sgauge_\idxall{A}(X)$
\[\label{eq:JtauAsigma}
\genpath{J}(\tau) \sgauge(\sigma)
=
\der \sigma\,\gen{J}X^{\idxall{A}}(\tau)
\lrbrk{
\sgauge_{\idxall{A}}(\sigma)\partial_\sigma \delta(\tau-\sigma)
+\partial_\sigma X^{\idxall{B}}(\sigma)\del_{\idxall{A}}\sgauge_{\idxall{B}}(\sigma)\delta(\tau-\sigma)
}.
\]
It yields the derivative of a delta-function which should be integrated by parts.
This leads to boundary terms which will turn out to be troublesome,
in particular for the Yangian action. 
To understand this problem better, 
let us consider the integral of such a term
over the bounded domains 
$\sigma_-<\sigma<\sigma_+$ and $\tau_-<\tau<\tau_+$
\[\label{eq:deltaprimeintegral}
I:=
\int \der\tau \, \der\sigma 
\bigsbrk{\theta(\tau-\tau_-)-\theta(\tau-\tau_+)}
\bigsbrk{\theta(\sigma-\sigma_-)-\theta(\sigma-\sigma_+)}
\partial_\sigma\delta(\tau-\sigma)\,
f(\tau,\sigma).
\]
Here we have implemented the boundaries of the integration domains by means
of the unit step function $\theta(\tau)$.
Integration by parts yields $I=I_1+I_2$ with 
\begin{align}
I_1&=
-\int \der\tau \, \der\sigma 
\bigsbrk{\theta(\tau-\tau_-)-\theta(\tau-\tau_+)}
\bigsbrk{\theta(\sigma-\sigma_-)-\theta(\sigma-\sigma_+)}
\delta(\tau-\sigma)\,
\partial_\sigma f(\tau,\sigma)
\nln &=
-\int \der\sigma
\bigsbrk{\theta(\sigma-\tau_-)-\theta(\sigma-\tau_+)}
\bigsbrk{\theta(\sigma-\sigma_-)-\theta(\sigma-\sigma_+)}
(\partial_\sigma f)(\sigma,\sigma),
\\
I_2&=
-
\int \der\tau \, \der\sigma 
\bigsbrk{\theta(\tau-\tau_-)-\theta(\tau-\tau_+)}
\bigsbrk{\delta(\sigma-\sigma_-)-\delta(\sigma-\sigma_+)}
\delta(\tau-\sigma)\,
f(\tau,\sigma)
\nln &=
-
\int \der\tau \, 
\bigsbrk{\theta(\tau-\tau_-)-\theta(\tau-\tau_+)}
\bigsbrk{\delta(\tau-\sigma_-)-\delta(\tau-\sigma_+)}
f(\tau,\tau)
\nln &=
-
\bigsbrk{\theta(\sigma_--\tau_-)-\theta(\sigma_--\tau_+)}
f(\sigma_-,\sigma_-)
+
\bigsbrk{\theta(\sigma_+-\tau_-)-\theta(\sigma_+-\tau_+)}
f(\sigma_+,\sigma_+)
\nln &=
-\theta(\sigma_--\tau_-)
f(\sigma_-,\sigma_-)
+\theta(\tau_+-\sigma_+)
f(\sigma_+,\sigma_+).
\end{align}
In the last line we have assumed that $\sigma_-,\tau_-< \sigma_+, \tau_+$.
The final result implies that the above action $\genpath{J} \swilson$ on the Wilson line
is ambiguous because, 
naively, the integration domains for $\genpath{J}(\tau)$ and $\sgauge(\sigma)$ should coincide. 
Equating $\sigma_\pm=\tau_\pm$ leads to the ill-defined
value $\theta(0)$ at both integration boundaries.
While this problem might be avoidable for conformal transformations
of closed Wilson loops, it unfortunately does affect the Yangian action.
In fact, these problems are reminiscent of or even related
to the ambiguities in defining the Yangian algebra 
in a relativistic non-linear sigma model 
identified in \cite{Maillet:1985fn}.

In our case, one might be able to resolve the ambiguity,
and thus cure the above problems
by specifying more precisely how to treat the endpoints.
Nevertheless, we will refrain from using this framework
and instead use operator insertions to represent conformal and
Yangian transformations.

\paragraph{Action by insertion.}

An alternative form of the conformal action on a Wilson line is by operator insertions
\[\label{eq:confold}
\genfield{J}\swilson
= \int \der\tau\, \swilson_{[0,\tau]} \, (\genfield{J}\sgauge)(\tau)\,\swilson_{[\tau,1]}
= \swilson[\genfield{J}\sgauge]
.
\]
Here, $\genfield{J}\sgauge$ is the action of the global conformal generator $\gen{J}$
on the one-form field $\sgauge$, see below.
The conformal variation $\genfield{J}\swilson$  of a Wilson line $\swilson$
is thus a Wilson line $\swilson[Q]$
with one insertion of the one-form local operator $Q(\tau)=\genfield{J}\sgauge(X(\tau))$
\[\label{eq:oneinsertion}
\swilson[Q] := \int \swilson_{[0,\tau]} \, Q(\tau)\,\swilson_{[\tau,1]}.
\]

The operator $\genfield{J}$ acts on the field $\sgauge_\idxall{A}$
in analogy to the action on the path in \eqref{eq:actbypath} 
\[
\genfield{J} =\int \der^d X\,\bigbrk{
-\gen{J}X^{\idxall{B}}\,\del_\idxall{B}\sgauge_\idxall{A}
-\del_\idxall{A} (\gen{J} X^\idxall{B}) \,\sgauge_\idxall{B}
}
\frac{\delta}{\delta\sgauge_\idxall{A}(X)}
\,.
\]
Effectively, it acts by transforming the argument $X^\idxall{B}$ as well as the index $_\idxall{A}$
\[
\genfield{J} \sgauge_\idxall{A}
=
-\gen{J}X^{\idxall{B}}\,\del_\idxall{B}\sgauge_\idxall{A}
-\del_\idxall{A} (\gen{J} X^\idxall{B}) \,\sgauge_\idxall{B}
.
\label{eqn:defgenonfields}
\]
Its action $\genfield{J}\sgauge$ equals \emph{minus}%
\footnote{\label{fn:theminus}%
The minus sign in the definition is required 
to match the Lie algebra of field operators $\genfield{J}$ (which act on the field variables)
with the algebra of derivative operators $\gen{J}$ (which act on coordinates).
Note that this difference effectively flips the order or operators:
$\genfield{J}_1\genfield{J}_2\Psi = -\genfield{J}_1\gen{J}_2\Psi
= -\gen{J}_2\genfield{J}_1\Psi = \gen{J}_2\gen{J}_1\Psi$.}
the action of the derivative operator
(conformal Killing vector)
$\gen{J}=\gen{J}X^\idxall{A} \del_\idxall{A}$ on the one-form field $\sgauge$
(by means of the Lie derivative)
\begin{align}
\genfield{J}\sgauge&
=
-\der X^\idxall{A} \,\gen{J}\sgauge_\idxall{A}
- \gen{J}(\der X^\idxall{B}) \,\sgauge_\idxall{B}
\nln&
=
-\der X^\idxall{A} \,\gen{J}X^{\idxall{B}}\,\del_\idxall{B}\sgauge_\idxall{A}
-\der (\gen{J} X^\idxall{B}) \,\sgauge_\idxall{B}
.
\end{align}
The conformal algebra is expressed in terms
of the structure constants
$\comm{\gen{J}^\idxconf{a}}{\gen{J}^{\idxconf{b}}} = 
f^{\idxconf{a}\idxconf{b}}{}_{\idxconf{c}}\gen{J}^{\idxconf{c}}$.
By construction the conformal action on the field satisfies the same algebra
\[\label{eq:confplain}
\comm{\genfield{J}^\idxconf{a}}{\genfield{J}^{\idxconf{b}}} = 
f^{\idxconf{a}\idxconf{b}}{}_{\idxconf{c}}\genfield{J}^{\idxconf{c}}.
\]
%

\paragraph{Comparison.}

Let us compare the two above definitions of the conformal action.
To compute the action on the former, we continue at \eqref{eq:JtauAsigma},
use translation invariance of the delta-function,
$\partial_\sigma \delta(\tau-\sigma)=- \partial_\tau \delta(\tau-\sigma)$,
and subsequently perform integration by parts on $\tau$
\begin{align}
\label{eq:comparepathfield}
\genpath{J}\swilson &=
\int \der\tau \, \der\sigma \,
\gen{J}X^{\idxall{A}}(\tau)
\swilson_{[0,\sigma]}
\lrbrk{
\sgauge_{\idxall{A}}(\sigma)\partial_\sigma \delta(\tau-\sigma)
+\partial_\sigma X^{\idxall{B}}(\sigma)\del_{\idxall{A}}\sgauge_{\idxall{B}}(\sigma)\delta(\tau-\sigma)
}
\swilson_{[\sigma,1]}
\nln &\simeq
\int \der\tau \, \der\sigma \,\partial_\tau \bigbrk{\gen{J}X^{\idxall{A}}(\tau)}
\delta(\tau-\sigma)
\swilson_{[0,\sigma]}
\sgauge_{\idxall{A}}(\sigma)
\swilson_{[\sigma,1]}
\nln &\qquad
+\int \der\tau \, \der\sigma \,\gen{J}X^{\idxall{A}}(\tau)
\delta(\tau-\sigma)
\swilson_{[0,\sigma]}
\partial_\sigma  X^{\idxall{B}}(\sigma)\del_{\idxall{A}}\sgauge_{\idxall{B}}(\sigma)
\swilson_{[\sigma,1]}
\nln &=
\int \der\tau \,
\swilson_{[0,\tau]}
\lrbrk{ 
\partial_\tau(\gen{J}X^{\idxall{A}})\sgauge_{\idxall{A}}
+\gen{J}X^{\idxall{A}}\partial_\tau X^{\idxall{B}}\del_{\idxall{A}}\sgauge_{\idxall{B}}
}
\swilson_{[\tau,1]}
\nln
&=-\swilson[\genfield{J}\sgauge]
= -\genfield{J}\swilson
.
\end{align}
We thus obtain the equivalence to the action on fields,%
\footnote{The minus sign in \eqref{eq:comparepathfield} 
is equivalent to the one in \eqref{eqn:defgenonfields} 
discussed in \namedref{footnote}{fn:theminus}.
Converting $\genpath{J}$ to $\genfield{J}$ effectively flips the ordering:
$\genpath{J}_1\genpath{J}_2\swilson 
=-\genpath{J}_1\genfield{J}_2\swilson 
=-\genfield{J}_2\genpath{J}_1\swilson 
=\genfield{J}_2\genfield{J}_1\swilson $.
The sign thus ensures consistency of the algebras.
}
modulo boundary terms which have been discussed above following
\eqref{eq:deltaprimeintegral}.

\subsection{Gauge Covariance}

The above form of the conformal action makes direct reference
to the gauge potential $\sgauge$ which is not gauge-covariant by itself.
This will later lead to problems in the definition of the Yangian
action. Let us therefore consider the gauge transformations in detail.

\paragraph{Gauge transformation.}

An infinitesimal gauge transformation $U=1+\Lambda+\order\brk{\Lambda^2}$ acts 
by the operator $\gaugegen[\Lambda]$ on
the gauge field $\sgauge$ as
\[\label{eq:totalder}
\gaugegen[\Lambda] \sgauge:=\cder\Lambda=\der\Lambda + \comm{\sgauge}{\Lambda}.
\]
For a plain Wilson line we therefore obtain
an insertion of the total covariant derivative which can be integrated
and leads to two boundary terms
\[
\label{eq:WilsonGauge}
\gaugegen[\Lambda] \swilson 
= \swilson[\cder\Lambda]
= -\Lambda(0)\swilson+\swilson\Lambda(1).
\]
This is in agreement with the finite gauge transformation of a Wilson line.
Consequently, a closed Wilson loop is a gauge-invariant quantity
\[
\gaugegen[\Lambda] \tr \swilson = \tr \comm{\swilson}{\Lambda} = 0.
\]

Next, let us consider a Wilson line $\swilson[Q]$ with some insertion $Q$.
The gauge transformation can hit a gauge field of the Wilson line
which may be to the left or to the right of the insertion,
or it can hit the insertion itself. We thus obtain three terms
\[
\gaugegen[\Lambda] \swilson[Q] = \swilson[\cder\Lambda;Q]+\swilson\bigsbrk{\gaugegen[\Lambda] Q}+\swilson[Q;\cder\Lambda].
\]
Here and in the following $\swilson[Q;Q']$ represents a Wilson loop with
two \emph{ordered} insertions $Q$ and $Q'$
\[\label{eq:twoinsertions}
\swilson[Q;Q'] := \int_{\tau<\tau'}
\swilson_{[0,\tau]}\, Q(\tau)\, \swilson_{[\tau,\tau']}\, Q'(\tau')\,\swilson_{[\tau',1]}.
\]
The above terms can be integrated to the expression
\begin{align}
\gaugegen[\Lambda] \swilson[Q] &=
-\Lambda(0)\swilson[Q]
+\swilson[\Lambda Q]
+\swilson\bigsbrk{\gaugegen[\Lambda] Q}
-\swilson[Q\Lambda]
+\swilson[Q]\Lambda(1)
\nln &=
-\Lambda(0)\swilson[Q]+\swilson[Q]\Lambda(1)
+\swilson\bigsbrk{\gaugegen[\Lambda] Q-\comm{Q}{\Lambda}}
.
\end{align}
In particular, the last term drops out if
the insertion is gauge-covariant, $\gaugegen[\Lambda] Q = \comm{Q}{\Lambda}$.
In that case, the Wilson line with insertion transforms according to the
same rule as the Wilson line without insertion
\eqref{eq:WilsonGauge};
moreover the Wilson loop with insertion is gauge-invariant.

\paragraph{Conformal transformation.}

Now the conformal action makes use of insertions
of the kind $\genfield{J}\sgauge$.
The latter is \emph{not} a gauge-covariant insertion; instead it transforms as
\[
\label{eq:GaugeGenInsertion}
\gaugegen[\Lambda](\genfield{J}\sgauge)=\comm{\genfield{J}\sgauge}{\Lambda}-\cder(\gen{J}\Lambda).
\]
One therefore obtains two extra boundary terms
in the gauge transformation of the conformal transformation
of the Wilson line:
\[\label{genJgauge}
\gaugegen[\Lambda] \swilson\sbrk{\genfield{J}\sgauge}
= -\Lambda(0)\swilson\sbrk{\genfield{J}\sgauge}
+\swilson\sbrk{\genfield{J}\sgauge}\Lambda(1)
+(\gen{J}\Lambda)(0) \swilson
-\swilson (\gen{J}\Lambda)(1)
.
\]
The additional boundary terms are interpreted as follows:
Under a gauge transformation, the Wilson line changes
by the gauge parameter field $\Lambda$
evaluated at its two ends.
Under a conformal transformation,
the gauge parameter field effectively changes
by $\gen{J}\Lambda$.%
\footnote{In the above discussion of the action on the path, 
this corresponds to including
the endpoints in the conformal transformation.}
All the boundary terms conveniently cancel for a closed Wilson loop
\[
\gaugegen[\Lambda] \tr \swilson\sbrk{\genfield{J}\sgauge}
= \tr\bigcomm{\swilson\sbrk{\genfield{J}\sgauge}}{\Lambda}
-\tr\comm{\swilson}{\gen{J}\Lambda}
=0.
\]
Consequently, the conformal transformation of a
gauge-invariant Wilson loop is still gauge-invariant.

\paragraph{Compensating gauge transformation.}

The extra terms in \eqref{genJgauge} can be avoided by
supplementing each conformal transformation $\genfield{J}$
by a compensating gauge transformation.
In the case of gauge-invariant observables,
such an additional gauge transformation will not make a difference.
We choose a gauge transformation parametrized by the field
\[
\Lambda_{\gen{J}} = \gen{J}X^{\idxall{A}}\, \sgauge_{\idxall{A}}.
\]
This is just the natural shift of gauge for a shift 
of coordinates $X^{\idxall{A}}$ by $\gen{J}X^{\idxall{A}}$.
Noting that $\Lambda_{\gen{J}}$ depends on the gauge potential, 
this field changes under an additional gauge transformation $\Lambda$ by
\[
\gaugegen[\Lambda] \Lambda_{\gen{J}} = \gen{J}X^{\idxall{A}}\, \cdel_{\idxall{A}} \Lambda,
\]
which by itself almost cancels
the additional terms in \eqref{genJgauge}.

We thus introduce the generator $\gaugegen[\gen{J}X^{\idxall{A}}\sgauge_{\idxall{A}}]$ of
compensating gauge transformations 
which acts on the gauge potential by 
\[
\gaugegen[\gen{J}X^{\idxall{B}}\sgauge_{\idxall{B}}] \sgauge_\idxall{A} 
= \cdel_\idxall{A} \bigbrk{\gen{J}X^{\idxall{B}}\, \sgauge_{\idxall{B}}}.
\]
In combination with the plain conformal transformation we obtain 
a pretty combination of terms
\begin{align}
\lrbrk{\genfield{J}+\gaugegen[\gen{J}X^{\idxall{B}}\sgauge_{\idxall{B}}]} \sgauge_\idxall{A}
&=
-\gen{J}X^{\idxall{B}}\del_\idxall{B}\sgauge_\idxall{A}
-\del_\idxall{A} (\gen{J} X^\idxall{B}) \,\sgauge_\idxall{B}
 + \cdel_\idxall{A} \bigbrk{\gen{J}X^{\idxall{B}} \sgauge_{\idxall{B}}}
\nln &=
-\gen{J}X^{\idxall{B}}\del_\idxall{B}\sgauge_\idxall{A}
+ \gen{J}X^{\idxall{B}} \del_{\idxall{A}}\sgauge_{\idxall{B}}
+ \gen{J}X^{\idxall{B}}\bigcomm{ \sgauge_{\idxall{A}}}{ \sgauge_{\idxall{B}}}
\nln &=
-\gen{J}X^{\idxall{B}}  \sfstr_{\idxall{B}\idxall{A}}.
\end{align}
Here $\sfstr_{\idxall{A}\idxall{B}}
=\del_{\idxall{A}}
\sgauge_{\idxall{B}}-\del_{\idxall{B}} \sgauge_{\idxall{A}}+\comm{\sgauge_{\idxall{A}}}{\sgauge_{\idxall{B}}}$
is the gauge field strength associated to $\sgauge$, and it is manifestly
gauge-covariant.
The violation of gauge covariance of $\genfield{J}\sgauge$ 
is fully compensated by the gauge transformation $\gaugegen[\gen{J}X^{\idxall{A}}\sgauge_{\idxall{A}}]\sgauge$.

\subsection{Gauge-Covariant Conformal Action}

It therefore makes sense to introduce a composite conformal action
\[
\label{eq:Jhatdef}
\genfull{J}:=\genfield{J} + \gaugegen[\gen{J}X^{\idxall{A}}\sgauge_{\idxall{A}}].
\]
On covariant fields it acts as (minus) the gauge-covariant conformal 
transformation $\gengauge{J} := \gen{J}X^\idxall{A} \cdel_\idxall{A}$.
On the gauge field itself (which is not a covariant field) it also 
produces a gauge-covariant combination, namely the field strength
\[\label{eq:fullconfacta}
\genfull{J}\sgauge_\idxall{A}=\gen{J}X^{\idxall{B}}  \sfstr_{\idxall{A}\idxall{B}}.
\]

\paragraph{Action on the Wilson line.}

The gauge-covariant conformal transformation thus acts on a
Wilson line by the insertion of a gauge field strength
\[\label{eq:confnew}
\genfull{J} \swilson
=
\swilson\sbrk{\genfull{J}\sgauge}
=
\swilson\bigsbrk{-\gen{J} X^{\idxall{A}} \der X^{\idxall{B}} \sfstr_{\idxall{A}\idxall{B}}}
.
\]
This form is to be compared to the straight conformal transformation
$\genfield{J} \swilson := \swilson\sbrk{\genfield{J}\sgauge}$ in \eqref{eq:confold}.
The difference between the two prescriptions is a gauge transformation
which acts as the insertion of a total covariant derivative operator
into the Wilson line
\[
\genfull{J} \swilson-\genfield{J}\swilson
=
\gaugegen[\gen{J}X^{\idxall{A}}\sgauge_{\idxall{A}}]\swilson
=
\swilson\bigsbrk{\cder \brk{\gen{J}X^{\idxall{A}} \sgauge_{\idxall{A}}}}
=
-\brk{\gen{J}X^{\idxall{A}} \sgauge_{\idxall{A}}}(0)\swilson + \swilson\brk{\gen{J}X^{\idxall{A}} \sgauge_{\idxall{A}}}(1).
\]
The difference thus has a simple structure.
In particular, this shows that the two conformal actions
on closed Wilson loops are equal
\[
\genfield{J}\tr \swilson=\genfull{J} \tr \swilson,
\]
which was to be expected because the closed Wilson loop is gauge-invariant.
For the purpose of conformal transformations
it therefore does not matter which of the two actions 
$\genfield{J}$ in \eqref{eq:confold} vs.\ $\genfull{J}$ in \eqref{eq:confnew}
is chosen. However, their local actions are different, 
and this difference is very important 
for the Yangian action to be discussed below.

In fact, it turns out that in concrete calculations,
the action of $\genfield{J}$ leads
to much more convenient expressions
and far fewer boundary terms to be considered
than $\genfull{J}$.
Conversely, the action of $\genfull{J}$ has the benefit
of maintaining gauge covariance.
The latter will prove to be the stronger issue
when it comes to the Yangian action.

\paragraph{Gauge-covariant algebra.}

Let us consider the algebra of the gauge-covariant conformal actions. 
When making one of the generators in \eqref{eq:confplain}
gauge-covariant we obtain the following
mixed algebra 
\[\label{eq:confmixed}
\comm{\genfield{J}^\idxconf{a}}{\genfull{J}^{\idxconf{b}}} = 
f^{\idxconf{a}\idxconf{b}}{}_{\idxconf{c}}\genfull{J}^{\idxconf{c}}.
\]
It is straight-forward to verify this relation by acting on the fundamental field
$\sgauge$.
For the algebra of gauge-covariant actions there is, however, 
an additional term on the right hand side
\[
\label{eq:confcomm}
\comm{\genfull{J}^\idxconf{a}}{\genfull{J}^\idxconf{b}} \sgauge
=
f^{\idxconf{a}\idxconf{b}}{}_\idxconf{c} \genfull{J}^\idxconf{c} \sgauge
- \cder \sconfrem^{\idxconf{a}\idxconf{b}},
\]
with
\[
\sconfrem^{\idxconf{a}\idxconf{b}}
=\gen{J}^\idxconf{a} X^{\idxall{A}}\, \gen{J}^\idxconf{b} X^{\idxall{B}}\, \sfstr_{\idxall{A} \idxall{B}}
.
\]
The first term represents the conformal algebra,
the second term is a gauge transformation with gauge parameter
$-\sconfrem^{\idxconf{a}\idxconf{b}}$
\[
\comm{\genfull{J}^\idxconf{a}}{\genfull{J}^\idxconf{b}} 
=
f^{\idxconf{a}\idxconf{b}}{}_\idxconf{c} \genfull{J}^\idxconf{c} 
- \gaugegen[\sconfrem^{\idxconf{a}\idxconf{b}}].
\]

In fact, this result is not surprising because our action is a combination
of a conformal and a gauge transformation.
The algebra thus closes onto a conformal transformation
and a gauge transformation.
On gauge-invariant objects, the algebra reduces to the
conformal algebra.

Furthermore, the appearance of gauge transformations
in the algebra of spacetime transformations
is a standard feature of gauge theories with extended supersymmetry.
For instance, in components there is the well-known
algebra relation among the chiral supercharges
\[
\acomm{\gen{Q}}{\gen{Q}}\gauge_{\mu} \sim \cdel_{\mu}\scalar.
\]
The resulting term can be interpreted as a gauge transformation
of $\gauge$ by $\scalar$
(both of which are top components of 
the superspace fields $\sgauge$ and $\sconfrem$, respectively).
Therefore it is very natural to use the
composite conformal action $\genfull{J}$
instead of the plain one $\genfield{J}$.

\paragraph{Comparison.}

Let us finally compare the gauge-covariant action to the action on the path.
The derivation is analogous to \eqref{eq:comparepathfield}
except that we now perform the integration by parts 
directly on $\sigma$ instead of $\tau$
\begin{align}
\genpath{J}\swilson &=
\int \der\tau \, \der\sigma \,
\gen{J}X^{\idxall{A}}(\tau)
\swilson_{[0,\sigma]}
\lrbrk{
\sgauge_{\idxall{A}}(\sigma)\partial_\sigma \delta(\tau-\sigma)
+\partial_\sigma X^{\idxall{B}}(\sigma)\del_{\idxall{A}}\sgauge_{\idxall{B}}(\sigma)\delta(\tau-\sigma)
}
\swilson_{[\sigma,1]}
\nln &\simeq
-\int \der\tau \, \der\sigma \,\gen{J}X^{\idxall{A}}(\tau)
\delta(\tau-\sigma)\partial_\sigma
\lrbrk{
\swilson_{[0,\sigma]}
\sgauge_{\idxall{A}}(\sigma)
\swilson_{[\sigma,1]}
}
\nln &\qquad
+\int \der\tau \, \der\sigma \,\gen{J}X^{\idxall{A}}(\tau)\delta(\tau-\sigma)
\swilson_{[0,\sigma]}
\partial_\sigma  X^{\idxall{B}}(\sigma)\del_{\idxall{A}}\sgauge_{\idxall{B}}(\sigma)
\swilson_{[\sigma,1]}
\nln &=
-\int \der\tau \,
\swilson_{[0,\tau]}
\gen{J}X^{\idxall{A}}
\partial_\tau X^\idxall{B}
\bigbrk{ 
\lrbrk{ 
\sgauge_{\idxall{B}}
\sgauge_{\idxall{A}}
+\del_\idxall{B}\sgauge_{\idxall{A}}
-
\sgauge_{\idxall{A}}
\sgauge_\idxall{B}
}
-\del_{\idxall{A}}\sgauge_{\idxall{B}}
}
\swilson_{[\tau,1]}
\nln
&=\swilson\bigsbrk{\gen{J}X^{\idxall{A}}
\partial_\tau X^\idxall{B}
\sfstr_{\idxall{A}\idxall{B}}
}
= -\swilson[\genfull{J}\sgauge]
= -\genfull{J}\swilson
.
\end{align}
The difference w.r.t.\ \eqref{eq:comparepathfield}
arises due to the different boundary terms that were discarded. 
This illustrates once again that the conformal action on the path 
$\genpath{J}$ defined in \eqref{eq:actbypath}
has to be used with care due to boundary terms in connection to distributions.

\subsection{Yangian Action on the Wilson Line}

Next we introduce the action of the Yangian level-one generators $\genY{J}$.

\paragraph{Naive action.}

In the framework of quantum algebra, the action of the level-one generator $\genY{J}$
on tensor product objects is defined by the coproduct
\[\label{eq:copro}
\copro(\genY{J}^{\idxconf{c}}) =
\genY{J}^\idxconf{c}\otimes 1+ 1\otimes \genY{J}^\idxconf{c}
+f^\idxconf{c}{}_{\idxconf{b}\idxconf{a}}\gen{J}^\idxconf{a} \otimes \gen{J}^\idxconf{b}.
\]
Wilson line objects can be viewed as continuous tensor products
of infinitesimal line elements. 
In terms of the action on the paths this action can be phrased as
\[
\label{eq:Jhatpathaction}
\genYpath{J}^\idxconf{c} =
\int \der\tau\,
\genYpath{J}^\idxconf{c} (\tau)
+
\int_{\tau<\tau'} \der\tau\,\der\tau'\,
f^\idxconf{c}{}_{\idxconf{b}\idxconf{a}}\,
\genpath{J}^\idxconf{a} (\tau)\, \genpath{J}^\idxconf{b} (\tau')\, ,
\]
dropping the co-product symbol $\Delta$ on the continuous tensor
product along the Wilson line.
As emphasized above, this definition is subtle w.r.t.\ to boundary terms
which will play an important role. In particular, there is now a 
new boundary of the integration domain at coincident insertion points $\tau=\tau'$,
where UV-singularities can arise.
We therefore prefer to use the language of operator insertions,
see \secref{sec:insertions}, in which the Yangian action reads
\[
\label{genYJfirsttry}
\genYfield{J}^\idxconf{c} \swilson =
\swilson\bigsbrk{\genYfield{J}^\idxconf{c} \sgauge}
+
f^\idxconf{c}{}_{\idxconf{b}\idxconf{a}}\swilson\bigsbrk{\genfield{J}^\idxconf{a} \sgauge; \genfield{J}^\idxconf{b} \sgauge}.
\]
The former term is a local action $\genYfield{J} \sgauge$ of the level-one operator which we shall
discard for the time being because we do not yet know its form.
However, the second term
(to be denoted as $\genYfieldnl{J} \swilson$)
is an ordered bi-local insertion of local operators
\eqref{eq:twoinsertions}
combined with the structure constants $f^\idxconf{a}{}_{\idxconf{b}\idxconf{c}}$ of the conformal algebra.
The second term of the Yangian action therefore is fully determined
in terms of the conformal action.

As we have seen above, there are two alternative definitions
$\genfield{J}$ and $\genfull{J}$ for the local conformal action.
In contradistinction to conformal transformations,
the level-one action crucially depends on this choice.

As an aside, we would like to emphasize that the level-one action
on the Wilson line is not a plain Wilson line,
but a Wilson line with two operator insertions.
As such it is in a different class of observables,
potentially with a different divergence structure.
The level-one action can be viewed as the second-order
variational operator \eqref{eq:Jhatpathaction} on path space
which cannot be decomposed into two consecutive first-order variations.

\paragraph{Gauge transformations.}

Let us apply a gauge transformation $\gaugegen[\Lambda]$ to the
bi-local contribution of the above level-one action
\begin{align}\label{genYJfirsttrygauge}
\gaugegen[\Lambda]  \genYfieldnl{J}^\idxconf{c} \swilson & =
f^\idxconf{c}{}_{\idxconf{b}\idxconf{a}}
\gaugegen[\Lambda]  \swilson\bigsbrk{\gen{J}^\idxconf{a} \sgauge; \gen{J}^\idxconf{b} \sgauge}
\nln &=
-\Lambda(0)\genYfieldnl{J}^\idxconf{c} \swilson
+\genYfieldnl{J}^\idxconf{c} \swilson \Lambda(1)
\nln &\qquad
+f^\idxconf{c}{}_{\idxconf{b}\idxconf{a}}
\swilson\bigsbrk{\cder\brk{\gen{J}^\idxconf{a} \Lambda}; \gen{J}^\idxconf{b} \sgauge}
+f^\idxconf{c}{}_{\idxconf{b}\idxconf{a}}
\swilson\bigsbrk{\gen{J}^\idxconf{a} \sgauge;\cder\brk{\gen{J}^\idxconf{b} \Lambda}}
\nln &=
-\Lambda(0)\genYfieldnl{J}^\idxconf{c} \swilson
+\genYfieldnl{J}^\idxconf{c} \swilson \Lambda(1)
\nln &\qquad
+f^\idxconf{c}{}_{\idxconf{b}\idxconf{a}}
\swilson\bigsbrk{
\gen{J}^\idxconf{a} \Lambda\, \gen{J}^\idxconf{b} \sgauge
-\gen{J}^\idxconf{a} \sgauge \,\gen{J}^\idxconf{b} \Lambda
}
\nln &\qquad
-f^\idxconf{c}{}_{\idxconf{b}\idxconf{a}}
\gen{J}^\idxconf{a} \Lambda(0) \swilson\sbrk{\gen{J}^\idxconf{b} \sgauge}
+f^\idxconf{c}{}_{\idxconf{b}\idxconf{a}}
\swilson\sbrk{\gen{J}^\idxconf{a} \sgauge}\gen{J}^\idxconf{b} \Lambda(1)
.
\end{align}
Beyond the usual terms, we find extra local terms
as well as terms which involve simultaneous insertions in the bulk and at the boundary.
Unfortunately, these terms do not cancel for a closed Wilson loop
\[\label{eq:firsttrygaugeyangianclosed}
\gaugegen[\Lambda]  \tr (\genYfieldnl{J}^\idxconf{c} \swilson) =
f^\idxconf{c}{}_{\idxconf{b}\idxconf{a}}
 \tr \swilson\bigsbrk{\acomm{\gen{J}^\idxconf{a} \Lambda}{\gen{J}^\idxconf{b} \sgauge}}
-2f^\idxconf{c}{}_{\idxconf{b}\idxconf{a}}
 \tr \bigbrk{\gen{J}^\idxconf{a} \Lambda \,\swilson\sbrk{\gen{J}^\idxconf{b} \sgauge}}
.
\]
Here we have used the anti-symmetry of the gauge algebra structure constants
$f^\idxconf{c}{}_{\idxconf{b}\idxconf{a}}$, which produce anti-commutators instead of commutators.
This result shows that the above definition of the Yangian action
does not respect gauge symmetry.
One might hope that additional terms in the definition of the level-one
action could repair gauge invariance.
It is conceivable that a suitable definition of the omitted local term in \eqref{genYJfirsttry}
will cancel the local contribution in \eqref{eq:firsttrygaugeyangianclosed}.
However, there is no obvious resolution for the second bulk-boundary term
in \eqref{eq:firsttrygaugeyangianclosed}.

Therefore, it is evident that the definition \eqref{genYJfirsttry}
does not respect gauge symmetry. It maps a gauge-invariant object
$\tr \swilson$ to something which is not gauge-invariant and which
consequently cannot serve as an observable in a gauge theory.

\paragraph{Gauge-covariant action.}

Here is where the alternative form of the local conformal action $\genfull{J}$ 
defined in \eqref{eq:Jhatdef}
comes into play. Omitting the local term, we may as well define the Yangian level-one
action as
\begin{align}
\label{eq:newyangact}
\genYfullnl{J}^\idxconf{c} \swilson
:=\mathord{}&
f^\idxconf{c}{}_{\idxconf{b}\idxconf{a}}
\swilson\bigsbrk{\genfull{J}^\idxconf{a} \sgauge; \genfull{J}^\idxconf{b} \sgauge}
\nln
=\mathord{}&
f^\idxconf{c}{}_{\idxconf{b}\idxconf{a}}
\swilson\bigsbrk{\gen{J}^\idxconf{a} X^{\idxall{A}} \der X^{\idxall{B}} \sfstr_{\idxall{A}\idxall{B}};
\gen{J}^\idxconf{b} X^{\idxall{C}} \der X^{\idxall{D}} \sfstr_{\idxall{C}\idxall{D}}}.
\end{align}
This expression is gauge-covariant by construction. We therefore have
\[
\gaugegen[\Lambda] \tr \bigbrk{\genYfullnl{J}^\idxconf{c} \swilson} = 0.
\]
Note that the difference between the two definitions can
be written in terms of local and bulk-boundary terms.
These are precisely the ones to repair gauge covariance.

\subsection{Local Terms and Renormalization}
\label{sec:localrenorm}

So far we have ignored a potential local term in the level-one action
\[
\genYfull{J}^\idxconf{c} \swilson =
\swilson\bigsbrk{\genYfull{J}^\idxconf{c} \sgauge}
+
f^\idxconf{c}{}_{\idxconf{b}\idxconf{a}}
\swilson\bigsbrk{\genfull{J}^\idxconf{a} \sgauge; \genfull{J}^\idxconf{b} \sgauge}.
\]
A local term may become important in case the two insertions
generate additional UV-divergences which are not present
in the original Wilson line and which need to be renormalized.
One can see this as follows:
Suppose we impose a point splitting regularization for the
Yangian action. We could for instance define the
bi-local term with a local cut-off
\[
\swilson[Q;Q']_\varepsilon :=
\int_{\tau'>\tau+\varepsilon}
\swilson_{[0,\tau]}\, Q(\tau)\, \swilson_{[\tau,\tau']}\, Q'(\tau')\,\swilson_{[\tau,1]}.
\]
This expression is clearly less susceptible to UV-divergences
because the two insertions are kept at a minimum distance.
Then the missing term
\begin{align}
\swilson[Q;Q'] - \swilson[Q;Q']_\varepsilon  &=
\int_{\tau<\tau'<\tau+\varepsilon}
\swilson_{[0,\tau]}\, Q(\tau)\, \swilson_{[\tau,\tau']}\, Q'(\tau')\,\swilson_{[\tau',1]}
\nln
&\simeq
\int_0^1
\int_{0}^{\varepsilon}
\swilson_{[0,\tau]}\, Q(\tau)\, \swilson_{[\tau,\tau+\sigma]}\, Q'(\tau+\sigma)\,\swilson_{[\tau+\sigma,1]}
\nln
&=
\swilson[R_\varepsilon]
\end{align}
can formally be written as a Wilson line with an insertion of
the non-local operator
\[
R_\varepsilon(\tau)=
\int_{0}^{\varepsilon}
Q(\tau)\, \swilson_{[\tau,\tau+\sigma]}\, Q'(\tau+\sigma) \swilson_{[\tau+\sigma,\tau]}.
\]
For sufficiently small $\varepsilon$ this term is almost local,
i.e.\ the operator product expansion allows to expand it in terms of local operators.
Then it has the same form as the local term in the level-one action.
We can thus write the renormalized level-one action as
\[
\genYfull{J}^\idxconf{c} \swilson =
\swilson\bigsbrk{(\genYfull{J}^\idxconf{c} \sgauge)_\varepsilon}
+
f^\idxconf{c}{}_{\idxconf{b}\idxconf{a}}
\swilson\bigsbrk{\genfull{J}^\idxconf{a} \sgauge; \genfull{J}^\idxconf{b} \sgauge}_\varepsilon.
\]
Here the local term fills the gap in the regularization of
the bi-local term.
Proper renormalization requires
that both terms are finite at finite $\varepsilon$,
and that their sum does not depend on $\varepsilon$.
Therefore the local term must depend on $\varepsilon$
and most likely be divergent in the limit $\varepsilon\to0$.
The latter issue is of no concern because the sum of the two
terms is independent of $\varepsilon$ by construction and thus finite.

\subsection{Superspace and Scalar Couplings}
\label{sec:superspacescalar}

The above construction was based on slightly simplified assumptions.
Let us now discuss how to generalize our results to
extended superspace and to scalar couplings.

\paragraph{Graded coordinates.}

The generalization to superspace consists of two steps.
One step is to introduce graded coordinates.
The above discussion did not make any assumptions on the set of
coordinates $X^{\idxall{A}}$ to be used.
There is nothing that prevents us from
referring to the $\superN=4$ superspace
coordinates $x,\theta,\bar\theta$ collectively by $X$
\[
X^{\idxall{A}}=(x^{\dot\alpha\beta},\theta^{a \beta},\bar\theta^{\dot\alpha}{}_b).
\]
The grading of superspace requires the introduction of various signs in the above.
This can be done according to a regular scheme.%
\footnote{\label{fn:super-contraction}Alternatively, the plain summation convention could
be refined to introduce the signs automatically. 
For example, 
$F^{\idxall{A}\idxall{B}\ldots\idxall{C}}{}_\idxall{A}
:=\sum_{\idxall{A}}(-1)^{|\idxall{A}||\idxall{B}|+\ldots+|\idxall{A}||\idxall{C}|}
F^{\idxall{A}\idxall{B}\ldots\idxall{C}}{}_\idxall{A}$
as well as for the opposite ordering of indices
$F_\idxall{A}{}^{\idxall{B}\ldots\idxall{C}\idxall{A}}
:=\sum_{\idxall{A}}(-1)^{|\idxall{A}|+|\idxall{A}||\idxall{B}|+\ldots+|\idxall{A}||\idxall{C}|}
F_\idxall{A}{}^{\idxall{B}\ldots\idxall{C}\idxall{A}}$
(all under the assumption that $F$ is a bosonic object).
}
We refrain from performing this step because it merely
clutters the expressions without adding to the understanding.
In the concrete one-loop calculations using the 
individual superspace coordinates in subsequent sections,
we will however make the signs explicit.

\paragraph{Superspace torsion.}

The second step is to introduce torsion.
This is normally done through superspace covariant derivatives $\sdel$.
We can write them collectively as
\[
\sdel_{\idxall{A}} = \del_{\idxall{A}}
+ \storsion^{\idxall{C}}_{\idxall{A}\idxall{B}} X^{\idxall{B}} \del_{\idxall{C}}.
\]
Here $\storsion^{\idxall{C}}_{\idxall{A}\idxall{B}}$ is the superspace torsion tensor.
It is (graded) anti-symmetric in the lower indices.
Furthermore, it is non-zero only if the upper index is bosonic 
and the lower indices are both fermionic.
Typically, it is proportional to a gamma-matrix for spacetime
times the identity matrix for internal directions.
It follows immediately that the contraction of two
torsion tensors vanishes
\[\storsion^{\idxall{A}}_{\idxall{B}\idxall{C}}\storsion^{\idxall{E}}_{\idxall{A}\idxall{D}}=0.\]
There is a corresponding superspace translation-invariant vielbein
\[
\sviel^{\idxall{A}}=\der X^{\idxall{A}}+\storsion^{\idxall{A}}_{\idxall{B}\idxall{C}}X^{\idxall{B}} \der X^{\idxall{C}}.
\]
It satisfies the identity
$\der \sviel^{\idxall{A}}=\storsion^{\idxall{A}}_{\idxall{B}\idxall{C}}\sviel^{\idxall{B}}\wedge \sviel^{\idxall{C}}$.
The covariant derivative and vielbein together make up the exterior
derivative
\[
\der = \sviel^{\idxall{A}} \sdel_{\idxall{A}}   = \der X^{\idxall{A}}\del_{\idxall{A}} .
\]

The main effect of torsion to the above construction lies in the interpretation
of the components $\sgauge_{\idxall{A}}$ of the gauge connection
and the components $\sfstr_{\idxall{A}\idxall{B}}$ of the field strength.
In superspace they are normally expanded
in terms of the vielbein $\sviel$ instead
\[
\sgauge=\sviel^{\idxall{A}} \sgauge^\text{cov}_{\idxall{A}},
\qquad
\sfstr=\half \sviel^{\idxall{A}} \wedge \sviel^{\idxall{B}} \sfstr^\text{cov}_{\idxall{A}\idxall{B}}.
\]
In particular, many of the components $\sfstr^\text{cov}_{\idxall{A}\idxall{B}}$ are forced to zero
by constraints of the superspace gauge theory, which will be discussed in \secref{sec:superspace} in more details.
However, one is free to expand in terms of the trivial vielbein $\der X$
\[
\sgauge=\der X^{\idxall{A}} \sgauge_{\idxall{A}},
\qquad
\sfstr=\half \der X^{\idxall{A}} \wedge \der X^{\idxall{B}} \sfstr_{\idxall{A}\idxall{B}}.
\]
Note that the relationship $\sfstr=\der \sgauge + \sgauge\wedge \sgauge$
between the gauge connection and field strength is
the same in both cases.
Only the expansion into components leads to different forms.

Our construction above made use of these plain components
$\sgauge_\idxall{A}$ and $\sfstr_{\idxall{A}\idxall{B}}$.
Therefore, all it takes to generalize the construction to
a space with torsion is to translate between the bases
\begin{align}
\sgauge_{\idxall{A}}&=
\sgauge^\text{cov}_{\idxall{A}}-\storsion^{\idxall{C}}_{\idxall{A}\idxall{B}}X^{\idxall{B}}
 \sgauge^\text{cov}_{\idxall{C}},
\nln
\sfstr_{\idxall{AB}}&=
\sfstr^\text{cov}_{\idxall{A}\idxall{B}}
-\storsion^{\idxall{D}}_{\idxall{A}\idxall{C}}X^{\idxall{C}} \sfstr^\text{cov}_{\idxall{D}\idxall{B}}
-\storsion^{\idxall{F}}_{\idxall{B}\idxall{E}}X^{\idxall{E}} \sfstr^\text{cov}_{\idxall{A}\idxall{F}}
+\storsion^{\idxall{D}}_{\idxall{A}\idxall{C}}X^{\idxall{C}}
\storsion^{\idxall{F}}_{\idxall{B}\idxall{E}}X^{\idxall{E}} \sfstr^\text{cov}_{\idxall{D}\idxall{F}}.
\end{align}
We can write the action of conformal actions either
for flat space or for superspace with torsion
\[
\genfull{J} \sgauge
=-(\gen{J} X^{\idxall{A}}) \der X^{\idxall{B}} \sfstr_{\idxall{A}\idxall{B}}
=-(\gen{J}^\text{cov} X^{\idxall{A}}) \sviel^{\idxall{B}} \sfstr^\text{cov}_{\idxall{A}\idxall{B}},
\]
where
\[
\gen{J}^\text{cov} X^{\idxall{A}} :=
\gen{J} X^{\idxall{A}}+\storsion^{\idxall{A}}_{\idxall{B}\idxall{C}}X^{\idxall{B}}\gen{J}X^{\idxall{C}}.
\]
Note that both forms are equivalent.
Therefore one can effectively perform the following replacement
in the above relations 
\begin{align}
\sgauge_{A}&\to \sgauge^\text{cov}_{\idxall{A}},
&
\der X^{\idxall{A}}&\to \sviel^{\idxall{\idxall{A}}} ,
\nln
\sfstr_{\idxall{A}\idxall{B}}&\to \sfstr^\text{cov}_{\idxall{A}\idxall{B}},
&
\gen{J}X^{\idxall{A}} &\to\gen{J}^\text{cov} X^{\idxall{A}} .
\end{align}
The replacement makes torsion more evident,
but it does not change the content of the relations.

\paragraph{Scalar couplings.}

The Maldacena--Wilson line combines
an ordinary Wilson line with couplings to the scalar fields.
We can write this as
\[
\swilson = \pathord \exp \int_\gamma (\sgauge+\sscalar),
\]
where $\sgauge$ is the ordinary gauge connection and
$\sscalar(\tau)$ is a one-form $\sscalar=Y^{\idxall{M}} \sscalar_{\idxall{M}}$ composed from the
scalar fields $\sscalar_{\idxall{M}}(X(\tau))$ and a coefficient one-form $Y^{\idxall{M}}(\tau)$
describing the path in an internal space corresponding
to the scalar fields; we shall call it the scalar connection.

Let us collect some relevant identities for the scalar connection:
First of all, $\sscalar$ is gauge-covariant
\[
\gaugegen[\Lambda] \sscalar = \comm{\sscalar}{\Lambda}.
\]
Conformal transformations should now also include the index of the scalar field
\begin{align}
 \genfield{J} \sscalar_\idxall{M}
&:= -(\gen{J}X^{\idxall{A}}) \del_{\idxall{A}} \sscalar_\idxall{M} - \gen{J}_\idxall{M}{}^\idxall{N} \sscalar_{\idxall{N}},
\nln
 \genfull{J} \sscalar_\idxall{M}
&:= -(\gen{J}X^{\idxall{A}}) \cdel_{\idxall{A}} \sscalar_\idxall{M} - \gen{J}_\idxall{M}{}^\idxall{N} \sscalar_{\idxall{N}}.
\end{align}
Here, the matrix $\gen{J}_\idxall{M}{}^\idxall{N}$ 
mainly represents the internal rotation symmetry
which does not act on $X$. Note that
the matrix must depend on $X$ in order for the 
supersymmetries and boosts to close properly.
As for the gauge field, the conformal action 
is defined as though the conformal transformation acts on the path.
Here the path includes the scalar components $Y$, and the above transformation matrix 
can be defined via $\gen{J}Y^\idxall{M} =: Y^\idxall{N} \gen{J}_\idxall{N}{}^\idxall{M}$.
The resulting transformation for the scalar connection $\sscalar$ is thus
\[\genfull{J} \sscalar
:= -(\gen{J}X^{\idxall{A}}) \cdel_{\idxall{A}} \sscalar - (\gen{J}Y^\idxall{M}) \sscalar_{\idxall{M}}
= -(\gen{J}^\text{cov}X^{\idxall{A}}) \scdel_{\idxall{A}} \sscalar - (\gen{J}Y^{\idxall{M}}) \sscalar_{\idxall{M}}.
\]
Note that in this expression, $\scdel_{\idxall{A}}$
is assumed to be both gauge and superspace covariant.

Finally, the inclusion of a scalar into the Wilson loop
changes the relationship \eqref{eq:totalder} between 
total derivatives and boundary terms.
Namely, the coupling to the scalar needs to be taken into account:
\[
\swilson\bigsbrk{\cder\Lambda+\comm{\sscalar}{\Lambda}}=-\Lambda(0)\swilson+\swilson\Lambda(1).
\]
This expression confirms the rule that it is usually sufficient to
introduce the scalar connection everywhere by the replacement
(even within covariant derivatives)
\[
\sgauge\to \sgauge+\sscalar.
\]
For the remainder of this section and the following one, 
we shall largely disregard the scalar connection 
because it clutters the relations somewhat 
while it can be reintroduced straight-forwardly.

\subsection{Consistency}
\label{sec:consistency}

An important question to answer is whether the level-one action 
is consistent with the underlying gauge theory. This mainly
concerns the question whether the level-one action on 
two equivalent objects yields two equivalent results. 
One aspect is the cyclicity of a trace which is not manifestly 
respected by the level-one action.
The other aspect, to be treated elsewhere, 
is that the superspace formulation requires constraints, 
see \secref{sec:supersp-geom},
hence two objects are equivalent if they differ by terms 
which vanish on the constraint surface.

Note that equivalences in field theory are often related to symmetries. 
For instance, the Noether currents and charges due to global symmetries
are conserved on shell; conservation is the statement that 
the divergence of the current is equivalent to zero. 
Above, however, we did not formulate the conformal symmetries 
in terms of Noether charges, but rather in terms of their action 
on the path or on the fields. 
In this formulation the symmetry is not an equivalence
because it relates classically inequivalent objects
(whose quantum expectation values agree).
The two formulations of symmetries merely become related in the quantum theory.
In order to deal with our formulation of conformal symmetry
we should consider the algebra with the level-one action. 
This will be the subject of \secref{sec:Yalgebra}.

\paragraph{Cyclicity.}

An important feature for closed Wilson loops is cyclicity.
The following argument towards cyclicity is analogous to the one
used in \cite{Drummond:2009fd} with some additions concerning gauge symmetry.

Let us compare the level-one action with two distinct reference points
$0$ and $\tau$ on a closed Wilson loop
\begin{align}
\Delta \genYfull{J}^\idxconf{a} \tr \swilson
&=
\genYfull{J}^\idxconf{a} \tr \swilson_{[\tau,1+\tau]}
-
\genYfull{J}^\idxconf{a} \tr \swilson_{[0,1]}
\nln&=
-2f^\idxconf{a}{}_{\idxconf{c}\idxconf{b}}
\tr \genfull{J}^\idxconf{b} \swilson_{[0,\tau]} \genfull{J}^\idxconf{c} \swilson_{[\tau,1]}
\nln&=
-2f^\idxconf{a}{}_{\idxconf{c}\idxconf{b}}
\genfull{J}^\idxconf{b} \bigsbrk{\tr \swilson_{[0,\tau]} \genfull{J}^\idxconf{c} \swilson_{[\tau,1]}}
+2f^\idxconf{a}{}_{\idxconf{c}\idxconf{b}}
\tr \swilson_{[0,\tau]} \genfull{J}^\idxconf{b} \genfull{J}^\idxconf{c} \swilson_{[\tau,1]}
\nln&=
-2f^\idxconf{a}{}_{\idxconf{c}\idxconf{b}}
\genfield{J}^\idxconf{b} \bigsbrk{\tr \swilson_{[0,\tau]} \genfull{J}^\idxconf{c} \swilson_{[\tau,1]}}
\nln&\qquad
+f^\idxconf{a}{}_{\idxconf{c}\idxconf{b}}
f^{\idxconf{b}\idxconf{c}}{}_\idxconf{d}\tr \swilson_{[0,\tau]} \genfull{J}^\idxconf{d} \swilson_{[\tau,1]}
-f^\idxconf{a}{}_{\idxconf{c}\idxconf{b}}
\tr \swilson_{[0,\tau]} \swilson_{[\tau,1]}[\cder \sconfrem^{\idxconf{b}\idxconf{c}}].
\end{align}
Here we evaluate the Yangian actions and write the difference
as a product of two conformal generators acting
on a part of the Wilson loop.
We then pull out the  generator 
$\genfull{J}^\idxconf{b}$ and let it act on everything.
The compensating term can be simplified using the conformal algebra \eqref{eq:confcomm}.
The result is clearly not zero, so the reference point matters in general.

However, all three residual terms have special properties.
The first term is a conformal variation of a gauge-invariant object.
Therefore, it vanishes within an expectation value.
The second term contains the combination
\[
f^\idxconf{a}{}_{\idxconf{c}\idxconf{b}}f^{\idxconf{b}\idxconf{c}}{}_\idxconf{d} \sim h \delta^\idxconf{a}_\idxconf{d},
\]
where $h$ is the dual Coxeter number of the conformal algebra.
For the $\superN=4$ superconformal algebra $\alg{psu}(2,2|4)$ this number is zero, $h=0$.
The final term contains the combination
\[\label{eq:gidentfirst}
f^\idxconf{a}{}_{\idxconf{c}\idxconf{b}}\sconfrem^{\idxconf{b}\idxconf{c}} \simeq0.
\]
This combination is zero for $\superN=4$ super Yang--Mills theory
as will be shown below in \secref{sec:g-identity-new}.
Therefore the Yangian level-one action respects cyclicity of the trace.

\paragraph{Gauge covariance.}

A closely related issue is gauge covariance. 
We have already seen that the level-one action $\genYfull{J}$ on a gauge-covariant
object is again gauge-covariant, see \eqref{eq:newyangact}.
How about the level-one action on the gauge variation of some object $\Psi$,
i.e.\ $\genYfull{J}\gaugegen[\Lambda]\Psi$?

To evaluate this expression, we need to recall the coproduct rule \eqref{eq:copro}. 
When acting with the level-one action $\genYfull{J}$ 
on a product of two gauge-covariant objects
$\Phi$ and $\Psi$ (their order being determined by the flow of gauge indices) 
we have to use the non-trivial color-ordered rule
\[\label{eq:coprofields}
\genYfull{J}^\idxconf{b} (\Phi\Psi) := 
\genYfull{J}^\idxconf{b} \Phi\,\Psi
+\Phi \,\genYfull{J}^\idxconf{b}\Psi
+f^{\idxconf{b}}{}_{\idxconf{d}\idxconf{c}} 
\genfull{J}^\idxconf{c} \Phi\, \genfull{J}^\idxconf{d}\Psi.
\]
We stress that this action of the level-one generator understands the tensor product
ordering to be encoded in the color ordering of the fields $\Phi$ and $\Psi$ in their matrix product. 
It departs from the ordering along the path appearing in the Wilson line as defined in \eqref{eq:Jhatpathaction}, 
but is equivalent 
to it. The color-ordered rule above is also consistent with the action of level-one Yangian
generators on color-ordered scattering amplitudes \cite{Drummond:2009fd}.
From these perspectives we therefore consider the rule \eqref{eq:coprofields} as the
natural one, which might find applications in $\superN=4$ SYM
 beyond scattering amplitudes and Wilson loops.

Using \eqref{eq:coprofields} we then find
\begin{align}\label{eq:levelonegauge}
\genYfull{J}^\idxconf{b}\gaugegen[\Lambda]\Psi
&=
\genYfull{J}^\idxconf{b}(\Psi\Lambda-\Lambda\Psi)
\nln
&=
\comm{\genYfull{J}^\idxconf{b}\Psi}{\Lambda}
+\comm{\Psi}{\genYfull{J}^\idxconf{b}\Lambda}
+f^{\idxconf{b}}{}_{\idxconf{d}\idxconf{c}} 
\acomm{\genfull{J}^\idxconf{c} \Psi}{\genfull{J}^\idxconf{d}\Lambda}
\nln
&=
\gaugegen[\Lambda]\genYfull{J}^\idxconf{b}\Psi
+\gaugegen[\genYfull{J}^\idxconf{b}\Lambda]\Psi
+f^{\idxconf{b}}{}_{\idxconf{d}\idxconf{c}} 
\genfull{J}^\idxconf{c} \acomm{\Psi}{\genfull{J}^\idxconf{d}\Lambda}
-\acomm{\Psi}{f^{\idxconf{b}}{}_{\idxconf{d}\idxconf{c}} \genfull{J}^\idxconf{c}\genfull{J}^\idxconf{d}\Lambda}
.
\end{align}
The result is a combination of two gauge transformations,
a conformal action
(note that the level-one action has turned a commutator into an anti-commutator
by means of the anti-symmetry of the structure constants)
and a term containing the combination 
\[
f^{\idxconf{b}}{}_{\idxconf{d}\idxconf{c}} \genfull{J}^\idxconf{c}\genfull{J}^\idxconf{d}
=
\half
f^{\idxconf{b}}{}_{\idxconf{d}\idxconf{c}} \comm{\genfull{J}^\idxconf{c}}{\genfull{J}^\idxconf{d}}
=
\half
f^{\idxconf{b}}{}_{\idxconf{d}\idxconf{c}} 
f^{\idxconf{c}\idxconf{d}}{}_{\idxconf{e}}
\genfull{J}^\idxconf{e}
-
\half
f^{\idxconf{b}}{}_{\idxconf{d}\idxconf{c}} 
\gaugegen[\sconfrem^{\idxconf{c}\idxconf{d}}]
\simeq 0.
\]
We hence find the commutator
\[
\comm[big]{\genYfull{J}^\idxconf{b}}{\gaugegen[\Lambda]} \Psi
 = \gaugegen[\genYfull{J}^\idxconf{b}\Lambda]\Psi
+f^{\idxconf{b}}{}_{\idxconf{d}\idxconf{c}} 
\genfull{J}^\idxconf{c} \acomm{\Psi}{\genfull{J}^\idxconf{d}\Lambda}\, .
\]
So the Yangian and gauge transformations close into a gauge transformation and a peculiar
conformal action term. In certain circumstances, e.g.\ inserted into correlation functions,
the latter term may vanish.

\section{Yangian Algebra}
\label{sec:Yalgebra}

Above we have argued that the Yangian generators should be formulated
in terms of gauge-covariant conformal transformations. 
These have an impact on the algebra, and we need to see if and how
the Yangian algebra is still satisfied.

\subsection{Mixed Level-One Algebra}

In a Yangian algebra, the level-one generators transform
in the adjoint representation of the conformal algebra
$\comm{\gen{J}^\idxconf{a}}{\genY{J}^{\idxconf{b}}} = 
f^{\idxconf{a}\idxconf{b}}{}_{\idxconf{c}}\genY{J}^{\idxconf{c}}$.
To maintain gauge-covariance we are forced to represent
$\genY{J}$ by the gauge-covariant level-one action $\genYfull{J}$.
For $\gen{J}$ there is, however, a choice between 
the plain $\genfield{J}$ and the gauge-covariant $\genfull{J}$.

As there are no additional terms in the mixed algebra \eqref{eq:confmixed} 
it is most convenient to consider $\genfield{J}$ first. 
When acting on the Wilson line with the (known) bi-local part of the level-one generator 
we can confirm the adjoint property
\[
\label{eqn:semistaredYalgebra}
\bigcomm{\genfield{J}^\idxconf{a}}{\genYfullnl{J}^\idxconf{b}}\swilson
=f^{\idxconf{a}\idxconf{b}}{}_\idxconf{c}\genYfullnl{J}^\idxconf{c} \swilson.
\]
Let us briefly illustrate this calculation step by step.
The first contribution to the commutator reads
\begin{align}
\genfield{J}^\idxconf{a} \genYfullnl{J}^\idxconf{b} \swilson
&=
f^{\idxconf{b}}{}_{\idxconf{d}\idxconf{c}}
\genfield{J}^\idxconf{a} 
\swilson[\genfull{J}^\idxconf{c}\sgauge;\genfull{J}^\idxconf{d}\sgauge]
\nln
&=
f^{\idxconf{b}}{}_{\idxconf{d}\idxconf{c}}
\bigbrk{
\swilson[\genfield{J}^\idxconf{a}\sgauge;\genfull{J}^\idxconf{c}\sgauge;\genfull{J}^\idxconf{d}\sgauge]
+\swilson[\genfull{J}^\idxconf{c}\sgauge;\genfield{J}^\idxconf{a}\sgauge;\genfull{J}^\idxconf{d}\sgauge]
+\swilson[\genfull{J}^\idxconf{c}\sgauge;\genfull{J}^\idxconf{d}\sgauge;\genfield{J}^\idxconf{a}\sgauge]
}
\nln&\qquad
+f^{\idxconf{b}}{}_{\idxconf{d}\idxconf{c}}
\bigbrk{
\swilson[\genfield{J}^\idxconf{a}\genfull{J}^\idxconf{c}\sgauge;\genfull{J}^\idxconf{d}\sgauge]
+\swilson[\genfull{J}^\idxconf{c}\sgauge;\genfield{J}^\idxconf{a}\genfull{J}^\idxconf{d}\sgauge]
}.
\end{align}
For the other contribution we need to take into account the coproduct rule \eqref{eq:coprofields}. 
By iteration of the rule we find 6 terms for 3 objects. 
Dropping the action of $\genYfull{J}$ on a single field
we obtain the other contribution to the commutator
\begin{align}
 \genYfullnl{J}^\idxconf{b} \genfield{J}^\idxconf{a}\swilson
&=
 \genYfullnl{J}^\idxconf{b} \swilson[\genfield{J}^\idxconf{a}\sgauge]
\nln
&=
f^{\idxconf{b}}{}_{\idxconf{d}\idxconf{c}} 
\bigbrk{
\swilson[\genfull{J}^\idxconf{c}\sgauge; \genfull{J}^\idxconf{d}\sgauge; \genfield{J}^\idxconf{a}\sgauge]
+\swilson[\genfull{J}^\idxconf{c}\sgauge; \genfield{J}^\idxconf{a}\sgauge; \genfull{J}^\idxconf{d}\sgauge]
+\swilson[\genfield{J}^\idxconf{a}\sgauge; \genfull{J}^\idxconf{c}\sgauge; \genfull{J}^\idxconf{d}\sgauge]
}
\nln&\qquad
+
f^{\idxconf{b}}{}_{\idxconf{d}\idxconf{c}} 
\bigbrk{
\swilson[\genfull{J}^\idxconf{c}\sgauge; \genfull{J}^\idxconf{d}\genfield{J}^\idxconf{a}\sgauge]
+\swilson[\genfull{J}^\idxconf{c}\genfield{J}^\idxconf{a}\sgauge; \genfull{J}^\idxconf{d}\sgauge]
}.
\end{align}
The terms on the first line are equal for both contributions and cancel
immediately.
Using the mixed algebra relation \eqref{eq:confmixed} and
the Jacobi identities of the structure constants the commutator of
the terms of the second line combine
\begin{align}
\bigcomm{\genfield{J}^\idxconf{a}}{ \genYfullnl{J}^\idxconf{b}} \swilson
&=
f^{\idxconf{b}}{}_{\idxconf{d}\idxconf{c}}
\bigbrk{
\swilson\bigsbrk{\comm{\genfield{J}^\idxconf{a}}{\genfull{J}^\idxconf{c}}\sgauge;\genfull{J}^\idxconf{d}\sgauge}
+\swilson\bigsbrk{\genfull{J}^\idxconf{c}\sgauge;\comm{\genfield{J}^\idxconf{a}}{\genfull{J}^\idxconf{d}}\sgauge}
}
\nln
&=
f^{\idxconf{b}}{}_{\idxconf{d}\idxconf{c}}
\bigbrk{
f^{\idxconf{a}\idxconf{c}}{}_{\idxconf{e}}
 \swilson\bigsbrk{\genfull{J}^\idxconf{e}\sgauge;\genfull{J}^\idxconf{d}\sgauge}
+
f^{\idxconf{a}\idxconf{d}}{}_{\idxconf{e}}
\swilson\bigsbrk{\genfull{J}^\idxconf{c}\sgauge;\genfull{J}^\idxconf{e}\sgauge}
}
\nln
&=
f^{\idxconf{a}\idxconf{b}}{}_{\idxconf{c}}
f^{\idxconf{c}}{}_{\idxconf{e}\idxconf{d}}
 \swilson\bigsbrk{\genfull{J}^\idxconf{d}\sgauge;\genfull{J}^\idxconf{e}\sgauge}
=
f^{\idxconf{a}\idxconf{b}}{}_{\idxconf{c}}\genYfullnl{J}^\idxconf{c} \swilson.
\end{align}
Gladly, all violations of gauge covariance have canceled out
in the commutator.

\subsection{Gauge-Covariant Level-One Algebra}

Next we want to investigate the corresponding algebra of 
purely gauge-covariant actions.
By construction, the difference to the above mixed algebra 
is a commutator of a gauge transformation
with a gauge-covariant level-one generator. 
Therefore, one may expect the result to be the same up to a gauge transformation. 
However, the derivation requires more effort, involves some subtleties
and also relies on a special feature of $\superN=4$ super Yang--Mills theory. 
Let us therefore go though it in detail.

First we shall consider the algebra on a single gauge connection $\sgauge$.
This appears trivial at first sight, but a subtlety resolves an issue at a different place.
We assume that the action of $\genYfull{J}^{\delta}$ on $\sgauge$ is trivial%
\footnote{In fact, there may be terms non-linear in the fields \cite{BG};
these are not relevant for the Wilson loop expectation value at one loop and
therefore we shall ignore them here.}
\[
\genYfull{J}^{\delta}\sgauge=0;
\]
this is consistent with considering just the bi-local action on the Wilson line.
The commutator then reads
\[
\bigcomm{\genfull{J}^\idxconf{a}}{\genYfull{J}^\idxconf{b}}\sgauge_\idxall{A}
=
-\genYfull{J}^\idxconf{b}\genfull{J}^\idxconf{a}\sgauge_\idxall{A}
=
\gen{J}^\idxconf{a}X^\idxall{B}\,\genYfull{J}^\idxconf{b}\sfstr_{\idxall{B}\idxall{A}}.
\]
Here it is tempting to discard the term $\genYfull{J}\sfstr$ because
$\sfstr$ consists of $\sgauge$ only on which the action is trivial.
However, thanks to the coproduct rule \eqref{eq:coprofields},
the action on the field strength is non-trivial 
because the latter is a non-linear combination of gauge potentials
\begin{align}\label{eq:leveloneonf}
\genYfull{J}^\idxconf{b}\sfstr_{\idxall{A}\idxall{B}}
&=
\genYfull{J}^\idxconf{b}(\sgauge_{\idxall{A}}\sgauge_{\idxall{B}}-\sgauge_{\idxall{B}}\sgauge_{\idxall{A}}) 
= 
f^{\idxconf{b}}{}_{\idxconf{d}\idxconf{c}} 
\genfull{J}^\idxconf{c}\sgauge_{\idxall{A}}\, \genfull{J}^\idxconf{d}\sgauge_{\idxall{B}}
-f^{\idxconf{b}}{}_{\idxconf{d}\idxconf{c}} 
\genfull{J}^\idxconf{c}\sgauge_{\idxall{B}}\, \genfull{J}^\idxconf{d}\sgauge_{\idxall{A}}
\nln
&=
f^{\idxconf{b}}{}_{\idxconf{d}\idxconf{c}} 
\bigacomm{\genfull{J}^\idxconf{c}\sgauge_{\idxall{A}}}{\genfull{J}^\idxconf{d}\sgauge_{\idxall{B}}}
.
\end{align}
For the above algebra on the gauge connection this implies
\[\label{eq:leveloneona}
\bigcomm{\genfull{J}^\idxconf{a}}{\genYfull{J}^\idxconf{b}}\sgauge_\idxall{A}
=
\gen{J}^\idxconf{a}X^\idxall{B}\,
f^{\idxconf{b}}{}_{\idxconf{d}\idxconf{c}} 
\bigacomm{\genfull{J}^\idxconf{c}\sgauge_{\idxall{B}}}{\genfull{J}^\idxconf{d}\sgauge_{\idxall{A}}}
=
f^{\idxconf{b}}{}_{\idxconf{d}\idxconf{c}} 
\bigacomm{\sconfrem^{\idxconf{a}\idxconf{c}}}{\genfull{J}^\idxconf{d}\sgauge_{\idxall{A}}}
.
\]
As it stands, the result does not agree with the expected algebra.
Nevertheless, we can rearrange the terms by extending the 
remaining conformal action over the whole term and then correct for the
discrepancy
\[
\bigcomm{\genfull{J}^\idxconf{a}}{\genYfull{J}^\idxconf{b}}\sgauge_\idxall{A}
=
f^{\idxconf{b}}{}_{\idxconf{d}\idxconf{c}} 
\genfull{J}^\idxconf{d}
\bigacomm{\sconfrem^{\idxconf{a}\idxconf{c}}}{\sgauge_{\idxall{A}}}
-
f^{\idxconf{b}}{}_{\idxconf{d}\idxconf{c}}
\bigacomm{ \genfull{J}^\idxconf{d}\sconfrem^{\idxconf{a}\idxconf{c}}}{\sgauge_{\idxall{A}}}
.
\]
The first term is acceptable because it is a conformal action 
on some composite object. 
Within quantum correlation functions of a conformal theory, 
this term does not contribute.%
\footnote{In fact, this statement would require 
the objects to be gauge-invariant which clearly does not apply here. 
However, once embedded into a gauge-invariant observable,
the statement becomes true.
See below for the implementation within a Wilson loop.}
Furthermore, the term could be 
represented as the action of some additional generator on $\sgauge$.
The second term we can transform further
using the conformal algebra
\begin{align}
f^{\idxconf{b}}{}_{\idxconf{d}\idxconf{c}}
\genfull{J}^\idxconf{d}\sconfrem^{\idxconf{a}\idxconf{c}}
&=
f^{\idxconf{b}}{}_{\idxconf{d}\idxconf{c}}
\gen{J}^\idxconf{a}X^\idxall{A}
\genfull{J}^\idxconf{d} \genfull{J}^\idxconf{c} \sgauge_\idxall{A}
=
\half f^{\idxconf{b}}{}_{\idxconf{d}\idxconf{c}}
\gen{J}^\idxconf{a}X^\idxall{A}
(
f^{\idxconf{d}\idxconf{c}}{}_\idxconf{e}\genfull{J}^\idxconf{e}\sgauge_\idxall{A}
-\cdel_\idxall{A}\sconfrem^{\idxconf{d}\idxconf{c}}
 )
\nln
&=
-\half f^{\idxconf{b}}{}_{\idxconf{d}\idxconf{c}}
f^{\idxconf{c}\idxconf{d}}{}_\idxconf{e}
\sconfrem^{\idxconf{a}\idxconf{e}}
+
\half f^{\idxconf{b}}{}_{\idxconf{d}\idxconf{c}}
\gen{J}^\idxconf{a}X^\idxall{A}
\cdel_\idxall{A}\sconfrem^{\idxconf{c}\idxconf{d}}\simeq 0.
\end{align}
As we have argued in the discussion on cyclicity in \secref{sec:consistency},
the resulting two terms vanish for $\superN=4$ super Yang--Mills theory.
Thus we can write the algebra on $\sgauge$ as 
\[
\comm{\genfull{J}^\idxconf{a}}{\genYfull{J}^\idxconf{b}}\sgauge_\idxall{A}
\simeq 
f^{\idxconf{b}}{}_{\idxconf{d}\idxconf{c}} 
\genfull{J}^\idxconf{d}
\acomm{\sconfrem^{\idxconf{a}\idxconf{c}}}{\sgauge_{\idxall{A}}}.
\]

One can show that the above algebra acting on some field $\Psi$ then takes the form
\[
\bigcomm{\genfull{J}^\idxconf{a}}{\genYfull{J}^\idxconf{b}}\Psi
=
f^{\idxconf{a}\idxconf{b}}{}_{\idxconf{c}} \genYfull{J}^\idxconf{c}\Psi
+
f^{\idxconf{b}}{}_{\idxconf{d}\idxconf{c}} 
\genfull{J}^\idxconf{d}
\bigacomm{\sconfrem^{\idxconf{a}\idxconf{c}}}{\Psi}
-
\bigacomm{ f^{\idxconf{b}}{}_{\idxconf{d}\idxconf{c}}\genfull{J}^\idxconf{d}\sconfrem^{\idxconf{a}\idxconf{c}}}{\Psi}.
\]
This result in fact follows from the algebra of the
level-one action with 
a plain conformal action \eqref{eqn:semistaredYalgebra}
and with a conformal compensating gauge transformation \eqref{eq:levelonegauge}.
We can also show that it is consistent with the coproduct rule \eqref{eq:coprofields}
in the sense
\begin{align}
\bigcomm{\genfull{J}^\idxconf{a}}{\genYfull{J}^\idxconf{b}}(\Phi\Psi)
&=
f^{\idxconf{a}\idxconf{b}}{}_{\idxconf{c}} \genYfull{J}^\idxconf{c}(\Phi\Psi)
+
f^{\idxconf{b}}{}_{\idxconf{d}\idxconf{c}} 
\genfull{J}^\idxconf{d}
\bigacomm{\sconfrem^{\idxconf{a}\idxconf{c}}}{\Phi\Psi}
-
\bigacomm{f^{\idxconf{b}}{}_{\idxconf{d}\idxconf{c}} \genfull{J}^\idxconf{d}\sconfrem^{\idxconf{a}\idxconf{c}}}{\Phi\Psi}
.
\end{align}

Let us finally address the algebra acting on the Wilson line.
One finds the analogous result
\begin{align}
\label{eqn:commutator_zero_one}
\bigcomm{\genfull{J}^\idxconf{a}}{\genYfull{J}^\idxconf{b}}\swilson
&=f^{\idxconf{a}\idxconf{b}}{}_\idxconf{c}\genYfull{J}^\idxconf{c} \swilson
+\swilson\bigsbrk{\comm{\genfull{J}^\idxconf{a}}{\genYfull{J}^\idxconf{b}}\sgauge}
- f^\idxconf{b}{}_{\idxconf{d}\idxconf{c}}
\swilson[\genfull{J}^\idxconf{c} \sgauge; \cder \sconfrem^{\idxconf{a}\idxconf{d}}]
- f^\idxconf{b}{}_{\idxconf{d}\idxconf{c}}
\swilson[\cder \sconfrem^{\idxconf{a}\idxconf{c}}; \genfull{J}^\idxconf{d} \sgauge]
.
\end{align}
The first term is just as expected. The second term
is the local action on a single gauge connection 
and the remaining ones are due to the compensating gauge transformations.
The total derivatives can be integrated
yielding two local terms and two bulk-boundary terms.
The former are precisely canceled by the algebra acting on the
gauge connection \eqref{eq:leveloneona}. We are left with
\begin{align}
\bigcomm{\genfull{J}^\idxconf{a}}{\genYfull{J}^\idxconf{b}}\swilson
&=f^{\idxconf{a}\idxconf{b}}{}_\idxconf{c}\genYfull{J}^\idxconf{c} \swilson
+ f^\idxconf{b}{}_{\idxconf{d}\idxconf{c}}
\bigbrk{
\swilson[\genfull{J}^\idxconf{d} \sgauge]\sconfrem^{\idxconf{a}\idxconf{c}}(1)
+ 
\sconfrem^{\idxconf{a}\idxconf{c}}(0)\swilson[\genfull{J}^\idxconf{d} \sgauge]
}
,
\end{align}
where we can pull the remaining conformal actions over the whole expression
at the expense of a term which is zero in $\superN=4$ super Yang--Mills theory
\[\label{eq:confyangfull}
\bigcomm{\genfull{J}^\idxconf{a}}{\genYfull{J}^\idxconf{b}}\swilson
\simeq f^{\idxconf{a}\idxconf{b}}{}_\idxconf{c}\genYfull{J}^\idxconf{c} \swilson
+ f^\idxconf{b}{}_{\idxconf{d}\idxconf{c}}
\genfull{J}^\idxconf{d}
\bigbrk{
\swilson\sconfrem^{\idxconf{a}\idxconf{c}}(1)
+ 
\sconfrem^{\idxconf{a}\idxconf{c}}(0)\swilson
}
.
\]
This shows that the gauge-covariant level-one algebra closes,
but only on-shell in $\superN=4$ super Yang--Mills theory.

Furthermore, the closure only holds within quantum expectation values
\[
\bigvev{
\bigcomm{\genfull{J}^\idxconf{a}}{\genYfull{J}^\idxconf{b}}\tr\swilson
}
\simeq f^{\idxconf{a}\idxconf{b}}{}_\idxconf{c} \bigvev{\genYfull{J}^\idxconf{c} \tr\swilson},
\]
where we use the conformal symmetry relation $\vev{\genfull{J} H}=0$ for any gauge-invariant
observable $H$
\[
f^\idxconf{b}{}_{\idxconf{d}\idxconf{c}}
\bigvev{
\genfull{J}^\idxconf{d}
\tr\bigbrk{
\swilson\sconfrem^{\idxconf{a}\idxconf{c}}(1)
+ 
\sconfrem^{\idxconf{a}\idxconf{c}}(0)\swilson
}}=0.
\]
One should mention that the presence of some other operator $\mathcal{O}$
besides the Wilson loop $\tr \swilson$ is expected to spoil the Yangian algebra.
This is because the insertion of $\mathcal{O}$ into the latter term
yields a non-vanishing contribution $\vev{\mathcal{O}\genfull{J} H}\neq 0$.
However, it is also not surprising because we can only expect Yangian symmetry
for objects with the planar topology of a disk. 
The planar Wilson loop has this topology, but any additional object
will lead to a non-trivial topology without Yangian symmetry.

Finally, let us emphasize that consistency of the Yangian algebra 
requires the Serre relations to hold. These are combinatorially
elaborate cubic relations among the Yangian generators. 
They have been shown to hold for the linear representation \cite{Dolan:2004ps},
which is an evaluation representation of the Yangian. 
The non-linear terms in the gauge-covariant action, 
however, very much complicate the derivation 
and actually might require the introduction of additional terms.
We refrain from attempting to verify the Serre relations,
and discuss their implications 
in the \hyperref[sec:conclusions]{conclusions}.

\subsection{Kappa-Symmetry}

Wilson loops with scalar coupling which are null in a ten-dimensional sense
enjoy eight additional translational symmetries in superspace
which is closely related to kappa-\hspace{0pt}symmetry
of string theory \cite{Beisert:2015jxa}.

\paragraph{Definition.}

We can formulate this kappa-symmetry as an action $\delta$ on the fields
in close analogy to the action $\genfull{J}$ of conformal symmetry 
\eqref{eq:fullconfacta} as follows%
\footnote{As before, we shall ignore the scalar couplings for reasons
of simplicity.
The derivation can either be understood in a ten-dimension sense
or assuming the path is null in four dimensions.}
\begin{align}
\kappagenfull \sgauge_\idxall{A} &:= \kappagen X^\idxall{B}\, \sfstr_{\idxall{A}\idxall{B}},
\end{align}
where $\kappagen X(\tau)$ is the displacement at
the point $\tau$ on the Wilson line.
The action on the Wilson line then reads
\[
\kappagenfull \swilson = \swilson[\delta_*\sgauge].
\]
The condition for kappa-symmetry is that the action annihilates 
the gauge connection
\[\label{eq:kappacond}
\kappagenfull \sgauge=\der X^\idxall{A}\,\kappagen X^\idxall{B}\, \sfstr_{\idxall{A}\idxall{B}}=0.
\]
There are two ways to meet this condition. 
Either $\kappagen X^\idxall{A}\sim \dot X^\idxall{B}$
in which case the symmetry is reparametrization invariance.
Or if $\dot X$ is null and
$\kappagen X^\idxall{A}\sim \tilde\storsion^\idxall{\idxall{A}\idxall{B}}_{\idxall{C}} K_\idxall{B} \dot X^\idxall{C}$
see section 2.3 of our paper \cite{Beisert:2015jxa} for details.

\paragraph{Algebra.}

Conformal symmetry respects kappa-symmetry. This statement follows from the commutator
\[
\comm{\genfull{J}}{\kappagenfull}\sgauge_\idxall{A}
=
(\comm{\gen{J}}{\kappagen}X^\idxall{B} )\sfstr_{\idxall{A}\idxall{B}}
-\cdel_\idxall{A}(\gen{J}X^\idxall{B}\,\kappagen X^\idxall{C}\,\sfstr_{\idxall{B}\idxall{C}}),
\]
which is straight-forward to derive. Now the first term
turns out to be another kappa-symmetry transformation because the condition \eqref{eq:kappacond}
for kappa-symmetry is respected by conformal transformations.
The second term is a total derivative representing a gauge transformation.
The general algebra reads
\[
\comm{\genfull{J}}{\kappagenfull}
=
\kappagenfull'-\gaugegen[\gen{J}X^\idxall{B}\,\kappagen X^\idxall{C}\,\sfstr_{\idxall{B}\idxall{C}}].
\]

Finally, let us consider the level-one Yangian action.
For the action of the commutator on a gauge connection we find
\[
\comm{\genYfull{J}^\idxconf{a}}{\kappagenfull}\sgauge_\idxall{A}
=-f^\idxconf{a}{}_{\idxconf{c}\idxconf{b}}
\acomm{\kappagen X^\idxall{B}\,\genfull{J}^\idxconf{b} \sgauge_{\idxall{B}}}{\genfull{J}^\idxconf{c}\sgauge_\idxall{A}}.
\]
This result is very reminiscent of \eqref{eq:leveloneona}.
For a kappa-symmetric Wilson line, we obtain
\begin{align}
\comm{\genYfull{J}^\idxconf{a}}{\kappagenfull}\swilson
&=
\swilson\bigsbrk{\comm{\genYfull{J}^\idxconf{a}}{\kappagenfull}\sgauge}
+f^\idxconf{a}{}_{\idxconf{c}\idxconf{b}}
\swilson\bigsbrk{\comm{\genfull{J}^\idxconf{b}}{\kappagenfull}\sgauge;\genfull{J}^\idxconf{c}\sgauge}
+f^\idxconf{a}{}_{\idxconf{c}\idxconf{b}}
\swilson\bigsbrk{\genfull{J}^\idxconf{b}\sgauge;\comm{\genfull{J}^\idxconf{c}}{\kappagenfull}\sgauge}
\nln
&=
-f^\idxconf{a}{}_{\idxconf{c}\idxconf{b}}
\swilson\bigsbrk{\acomm{\kappagen X^\idxall{B}\,\genfull{J}^\idxconf{b} 
\sgauge_{\idxall{B}}}{\genfull{J}^\idxconf{c}\sgauge_\idxall{A}}}
\nln&\qquad
+f^\idxconf{a}{}_{\idxconf{c}\idxconf{b}}
\swilson\bigsbrk{\cder(\kappagen X^\idxall{B}\,\genfull{J}^\idxconf{b} 
\sgauge_{\idxall{B}});\genfull{J}^\idxconf{c}\sgauge}
+f^\idxconf{a}{}_{\idxconf{c}\idxconf{b}}
\swilson\bigsbrk{\genfull{J}^\idxconf{b}\sgauge;\cder(\kappagen X^\idxall{B}\,\genfull{J}^\idxconf{c} \sgauge_{\idxall{B}})}
\nln
&=
-f^\idxconf{a}{}_{\idxconf{c}\idxconf{b}}
(\kappagen X^\idxall{B}\,\genfull{J}^\idxconf{b} \sgauge_{\idxall{B}})(0)
\swilson\bigsbrk{\genfull{J}^\idxconf{c}\sgauge}
-f^\idxconf{a}{}_{\idxconf{c}\idxconf{b}}
\swilson\bigsbrk{\genfull{J}^\idxconf{c}\sgauge}
(\kappagen X^\idxall{B}\,\genfull{J}^\idxconf{b} \sgauge_{\idxall{B}})(1)
\nln
&\simeq
-f^\idxconf{a}{}_{\idxconf{c}\idxconf{b}}\genfull{J}^\idxconf{c}
\lrbrk{
(\kappagen X^\idxall{B}\,\genfull{J}^\idxconf{b} \sgauge_{\idxall{B}})(0)\swilson
+\swilson(\kappagen X^\idxall{B}\,\genfull{J}^\idxconf{b} \sgauge_{\idxall{B}})(1)
}.
\end{align}
This parallels the considerations in \secref{sec:consistency}.
Here we have used the constraints of $\superN=4$ SYM and arrive at 
a conformal transformation acting on something. 
Again this result is analogous of \eqref{eq:confyangfull}
and vanishes within an expectation value.
This shows that the Yangian algebra respects
kappa-symmetry.

\section{\texorpdfstring{$\superN=4$}{N=4} SYM in Non-Chiral Superspace}
\label{sec:superspace}

In the following we collect the relevant conventions and results for non-chiral
$\superN=4$ superspace following our previous work \cite{Beisert:2015jxa}.
This superspace arises from the dimensional reduction of $\superN=1$ superspace
in ten dimensions.

\subsection{Superspace Geometry and Fields}
\label{sec:supersp-geom}

The non-chiral $\superN=4$ four-dimensional superspace is parametrized by the coordinates 
\((x^{\dot\alpha \alpha}, \theta^{a\alpha}, \bar{\theta}^{\dot\alpha}{}_{a})\).  We will 
frequently use a matrix notation similar to the one in
ref.~\cite{Beisert:2012gb}.  The bosonic coordinate  \(x^{\dot{\alpha} \alpha}\) is
represented by a \(2 \times 2\) matrix, while \(\theta^{a \alpha}\) is a \(4 \times
2\) and \(\bar{\theta}^{\dot{\alpha}}{}_{a}\)  a \(2 \times 4\)
matrix.

We define \(x^\pm = x \mp 2 \bar{\theta} \theta\), 
which have the nice property of transforming in a simpler way under supersymmetry than \(x\) does.  
The \(x^{\pm}\) satisfy the constraints
\[
  x = \sfrac 1 2 (x^+ + x^-), \qquad
  \bar{\theta} \theta = \sfrac 1 4 (x^- - x^+)\, ,
\]
which are preserved by the supersymmetry transformations. 
We work in mostly minus signature and impose the reality conditions
\[
  x^\ddagger = x, \qquad
  \theta^\ddagger = \bar{\theta}=-i\theta^{\dagger}, \qquad
  \bar{\theta}^\ddagger = \theta, \qquad
  (x^{\pm})^\ddagger = x^{\mp}.
\]
For Gra{\ss}mann variables \(\chi, \psi\) we use the convention
that \((\chi \psi)^\ddagger = -\psi^\ddagger \chi^\ddagger\), 
while for transpositions we use 
\((\chi \psi)^\trans = -\psi^\trans \chi^\trans\). 
There are three types of supersymmetric interval definitions that will prove useful
\begin{align}
  x_{12} &= x_1 - x_2 + 2 \bar{\theta}_2 \theta_1 - 2 \bar{\theta}_1 \theta_2,\\
  x_{12}^{+-} &= x_1^+ - x_2^- + 4 \bar{\theta}_2 \theta_1 =
     x_{12} - 2 \bar{\theta}_{12} \theta_{12},\\
  x_{12}^{-+} &= x_1^- - x_2^+ - 4 \bar{\theta}_1 \theta_2 =
     x_{12} + 2 \bar{\theta}_{12} \theta_{12}.
\end{align}
The supersymmetry covariant derivatives are defined as
\[
  D_{ \alpha a} = \partial_{\alpha a} + \sigma^{\mu}_{\alpha \dot{\alpha}}
  \bar{\theta}^{\dot{\alpha}}{}_{a} \partial_{\mu}, \qquad
  \bar{D}^{a}{}_{\dot{\alpha}} = \bar{\partial}^{a}{}_{\dot{\alpha}} +
  \theta^{a \alpha}\sigma_{\alpha \dot{\alpha}}^{\mu} \partial_{\mu}\, .
\]
We can furthermore define the dual vielbeine
\begin{equation}
  e:=e_\text{B} = \der x - 2 \der \bar{\theta} \theta + 2 \bar{\theta} \der \theta, \qquad
  e_\text{F} = \der \theta, \qquad
  \bar{e}_\text{F} = \der \bar{\theta} ,
\end{equation} 
whose differentials are%
\footnote{In the following we will mostly omit 
the wedge symbol in the wedge product of differential forms.}
\begin{equation}
 \der e = 4 \der\bar\theta \, \der\theta \, , \qquad
  \der \der \theta = 0\, , \qquad \der\der \bar{\theta} = 0.
\end{equation}

Differentials in superspace have two gradings: a form grading and a Gra{\ss}mann grading. 
 The superspace coordinates \(x\), \(\theta\) and \(\bar{\theta}\) 
have zero form grading while \(\der x\), \(\der \theta\) and \(\der \bar{\theta}\) have form grading one. 
 When permuting two such expressions, we pick up a sign from the differential grading and another from the Gra{\ss}mann grading.

The total differential in superspace can be written using normal or
supersymmetry covariant derivatives
\begin{equation}
  \der = \half \der x^{\dot{\alpha} \alpha} \partial_{\alpha \dot{\alpha}} +
   \der \theta^{a \alpha} \partial_{\alpha a} +
   \der \bar{\theta}^{\dot{\alpha}}{}_a \bar{\partial}^a{}_{\dot{\alpha}} =
   \half e^{\dot{\alpha} \alpha} \partial_{\alpha \dot{\alpha}} +
   \der\theta^{a \alpha} D_{\alpha a} +
   \der\bar{\theta}^{\dot{\alpha}}{}_a \bar{D}^a{}_{\dot{\alpha}}.
\end{equation}
The gauge connection is a one-form in superspace which may be
decomposed on the vielbein basis as
\begin{equation}
  \mathcal{A} = \sfrac{1}{2} e^{\dot{\alpha} \alpha} \sgauge^{\text{cov}}_{\alpha \dot{\alpha}} +
   \der\theta^{a \alpha} \sgauge^{\text{cov}}_{ \alpha a} +
   \der\bar{\theta}^{\dot{\alpha}}{}_a \sgauge^{\text{cov}\, a}{}_{\dot{\alpha}}.
\end{equation}
This allows us to introduce the  supersymmetry and gauge-covariant derivatives as
\[
  \scdel_{\alpha \dot{\alpha}} = \partial_{\alpha \dot{\alpha}} + \sgauge^{\text{cov}}_{\alpha \dot{\alpha}}, \qquad
  \scdel_{\alpha a} = D_{\alpha a} + \sgauge^{\text{cov}}_{\alpha a}, \qquad
  \scdelb{}^{a}{}_{\dot{\alpha}} = \bar{D}^a{}_{\dot{\alpha}} + \sgauge^{\text{cov} \, a}{}_{\dot{\alpha}}\, ,
\]
which satisfy the constraints
\begin{subequations}
\label{eqn:N=4constraints}
\begin{align}
\label{eqn:N=4constraints1}
  \{\scdel_{ \alpha a}, \scdel_{\beta b}\} &= -4 \epsilon_{\alpha \beta} \Phi_{a b}, 
\\
\label{eqn:N=4constraints1b}
  \{\scdel_{\alpha a}, \scdelb{}^{b}{}_{\dot{\beta}}\} &= 2 \delta_a^b \scdel_{\alpha \dot{\beta}}, \\
  \{\scdelb{}^{a}{}_{\dot{\alpha}}, \scdelb{}^b{}_{\dot{\beta}}\}
 &= -2 \epsilon_{\dot{\alpha} \dot{\beta}} \epsilon^{a b c d} \Phi_{c d}.
\label{eqn:N=4constraints2}
\end{align}
\end{subequations}
Here the scalar superfield $\Phi_{ab}$ arises whose leading component are the $\superN=4$ SYM scalar
fields $\phi_{ab}$. The scalar field in the constraints above arises through the dimensional reduction
where the 10d gauge field decomposes as $\sgauge^{\text{cov}}_{\hat\mu}(x^{\hat\mu},\theta^{\hat\alpha})\to 
\{\sgauge^{\text{cov}}_{\alpha\dot\alpha}(x,\theta,\bar\theta), \Phi_{ab}
(x,\theta,\bar\theta)\}$.
The components of the field strength arise from taking
commutators or anti-commutators of the gauge-covariant derivatives $\scdel$.  In fact these commutators
may all be expressed in terms of gauge-covariant derivatives acting on the
scalar fields, as follows
\begin{align}
  [\scdel_{\alpha a}, \scdel_{\beta \dot{\beta}}] &=
  \sfrac{2}{3} \epsilon_{\alpha \beta} [\scdelb{}^b{}_{\dot{\beta}}, \Phi_{a b}],
\\
  [\scdelb{}^a{}_{\dot{\alpha}}, \scdel_{\beta \dot{\beta}}] &=
  \sfrac{1}{3} \epsilon_{\dot{\alpha} \dot{\beta}} \epsilon^{abcd} [\scdel_{\beta b}, \Phi_{c d}],
\\
  [\scdel_{\alpha \dot{\alpha}}, \scdel_{\beta \dot{\beta}}] &=
  \sfrac{1}{24} \epsilon_{\dot{\alpha} \dot{\beta}} 
\epsilon^{abcd} \bigacomm{\scdel_{\alpha a}}{[\scdel_{ \beta b}, \Phi_{cd}]}
  +\sfrac{1}{12} \epsilon_{\alpha \beta} 
\bigacomm{\scdelb{}^{a}{}_{\dot{\alpha}}}{ [\scdelb{}^{b}{}_{\dot{\beta}}, \Phi_{ab}]}.
\end{align}

Let us now consider the linearized field strength
\(\mathcal{F}_{\text{lin}} = \der \mathcal{A}\).  Using the
rules of differential calculus described above, we have
\begin{align}
  \label{eq:flin}
  \mathcal{F}_{\text{lin}} &= \der \mathcal{A} =
  \sfrac{1}{2}\der e^{\dot{\alpha} \alpha} \sgauge^{\text{cov}}_{\alpha \dot{\alpha}} -
  \sfrac{1}{2} e^{\dot{\alpha} \alpha} \der \sgauge^{\text{cov}}_{\alpha \dot{\alpha}} -
  \der\theta^{a \alpha} \der \sgauge^{\text{cov}}_{\alpha a} -
  \der\bar{\theta}^{\dot{\alpha}}{}_a \der \sgauge^{\text{cov} \, a}{}_{\dot{\alpha}} 
\nonumber\\
  &=+2 (\der\theta   \epsilon  \der\theta^\trans)^{ab} \Phi_{ab}
  + (\der\bar{\theta}^\trans  \epsilon  \der\bar{\theta})_{ab}  \epsilon^{abcd} \Phi_{cd}
\nonumber\\&\qquad
+\sfrac{1}{3} (e  \epsilon  \der\theta^\trans)^{\dot{\alpha} a} \bar{D}^b{}_{\dot{\alpha}} \Phi_{ab}
  +\sfrac{1}{6} (e^\trans \epsilon  \der\bar{\theta})^{\alpha}_{\hphantom{\alpha} a} \epsilon^{abcd} D_{\alpha b} \Phi_{cd}
\nonumber\\&\qquad
+\sfrac{1}{96} (e  \epsilon  e^\trans)^{\dot{\alpha} \dot{\beta}} \bar{D}^a{}_{\dot{\alpha}} \bar{D}^b{}_{\dot{\beta}} \Phi_{ab}
  +\sfrac{1}{192} (e^\trans  \epsilon  e)^{\alpha \beta} \epsilon^{abcd} D_{\alpha a} D_{\beta b} \Phi_{cd},
\end{align} 
where we kept only those terms being linear in
the superfields.
The component containing \((\der\theta \der\bar{\theta})_{}\) has coefficient zero.  The
components containing \((\der\theta e)\) and \((\der\bar{\theta} e)\) are the
fermionic superfields while the component containing \((e e)\) is the
supersymmetrization of the bosonic field strength \(F_{\alpha
  \dot{\alpha}, \beta \dot{\beta}}\).  Also note that
\(\mathcal{F}_{\text{lin}}\) does not depend on \((\der\theta   \epsilon  \der\theta^\trans)\)
and \((\der\bar{\theta}^\trans   \epsilon  \der\bar{\theta})\) separately, but it
depends only on the combination \((\der\theta \epsilon \der\theta^\trans)^{ab} +
\tfrac 1 2 \epsilon^{abcd} (\der\bar{\theta}^\trans \epsilon
\der\bar{\theta})_{cd}\), as required by the duality of the scalar fields.
Finally we shall split up the linearized field strength into its chiral and anti-chiral parts,
\(\mathcal{F}_{\text{lin}} = \mathcal{F}^+ + \mathcal{F}^-\), with
\begin{subequations}
\label{eqn:sfstrF}
\begin{align}
\label{eqn:sfstrF-}
  \mathcal{F}^- &=
  \bigsbrk{2 ( \der\theta \epsilon  \der\theta^\trans)^{ab}  +
  \sfrac{1}{3}(e \epsilon  \der\theta^\trans)^{\dot{\alpha} a} \bar{D}^b{}_{\dot{\alpha}}+
  \sfrac{1}{96} (e \epsilon e^\trans)^{\dot{\alpha} \dot{\beta}} \bar{D}^a{}_{\dot{\alpha}} \bar{D}^b{}_{\dot{\beta}}} 
 \Phi_{ab},\\
  \label{eqn:sfstrF+}
  \mathcal{F}^+ &= 
 \bigsbrk{  2 (\der\bar{\theta}^\trans \epsilon \der\bar{\theta})_{ab}  +
  \sfrac{1}{3} (e^\trans \epsilon \der\bar{\theta})^{\alpha}_{\hphantom{\alpha} a} D_{\alpha b}  +
  \sfrac{1}{96} (e^\trans \epsilon e)^{\alpha \beta} D_{\alpha a} D_{\beta b} }\bar{\Phi}^{ab}.
\end{align}
\end{subequations}
Hence we have shown that all the components of the linearized 
field strength may be written as supersymmetry covariant derivatives of
the scalar superfields $\Phi_{ab}=\half \epsilon_{abcd}\, \bar{\Phi}^{cd}$.

\subsection{Two-Point Functions}
\label{sec:two-point-functions}

From the knowledge of the scalar-scalar two-point function, we
can compute the two-point functions of the field strengths.  
We find (see also ref.~\cite{Beisert:2015jxa} 
for inversion formulas which can be used to compute these two-point functions)
\begin{align}
  \bigvev{ \bar{\Phi}^{ab}(1) \Phi_{cd}(2)} &= - \frac {g^2}{\pi^2}
  \frac {(1-4 \theta_{12} x_{12}^{-+,-1} \bar{\theta}_{12})^a_{\hphantom{a} [c}
  (1-4 \theta_{12} x_{12}^{-+,-1} \bar{\theta}_{12})^b_{\hphantom{b} \vert d]}}{(x_{12}^{-+})^2},
\\
  \bigvev{ \mathcal{F}^+(1) \Phi_{ab}(2)} &= -\frac {2 g^2}{\pi^2} \epsilon^{\alpha \beta}
  \der_1 (x_{12}^{-+,-1} \bar{\theta}_{12})_{\alpha a}
  \der_1 (x_{12}^{-+,-1} \bar{\theta}_{12})_{\beta b},
\\
  \bigvev{ \mathcal{F}^+(1) \mathcal{F}^-(2)} &=
  \frac {g^2}{4 \pi^2} \Bigl(
  \tr\bigsbrk{\der_1 (x_{12}^{-+,-1} \der_2 x_{12}^{-+}) \der_1 (x_{12}^{-+,-1} \der_2 x_{12}^{-+})} 
\nonumber\\
  &\qquad
-\tr\bigsbrk{\der_1 (x_{12}^{-+,-1} \der_2 x_{12}^{-+})} 
 \tr\bigsbrk{\der_1 (x_{12}^{-+,-1} \der_2 x_{12}^{-+})}
\Bigr),
\\
  \bigvev{ \mathcal{F}^+(1) \mathcal{F}^+(2) } &= -\frac {2 g^2}{\pi^2}
  \epsilon^{\alpha \beta} \epsilon^{\gamma \delta} \Xi^{a b c d}(1,2)
  \der_1 (x_{12}^{-+,-1} \bar{\theta}_{12})_{\alpha c}
  \der_1 (x_{12}^{-+,-1} \bar{\theta}_{12})_{\beta d}
\nonumber\\
  &\qquad
\cdot\der_2 (x_{12}^{+-,-1} \bar{\theta}_{12})_{\gamma a}
  \der_2 (x_{12}^{+-,-1} \bar{\theta}_{12})_{\delta b}\, .
\end{align}
Note that in the above we have split off the color factors. 
In the last line we introduced the object
\begin{align}
  \Xi^{abcd}(1,2) &= (x_{12}^{+-})^2 \epsilon^{a'b'cd}
  (1 - 4 \theta_{12} x_{12}^{-+,-1} \bar{\theta}_{12})^a_{\hphantom{a} a'}
  (1 - 4 \theta_{12} x_{12}^{-+,-1} \bar{\theta}_{12})^b_{\hphantom{b} b'}
\nonumber\\
  &= (x_{12}^{-+})^2 \epsilon^{abc'd'}
  (1 + 4 \theta_{12} x_{12}^{+-,-1} \bar{\theta}_{12})^c_{\hphantom{c} c'}
  (1 + 4 \theta_{12} x_{12}^{+-,-1} \bar{\theta}_{12})^d_{\hphantom{d} d'}.
\end{align}
This quantity has the following symmetry properties
\[
  \Xi^{abcd}(1,2) = \Xi^{cdab}(2,1), \qquad
  \Xi^{abcd}(1,2) = -\Xi^{bacd}(1,2), \qquad
  \Xi^{abcd}(1,2) = -\Xi^{abdc}(1,2).
\]
These correlation functions are manifestly super Poincar\'e invariant.


\subsection{Superconformal Algebra}
\label{sec:supconfalgebra}
In preparation for the subsequent sections on the Yangian symmetry 
of super Maldacena--Wilson loops we will now briefly discuss how the superconformal algebra 
can be represented on the relevant non-chiral superspace. 
For the generators of superconformal transformations 
one finds the following differential operator expressions, see \cite{Beisert:2015jxa} 
for more details: 
\begin{subequations}
\label{eqn:sconfgen}
\begin{align}
\label{eqn:sconfgen1}
  \gen{P}_{\alpha \dot{\alpha}} &= -\partial_{\alpha \dot{\alpha}} \, ,
  \\ 
\gen{Q}_{\alpha a} &= \bar{\theta}^{\dot{\alpha}}{}_a \partial_{\alpha \dot{\alpha}} 
- \partial_{\alpha a}\, ,
 \\ 
\bar{\gen{Q}}^a{}_{\dot{\alpha}} &= \theta^{a \alpha} \partial_{\alpha \dot{\alpha}} 
- \bar{\partial}^a{}_{\dot{\alpha}}\, ,
\\
  \gen{L}^{\alpha}{}_{\beta} &= -x^{\dot{\gamma} \alpha} \partial_{\beta \dot{\gamma}} 
- 2  \theta^{c \alpha} \partial_{\beta c} 
+ \tfrac{1}{2} \delta^\alpha_\beta \bigl(x^{\dot{\gamma} \gamma} \partial_{\gamma \dot{\gamma}} 
+ 2  \theta^{c \gamma} \partial_{\gamma c} \bigr)\, ,
\\
\bar{\gen{L}}^{\dot{\alpha}}{}_{\dot{\beta}} &= -x^{\dot{\alpha} \gamma} \partial_{\gamma \dot{\beta}} 
- 2  \bar{\theta}^{\dot{\alpha}}{}_c \bar{\partial}^c{}_{\dot{\beta}} 
+ \tfrac{1}{2} \delta^{\dot{\alpha}}_{\dot{\beta}} \bigl(x^{\dot{\gamma} \gamma} \partial_{\gamma \dot{\gamma}} 
+ 2  \bar{\theta}^{\dot{\gamma}}{}_c \bar{\partial}^c{}_{\dot{\gamma}} \bigr)\, ,
\\
  \gen{D} &= -\tfrac{1}{2} \bigl(x^{\dot{\alpha} \alpha} \partial_{\alpha \dot{\alpha}}
+ \theta^{a \alpha} \partial_{\alpha a} 
+ \bar{\theta}^{\dot{\alpha}}{}_{a} \bar{\partial}^a{}_{\dot{\alpha}}
+ \tfrac{1}{2}   q^{ab} \partial_{ab} \bigr) \, ,
\\
\gen{S}^{a \alpha}&= -(x^+)^{\dot{\delta} \alpha} \theta^{a \delta} \partial_{\delta \dot{\delta}} 
+ 4 \theta^{c \alpha} \theta^{a \gamma} \partial_{\gamma c} + (x^-)^{\dot{\gamma} \alpha} \bar{\partial}^{a}{}_{\dot{\gamma}} 
+ 2  \theta^{c \alpha} q^{a d} \partial_{cd} - \theta^{a \alpha} q^{cd} \partial_{cd}\, , 
\\
\gen{\bar{S}}^{\dot{\alpha}}{}_a 
\ifjournal\else\makebox[0cm][l]{\hspace{-0.4em}\smash{\raisebox{+0.6ex}{\includegraphics[angle=20,height=5mm]{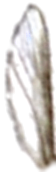}}}}\fi
&= -(x^-)^{\dot{\alpha} \gamma} \bar{\theta}^{\dot{\gamma}}{}_a \partial_{\gamma \dot{\gamma}}
-  4 \bar{\theta}^{\dot{\gamma}}{}_a \bar{\theta}^{\dot{\alpha}}{}_c \bar{\partial}^c{}_{\dot{\gamma}}    
 + (x^+)^{\dot{\alpha} \gamma} \partial_{\gamma a} - 2  \bar{\theta}^{\dot{\alpha}}{}_b q^{b e} \partial_{ae} \, , 
\\
\gen{K}^{\dot{\alpha} \alpha}&=2  \theta^{c \alpha} (x^+)^{\dot{\alpha} \gamma} \partial_{\gamma c} 
+ 2 (x^-)^{\dot{\gamma} \alpha} \bar{\theta}^{\dot{\alpha}}{}_c \bar{\partial}^c{}_{\dot{\gamma}}
 + \tfrac{1}{2}  (x^+)^{\dot{\gamma} \alpha} (x^+)^{\dot{\alpha} \gamma} \partial_{\gamma \dot{\gamma}}
\nonumber \\
 & \qquad 
+ \tfrac{1}{2}  (x^-)^{\dot{\gamma} \alpha} (x^-)^{\dot{\alpha} \gamma} \partial_{\gamma \dot{\gamma}}
 - 4 \theta^{c \alpha} \bar{\theta}^{\dot{\alpha}}{}_a q^{ad} \partial_{cd} 
+ \tfrac{1}{2} (x^+)^{\dot{\alpha} \alpha} q^{cd} \partial_{cd} \, ,\\ 
 \gen{R}^a{}_b&= 2 \bar{\theta}^{\dot{\gamma}}{}_b \bar{\partial}^a{}_{\dot{\gamma}} 
- 2 \theta^{a \gamma} \partial_{\gamma b} -  q^{ad} \partial_{bd}
 - \tfrac{1}{4} \delta^a_b \bigl(2 \bar{\theta}^{\dot{\gamma}}{}_c \bar{\partial}^c{}_{\dot{\gamma}} 
- 2 \theta^{c \gamma} \partial_{\gamma c} -  q^{cd} \partial_{cd} \bigr) \, .
 \label{eqn:sconfgen10}
\end{align}
\end{subequations}
Here, $\partial_{cd}$ denotes the derivative%
\footnote{As usual, we define the derivative 
with spinor indices as $\partial_{cd}=\Sigma_{cd}^i \partial_i$ with $\Sigma_{cd}^i$ 
being the six-dimensional Pauli matrices.} 
with respect to the coordinate $q^{cd}$ 
which enters the super Maldacena--Wilson loop via the coupling to the superscalars. 
This derivative acts on $q^{ab}$ as follows
\begin{align}
\partial_{cd} q^{ab} = 2 \bigl( \delta^a_c \delta^b_d - \delta^a_d \delta^b_c \bigr) \, .
\end{align}
The generators \eqref{eqn:sconfgen}
satisfy the usual superconformal algebra relations 
which we will now discuss in detail. 
We start by listing the commutation relations 
of the $\alg{su}(2)$, $\alg{su}(4)$, $\alg{su}(2)$ rotation generators 
$\gen{L}^{\alpha}{}_{\beta}$, $\gen{R}^a{}_b$, $\bar{\gen{L}}^{\dot{\alpha}}{}_{\dot{\beta}}$. 
Under one of these rotations, the indices of any generator $\gen{J}$ transform canonically according to   
\begin{align}
\bigl[\gen{L}^{\alpha}{}_{\beta} , \gen{J}^\gamma \bigr]
&= -2 \delta^\gamma_\beta \gen{J}^\alpha + \delta^\alpha_\beta \gen{J}^\gamma \, ,  
& \bigl[\gen{L}^{\alpha}{}_{\beta} , \gen{J}_\gamma \bigr]&= 2 \delta^\alpha_\gamma \gen{J}_\beta 
- \delta^\alpha_\beta \gen{J}_\gamma \, , 
\\
\bigl[\bar{\gen{L}}^{\dot{\alpha}}{}_{\dot{\beta}} , \gen{J}^{\dot{\gamma}} \bigr]&= -2 
\delta^{\dot{\gamma}}_{\dot{\beta}} \gen{J}^{\dot{\alpha}} + \delta^{\dot{\alpha}}_{\dot{\beta}} \gen{J}^{\dot{\gamma}} \, ,
 & \bigl[\bar{\gen{L}}^{\dot{\alpha}}{}_{\dot{\beta}} , \gen{J}_{\dot{\gamma}} \bigr]&= 2 
\delta^{\dot{\alpha}}_{\dot{\gamma}} \gen{J}_{\dot{\beta}} - \delta^{\dot{\alpha}}_{\dot{\beta}} \gen{J}_{\dot{\gamma}} \, ,
\\
\bigl[\gen{R}^a{}_b, \gen{J}^c \bigr]&= -2 \delta^c_b \gen{J}^a + \sfrac{1}{2}\delta^a_b \gen{J}^c \, , 
& \bigl[\gen{R}^a{}_b, \gen{J}_c \bigr]&= 2 \delta^a_c \gen{J}_b - \sfrac{1}{2}\delta^a_b \gen{J}_c \, .
\end{align}
The commutators involving the conformal dilatation generator $\gen{D}$ are given by
\begin{align}
\bigl[\gen{D}, \gen{J} \bigr] = \dim(\gen{J}) \, \gen{J} \, ,
\end{align}
with the non-vanishing dimensions
\begin{align}
\dim(\gen{P}) = -\dim(\gen{K})= 1 \, , 
\qquad 
\dim(\gen{Q}) =\dim(\bar{\gen{Q}})= -\dim(\gen{S}) = -\dim(\bar{\gen{S}})= \tfrac{1}{2} \, .
\end{align}
Finally, we need to compute the commutation relations among the translation generators 
$\gen{P}_{\alpha \dot{\alpha}}$, the boost generators $\gen{K}^{\dot{\alpha} \alpha}$
 and their superpartners $\gen{Q}_{\alpha a}$, $\bar{\gen{Q}}^a{}_{\dot{\alpha}}$, 
$\gen{S}^{a \alpha}$ and $\bar{\gen{S}}^{\dot{\alpha}}{}_a$. We find
\begin{align}
\bigl[\gen{Q}_{\beta b}, \gen{K}^{\dot{\alpha} \alpha} \bigr]
&= -2 \delta^\alpha_\beta \bar{\gen{S}}^{\dot{\alpha}}{}_b\, , 
&  \bigl[\bar{\gen{Q}}^b{}_{\dot{\beta}} , \gen{K}^{\dot{\alpha} \alpha} \bigr]
&= -2 \delta^{\dot{\alpha}}_{\dot{\beta}} \gen{S}^{b \alpha}\, , \\ 
\bigl[\gen{P}_{\alpha \dot{\alpha}}, \gen{S}^{b \beta} \bigr]
&= + 2 \delta^\beta_\alpha \bar{\gen{Q}}^b{}_{\dot{\alpha}}\, , 
& \bigl[\gen{P}_{\alpha \dot{\alpha}}, \bar{\gen{S}}^{\dot{\beta}}{}_b \bigr]
&=+ 2 \delta^{\dot{\beta}}_{\dot{\alpha}} \gen{Q}_{\alpha b}\, , \\
\bigl\{ \gen{Q}_{\alpha a}, \bar{\gen{Q}}^b{}_{\dot{\beta}} \bigr\}
&= + 2 \delta_a^b \gen{P}_{\alpha \dot{\beta}}\, , 
&  \bigl\{ \gen{S}^{a \alpha}, \bar{\gen{S}}^{\dot{\beta}}{}_b \bigr\}
&=-2 \delta^a_b \gen{K}^{\dot{\beta} \alpha}\, .
\end{align} 
The remaining non-vanishing commutators are given by
\begin{align}
\bigl[\gen{P}_{\beta \dot{\beta}}, \gen{K}^{\dot{\alpha} \alpha} \bigr]
&=2 \delta^{\dot{\alpha}}_{\dot{\beta}} \gen{L}^{\alpha}{}_{\beta} 
+ 2 \delta^\alpha_\beta \bar{\gen{L}}^{\dot{\alpha}}{}_{\dot{\beta}} 
+ 4 \delta^{\dot{\alpha}}_{\dot{\beta}} \delta^{\alpha}_{\beta} \gen{D}\, , 
\\
\bigl\{ \gen{Q}_{\alpha a}, \gen{S}^{b \beta} \bigr\}
&=-2 \delta^b_a \gen{L}^{\beta}{}_{\alpha} 
+  2 \delta^\beta_\alpha \gen{R}^b{}_a - 2 \delta^b_a \delta^{\beta}_{\alpha} \gen{D}\, , 
\\
\bigl\{ \bar{\gen{Q}}^a{}_{\dot{\alpha}}, \bar{\gen{S}}^{\dot{\beta}}{}_b \bigr\}
&=-2 \delta^a_b \bar{\gen{L}}^{\dot{\beta}}{}_{\dot{\alpha}} 
- 2 \delta^{\dot{\beta}}_{\dot{\alpha}} \gen{R}^a{}_b - 2 \delta^a_b \delta^{\dot{\beta}}_{\dot{\alpha}} \gen{D}\, . 
\end{align}
Note that the central charge vanishes exactly as can be 
read off from the anti-commutators between the generators 
of supertranslations and superboosts. 
Hence, the generators \eqref{eqn:sconfgen} 
form a representation of the algebra $\alg{psu}(2,2|4)$ as expected.

Finally, let us extend the superconformal algebra by the hypercharge generator $\gen{B}$ 
as well as the central charge $\gen{C}$. Together with the generators of $\alg{psu}(2,2|4)$ 
these generators span the algebra $\alg{u}(2,2|4)$.
The motivation for this extension is twofold. 
First, it will enable us to discuss level-one hypercharge symmetry \cite{Matsumoto:2007rh,Beisert:2011pn}, 
which we will do later on. Second, the hypercharge generator will 
help us to simplify the computation of the level-one momentum remainder term. 
On our superspace the generators $\gen{B}$ and $\gen{C}$ are represented by
\begin{align}
\gen{B} = \tfrac{1}{2}   \theta^{a \alpha} \partial_{\alpha a}
- \tfrac{1}{2} \bar{\theta}^{\dot{\alpha}}{}_a \bar{\partial}^a{}_{\dot{\alpha}}\, , \hspace{2cm} \gen{C} = 0\, .
\end{align}
While $\gen{C}$ commutes with everything, the hypercharge generator 
satisfies commutation relations which are similar to the ones satisfied 
by the dilatation generator $\gen{D}$. More precisely, one finds
\begin{align}
\bigcomm{\gen{B}}{\gen{J}} = \operatorname{hyp}(\gen{J}) \, \gen{J} \, ,
\end{align}
with the non-vanishing hypercharges
\begin{align}
\operatorname{hyp}(\bar{\gen{Q}})=-\operatorname{hyp}(\gen{Q}) 
=  \operatorname{hyp}(\gen{S}) = -\operatorname{hyp}(\bar{\gen{S}})= \tfrac{1}{2} \, .
\end{align}
As a closing comment we would like to stress again the point 
that all the generators mentioned above commute with the ten-dimensional light-likeness 
constraint $p^2-q^2=0$ modulo terms which vanish on the constraint surface, see \cite{Beisert:2015jxa}.

\subsection{G-Identity}
\label{sec:g-identity-new}

In this section we want to prove an identity that plays quite an important role
 in our discussion on Yangian symmetry of smooth super Wilson loops. 
It reads
\begin{align}
f^\idxconf{e}{}_{\idxconf{b}\idxconf{c}} \,   \sconfrem^{\idxconf{c} \idxconf{b}} \simeq 0 \, ,
\label{eqn:Gid1}
\end{align}
where $f^\idxconf{e}{}_{\idxconf{b}\idxconf{c}}$ are the structure constants 
of $\alg{u}(2,2|4)$
\footnote{Here and in the following we will not use different symbols 
for the structure constants of $\alg{u}(2,2|4)$ and $\alg{psu}(2,2|4)$ 
as it will be clear from the context which structure constants are meant.} 
and $\sconfrem^{\idxconf{c} \idxconf{b}}$ is 
(including all fermionic signs) given by
\begin{align}
\sconfrem^{\idxconf{c} \idxconf{b}} 
&=-(-1)^{|\idxall{A}| |\idxconf{b}|} \, \gen{J}^{\idxconf{c}}_{\text{cov}}  X^\idxall{A} \, 
\gen{J}^{\idxconf{b}}_{\text{cov}} X^\idxall{B}  \, \sfstr^\text{cov}_{\idxall{B}\idxall{A}}
 \nonumber \\ 
&=  - i[\gen{J}^\idxconf{c} ] \, i[\gen{J}^\idxconf{b}] \, 
\Bigl(-\sfrac{1}{2} \sviel^{\idxall{A}} \wedge \sviel^{\idxall{B}} \sfstr^\text{cov}_{\idxall{B}\idxall{A}}  \Bigr) 
\nonumber \\
&= -i[\gen{J}^\idxconf{c}] \, i[\gen{J}^\idxconf{b}] \, \sfstr \, .
\end{align}
Here $i[\gen{J}^\idxconf{b}] \, \sfstr$ denotes the 
contraction%
\footnote{Given a purely bosonic $p$-form 
$\omega=1/r! \, \, \omega_{ \mu_1 \ldots \mu_r} \der x^{\mu_1} \wedge \ldots \wedge \der x^{\mu_r}$ 
the interior product $i[X]$ between $\omega$
and the vector field $X=X^\mu \partial_\mu$ 
is defined as $i[X] \, \omega=1/(r-1)! \: X^\nu \omega_{ \nu \mu_2 \ldots \mu_r} 
\der x^{\mu_2} \wedge \ldots \wedge \der x^{\mu_r}$.
The generalization of this concept to forms on superspace 
is straightforward since the only difference is that
in superspace one has to take into account the Gra{\ss}mann 
degree of the various objects and the associated signs.}
of the vector field 
$\gen{J}^\idxconf{c}=\gen{J}^\idxconf{c}_{\text{cov}} X^{\idxall{A}} \, \sdel_{\idxall{A}}$ 
with the field strength two-form. Note that in relation \eqref{eqn:Gid1} 
we have tactically extended the underlying algebra 
from $\alg{psu}(2,2|4)$ to $\alg{u}(2,2|4)$.
The reason for this is, as explained above, 
that we want to discuss the invariance of our Wilson loop under level-one hypercharge transformations. 
An important point to note, however, is that once we have proven the G-identity for $\alg{u}(2,2|4)$ 
the $\alg{psu}(2,2|4)$ version follows immediately. This becomes obvious by noting that the index range of 
the summed indices in \eqref{eqn:Gid1} can actually be restricted to $\alg{psu}(2,2|4)$ 
as the differential operator corresponding to $\gen{C}$ vanishes exactly, see \secref{sec:supconfalgebra}. 

Our strategy to prove \eqref{eqn:Gid1} is the following: 
first, we will show that the G-identity holds for $\idxconf{e}$ corresponding 
to the hypercharge generator $\gen{B}$. In the next step we will then use 
the superconformal transformation properties of $\sconfrem^{\idxconf{c} \idxconf{b}}$ 
to argue that \eqref{eqn:Gid1} holds for any $\alg{u}(2,2|4)$ index $\idxconf{e}$. 
Note that for $\idxconf{e}$ corresponding to $\gen{C}$ the G-identity is trivial 
since the structure constants vanish in this case.

If we fix $\idxconf{e}$ to be equal to $\gen{B}$ equation \eqref{eqn:Gid1} becomes:
\begin{align}
f^{\idxconf{e}}{}_{\idxconf{b} \idxconf{c}} \, \sconfrem^{\idxconf{c} \idxconf{b}} 
\propto  \left( i \bigl[\gen{S}^{b \beta} \bigr]    i \bigl[ \gen{Q}_{\beta b} \bigr]  -
   i \bigl[\gen{\bar{S}}^{\dot{\beta}}{}_b \bigr]   i \bigl[\gen{\bar{Q}}^b{}_{\dot{\beta}} \bigr] \right) \sfstr \, .
   \label{eqn:GidB}
\end{align}
In order to evaluate this expression it is useful to first compute the contractions 
between the vector fields $\gen{Q}, \gen{\bar{Q}}, \gen{S}, \gen{\bar{S}}$ 
and the basis one-forms. We find
\begin{subequations}
\begin{align}
i \bigl[ \gen{Q}_{\beta b} \bigr]  \svielB^{\dot{\alpha} \alpha} &= 4 \delta^\alpha_\beta \bar{\theta}^{\dot{\alpha}}{}_b \, ,
&\qquad
i \bigl[\gen{Q}_{ \beta b} \bigr] \svielF^{a \alpha} &= -\delta^a_b \delta^\alpha_\beta \, ,
\\
i \bigl[ \gen{\bar{Q}}^b{}_{\dot{\beta}} \bigr]  \svielB^{\dot{\alpha} \alpha} 
&= 4 \delta^{\dot{\alpha}}_{\dot{\beta}} \theta^{b \alpha} \, ,
&\qquad
i \bigl[\gen{\bar{Q}}^b{}_{\dot{\beta}} \bigr]  \svielF^{a \alpha} &= 0  \, , \\
i \bigl[ \gen{S}^{b \beta} \bigr]  \svielB^{\dot{\alpha} \alpha} &= -4 (x^-)^{\dot{\alpha} \beta} \theta^{b \alpha}\, ,
&\qquad
i \bigl[\gen{S}^{b \beta} \bigr]  \svielF^{a \alpha} &= 4 \theta^{a \beta} \theta^{b \alpha}\, , 
\\
i \bigl[ \gen{\bar{S}}^{\dot{\beta}}{}_b \bigr]  \svielB^{\dot{\alpha} \alpha} 
&= - 4  \bar{\theta}^{\dot{\alpha}}{}_b (x^+)^{\dot{\beta} \alpha} \, ,
&\qquad
i \bigl[ \gen{\bar{S}}^{\dot{\beta}}{}_b \bigr]  \svielF^{a \alpha} &= \delta^a_b (x^+)^{\dot{\beta} \alpha} \, ,
\\
i \bigl[ \gen{Q}_{ \beta b} \bigr]  \svielC^{\dot{\alpha}}{}_a  &= 0  \, ,
&\qquad
i \bigl[\gen{\bar{Q}}^b{}_{\dot{\beta}} \bigr]  \svielC^{\dot{\alpha}}{}_a  
&= -\delta^b_a \delta^{\dot{\alpha}}_{\dot{\beta}} \, , \\
i \bigl[\gen{S}^{b \beta} \bigr]  \svielC^{\dot{\alpha}}{}_a  &= \delta^b_a (x^-)^{\dot{\alpha} \beta} \, ,
&\qquad
i \bigl[\gen{\bar{S}}^{\dot{\beta}}{}_b \bigr]  \svielC^{\dot{\alpha}}{}_a  
&= -4 \bar{\theta}^{\dot{\alpha}}{}_b \bar{\theta}^{\dot{\beta}}{}_a  \, .
\end{align}
\end{subequations}
Given these formulas we can now apply the double contraction expression of \eqref{eqn:GidB} 
to all the basis two-forms appearing in the decomposition 
of the field strength superfield \eqref{eqn:sfstrF}. 
Note that due to the constraints \eqref{eqn:N=4constraints} the field strength superfield 
contains only a restricted set of basis two-forms. 
In particular, the coefficient of the mixed fermionic two-form is set to zero 
by \eqref{eqn:N=4constraints1b}. 
A useful formula for computing \eqref{eqn:GidB} is 
\(i[X] i[Y] (\alpha \wedge \beta) = - (i[X] \alpha)(i[Y] \beta) + (i[Y] \alpha) (i[X] \beta)\), 
where \(X, Y\) are vector fields and \(\alpha, \beta\) are one-forms 
(this holds when all the quantities are Gra{\ss}mann even; if not, extra signs are needed).  
As an example of a
computation where fermionic signs appear, consider
\begin{align}
&\left( i \bigl[ \gen{S}^{c \gamma} \bigr] i \bigl[ \gen{Q}_{\gamma c} \bigr]  -
   i \bigl[ \gen{\bar{S}}^{\dot{\gamma}}{}_c \bigr]  
i \bigl[ \gen{\bar{Q}}^c{}_{\dot{\gamma}} \bigr] \right) 
\left( \svielF^{a \alpha} \, \epsilon_{\alpha \beta} \, \svielF^{b \beta}\right) 
\nonumber \\
   &= i \bigl[ \gen{S}^{c \gamma} \bigr] \Bigl[ \bigl( i \bigl[ \gen{Q}_{\gamma c} \bigr]  
\svielF^{a \alpha} \bigr) \epsilon_{\alpha \beta} \, \svielF^{b \beta} + \svielF^{a \alpha} 
\, \epsilon_{\alpha \beta} \bigl( i \bigl[ \gen{Q}_{\gamma c} \bigr]  \svielF^{b \beta} \bigr) \Bigr] 
\nonumber \\
 &=\epsilon_{\alpha \beta} \bigsbrk{4  \theta^{a \beta} \theta^{b \alpha} - 4  \theta^{a \beta} \theta^{b \alpha} } 
\nonumber \\
 &=0 \, .
\end{align}
In a similar way one can show that the double contraction expression in \eqref{eqn:GidB} 
also annihilates all the other basis two-forms appearing in \eqref{eqn:sfstrF}. 

Having proven the G-identity for $\idxconf{e}$ 
corresponding to $\gen{B}$ we now proceed and argue that 
it actually holds for any $\alg{u}(2,2|4)$ index $\idxconf{e}$. 
To do so, we use the following identity 
\begin{align}
\gengauge{J}^\idxconf{a} \sconfrem^{\idxconf{c} \idxconf{b}} 
=-\genfull{J}^\idxconf{a}\sconfrem^{\idxconf{c}\idxconf{b}}
+ f^{\idxconf{a}\idxconf{c}}{}_{\idxconf{d}} \sconfrem^{\idxconf{d}\idxconf{b}}
+ (-1)^{|\idxconf{a}| |\idxconf{c}|} f^{\idxconf{a}\idxconf{b}}{}_{\idxconf{d}} \sconfrem^{\idxconf{c}\idxconf{d}} \, ,
\label{eqn:Gconftraf}
\end{align}
where $\gengauge{J}^\idxconf{a} = \gen{J}^\idxconf{a}_{\text{cov}} X^\idxall{A} \scdel_\idxall{A}$ and $\genfull{J}^\idxconf{a}$ 
is a composite superconformal generator 
of field transformations which acts on $\sgauge^{\text{cov}}_{\idxall{A}}$ 
as defined in \eqref{eq:fullconfacta}
\begin{align}
\genfull{J}^\idxconf{a} \sgauge^{\text{cov}}_{\idxall{A}}
=- \gen{J}^\idxconf{a}_{\text{cov}} X^{\idxall{B}} \, \sfstr^{\text{cov}}_{\idxall{B} \idxall{A}} \, .
\label{eqn:Jstar_sgauge_cov}
\end{align}
Equation \eqref{eqn:Gconftraf}, simply stating that $\gengauge{J}^\idxconf{a}$ acts on $\sconfrem^{\idxconf{c}\idxconf{b}}$ 
by transforming the field as well as the conformal indices, 
can easily be proven by acting on $\sconfrem^{\idxconf{c}\idxconf{b}}$ 
with $\genfull{J}^\idxconf{a}$ and using the Bianchi identity 
as well as the superconformal commutation relation. 
If we contract equation \eqref{eqn:Gconftraf} with $f^{\idxconf{e}}{}_{\idxconf{b} \idxconf{c}}$ 
and rewrite the last two terms using a supersymmetric version of the Jacobi identity
\begin{align}
  f^{\idxconf{e}}{}_{\idxconf{b} \idxconf{c}} f^{\idxconf{a} \idxconf{c}}{}_{\idxconf{d}} 
+ (-1)^{|\idxconf{a}| |\idxconf{d}|}
f^{\idxconf{e}}{}_{\idxconf{c} \idxconf{d}} f^{\idxconf{a} \idxconf{c}}{}_{\idxconf{b}}
= f^{\idxconf{a} \idxconf{e}}{}_{\idxconf{c}} f^{\idxconf{c}}{}_{\idxconf{b} \idxconf{d}} \, ,
\label{eqn:susy_jacobi}
\end{align} 
we get 
\begin{align}
\bigl(\gengauge{J}^\idxconf{a}+ \genfull{J}^\idxconf{a} \bigr)  
f^{\idxconf{e}}{}_{\idxconf{b} \idxconf{c}} \,  \sconfrem^{\idxconf{c} \idxconf{b}}  
= f^{\idxconf{a} \idxconf{e}}{}_{\idxconf{c}} f^{\idxconf{c}}{}_{\idxconf{b} \idxconf{d}} 
\sconfrem^{\idxconf{d}\idxconf{b}} \, .
\end{align}

Now, the important thing that we need to recall is that the $\mathcal{N}=4$ 
SYM constraints \eqref{eqn:N=4constraints} are preserved by a $\alg{psu}(2,2|4)$ transformation. 
Thus, if $\idxconf{e}$ is fixed to $\gen{B}$ and we choose $\idxconf{a}$ 
to be equal to a $\alg{psu}(2,2|4)$ index the left-hand side of the above equation 
vanishes leaving us with
\begin{align}
f^{\idxconf{a} \idxconf{e}}{}_{\idxconf{c}} f^{\idxconf{c}}{}_{\idxconf{b} \idxconf{d}} 
\, \sconfrem^{\idxconf{d}\idxconf{b}}=0 \, .
\label{eqn:Gidtransf}
\end{align}
The last equation immediately allows for the conclusion that the G-identity \eqref{eqn:Gid1} 
also holds for $\idxconf{e}$ corresponding to $\gen{Q}$, $\gen{\bar{Q}}$, $\gen{S}$ and $\gen{\bar{S}}$. 
Having proven the G-identity for $\gen{B}$ 
as well as for the four former mentioned indices one can now argue 
that it holds for any $\alg{u}(2,2|4)$ index $\idxconf{e}$ 
by iterating the argument with the free indices properly chosen.

\section{Yangian Invariance at One Loop}
\label{sec:oneloop}

In this section we would like to demonstrate Yangian invariance
of the quantum expectation value of a generic smooth Maldacena--Wilson loop
at the first perturbative order.

At the one-loop level there is only one
term contributing to the expectation value of a Wilson loop:
The Wilson loop must be expanded to two fields
which are consequently connected by a propagator.
\[
\frac{1}{N}\, 
\bigvev{\tr \swilson}_{(1)} = \frac{N}{4}\, 
\int_{1,2}
\bigvev{(\sgauge+\sscalar)(1)\, (\sgauge+\sscalar)(2)}.
\]
All effects of a non-abelian gauge group are
irrelevant at this level, and we will effectively
assume an abelian gauge group.

The bi-local term of the Yangian action \eqref{eq:newyangact}
already contains two instances of the field strength $\mathcal{F}$
(or of the scalar field $\sscalar$ which is in fact included in $\mathcal{F}$).
\[
\frac{1}{N}\, 
\genYfullnl{J}^{\idxconf{c}}
\bigvev{\tr \swilson}_{(1)} = \frac{N}{2}\, 
f^\idxconf{c}{}_{\idxconf{b}\idxconf{a}}
\int_{1<2}
\bigvev{\genfull{J}^\idxconf{a}(\sgauge+\sscalar)(1)\, \genfull{J}^\idxconf{b}(\sgauge+\sscalar)(2)}.
\]
All higher powers of the fields are discarded
for the one-loop/abelian expectation value,
and the field strengths are effectively given by $\sfstr \to \sfstr_{\text{lin}}= \der \sgauge$.
In the following we will denote the linearized field strength by $\sfstr$ for simplicity.

\subsection{Symmetry of the Gauge Propagator}
\label{sec:sym_gauge_propagator}

We start by investigating the bi-local action on the gauge propagator%
\footnote{When the generator is written outside the correlator it is defined to act on all the fields inside.}
\[
C^\idxconf{c}_{12}=
\genYfullnl{J}^\idxconf{c} \bigvev{ \sgauge(1)\,\sgauge(2)}
=
f^\idxconf{c}{}_{\idxconf{b}\idxconf{a}}\bigvev{\genfull{J}^\idxconf{a}\sgauge(1)\, \genfull{J}^\idxconf{b}\sgauge(2)}.
\]
Now we perform a ``partial integration'' on one of the constituent conformal generators,
i.e.\ we extend its action to the other field and subtract the 
additional action on the second field
\begin{align}
f^\idxconf{c}{}_{\idxconf{b}\idxconf{a}} \bigvev{\genfull{J}^\idxconf{a} \sgauge(1)\, \genfull{J}^\idxconf{b}\sgauge(2)} 
&=
 f^\idxconf{c}{}_{\idxconf{b}\idxconf{a}} \genfull{J}^\idxconf{a}\bigvev{\sgauge(1)\, \genfull{J}^\idxconf{b}\sgauge(2)} 
-f^\idxconf{c}{}_{\idxconf{b}\idxconf{a}} \bigvev{\sgauge(1)\, \genfull{J}^\idxconf{a}\genfull{J}^\idxconf{b}\sgauge(2)}  \, .
\end{align}
The conformal generator in the first term acts on both fields
and thus corresponds to a symmetry action on an expectation value.
Symmetry implies that the expectation value 
of the physical degrees of freedom is zero, neglecting potential contact terms which are
not relevant here.
However, one has to watch out because one of the fields is a
gauge potential which contains unphysical degrees of freedom.
The latter are not constrained by conformal symmetry,
and we conclude that the result takes the form of a total derivative 
of some function $B_{12}$
\[
 f^\idxconf{c}{}_{\idxconf{b}\idxconf{a}} \genfull{J}^\idxconf{a}\bigvev{\sgauge(1)\, \genfull{J}^\idxconf{b}\sgauge(2)} 
=\der_1 B_{12}^{\idxconf{c}}\, ,
\]
which should hold at the leading perturbative order.
For the second term we use the graded anti-symmetry of the conformal indices
and apply the algebra relation \eqref{eq:confcomm}
\begin{align}
f^\idxconf{c}{}_{\idxconf{b}\idxconf{a}} 
\bigvev{\sgauge(1)\, \genfull{J}^\idxconf{a}\genfull{J}^\idxconf{b}\sgauge(2)} 
&=
\half f^\idxconf{c}{}_{\idxconf{b}\idxconf{a}} 
\bigvev{\sgauge(1)\, \gcomm{\genfull{J}^\idxconf{a}}{\genfull{J}^\idxconf{b}}\sgauge(2)} 
\nln
&=
\half 
f^\idxconf{c}{}_{\idxconf{b}\idxconf{a}}
f^{\idxconf{a}\idxconf{b}}{}_\idxconf{d}
\bigvev{\sgauge(1)\, \genfull{J}^\idxconf{d}\sgauge(2)} 
-
\half 
f^\idxconf{c}{}_{\idxconf{b}\idxconf{a}}
\bigvev{\sgauge(1)\, \cder \sconfrem^{\idxconf{a}\idxconf{b}}(2)} .
\end{align}
As in the argument towards cyclicity in \secref{sec:consistency}
both terms are zero; the former because the dual Coxeter number of
the superconformal algebra is zero%
\footnote{In fact, for $\alg{u}(2,2|4)$ 
the combination $f^\idxconf{c}{}_{\idxconf{b}\idxconf{a}} f^{\idxconf{a}\idxconf{b}}{}_\idxconf{d}$ is not zero, 
but rather proportional to $\delta^{\idxconf{c}}_{\gen{B}} \delta_{\idxconf{d}}^{\gen{C}}$. 
However, note that since $\gen{C}$ is always represented by zero (see \secref{sec:supconfalgebra}), 
this expression effectively vanishes as well.}, 
the latter identity was shown in \secref{sec:g-identity-new}. 
Combining the results we find $C^\idxconf{c}_{12}=\der_1 B^\idxconf{c}_{12}$. 
Equivalently, $C^\idxconf{c}_{12}=-\der_2 B^\idxconf{c}_{21}$. 
Altogether this implies that $C_{12}$ is in fact the double derivative
of some function $R_{12}$
\[\label{eq:propsym}
\genYfullnl{J}^\idxconf{c} \bigvev{ \sgauge(1)\,\sgauge(2)}=\der_1 \der_2 R^\idxconf{c}_{12}.
\]
This relation can be independently checked through an explicit calculation at leading
perturbative order, see \secref{sec:RemFunct}.

\subsection{Symmetry of the Wilson Loop}
\label{sec:symwil}

For the Maldacena--Wilson loop, we furthermore need
the bi-local action on correlators involving scalar fields
$\vev{\sgauge\sscalar}$ and $\vev{\sscalar\sscalar}$.
These follow from the above considerations by noting that
the scalar field is a particular component
of the field strength $\sscalar\in \sfstr = \der \sgauge$.
Namely \eqref{eq:propsym} implies 
\[
\genYfullnl{J}^\idxconf{c} \bigvev{ \sfstr(1)\,\sgauge(2)}
=\genYfullnl{J}^\idxconf{c} \bigvev{ \der\sgauge(1)\,\sgauge(2)}
=\der_1\der_1 \der_2 R^\idxconf{c}_{12}=0.
\]
Consequently also, 
\begin{align}
\genYfullnl{J}^\idxconf{c} \bigvev{ \sfstr(1)\,\sfstr(2)}&=0,
\\
\genYfullnl{J}^\idxconf{c} \bigvev{ \sscalar(1)\,\sgauge(2)}&=0,
\\
\genYfullnl{J}^\idxconf{c} \bigvev{ \sscalar(1)\,\sscalar(2)}&=0.
\end{align}
Now we can, in principle, use the bi-local symmetry 
of the gauge fields \eqref{eq:propsym} 
towards Yangian symmetry of the Wilson loop expectation value
\begin{align}
\label{eq:naivewilsoninvariance}
\genYfullnl{J}\bigvev{\tr\swilson}
&\sim
\int_{1<2}
\der_1 \der_2 R_{12}
=
 \int\eval{\der_2 R_{12}}_{1=2}
-\int \der_2 R_{02}
\nln
&=
\int\eval{\der_2 R_{12}}_{1=2}
-R_{00}+R_{00}
=
\int\eval{\der_2 R_{12}}_{1=2}.
\end{align}
This result is not zero, but all the bi-local and bulk-boundary 
contributions have gone away.
What remains is a local insertion 
which we could compensate by an appropriate local contribution 
to the Yangian action.
However, to show invariance is not as straight-forward because
the function $R_{12}$ is in fact UV-divergent when the two
points approach each other. 
In the above derivation we have been careless w.r.t.\ this option,
and we need to repeat the derivation in the presence of a regulator.
Before doing so, we should compute the remainder functions $R_{12}$.

\subsection{Remainder Functions}
\label{sec:RemFunct}

In the previous two sections we have argued that the only contribution 
to the remainder term $R^{\idxconf{d}}_{12}$ 
comes from the correlator of the product of two gauge fields 
to which the respective bi-local level-one generator $\genYfullnl{J}^\idxconf{d}$ 
has been applied
\begin{align}
 \genYfullnl{J}^\idxconf{d} \bigvev{\sgauge(1) \sgauge(2) }
=f^{\idxconf{d}}{}_{\idxconf{b}\idxconf{c}} 
\bigvev{ \genfull{J}^{\idxconf{c}} \sgauge(1) \, \genfull{J}^{\idxconf{b}} \sgauge(2) }
=\der_1 \der_2 R^\idxconf{d}_{12}\, .
\label{eqn:Jstar_sgauge_sgauge}
\end{align}
In this section we will determine all the remainder functions $R^\idxconf{d}_{12}$. 
For this we will first explicitly compute the level-one momentum 
and the level-one hypercharge remainder function and then determine 
the other remainder functions by algebra considerations. 
In principle, it would be sufficient to only compute the level-one hypercharge 
remainder function explicitly as all the other remainder terms can be derived from this one.%
\footnote{Note that this statement 
does not hold for the level-one momentum remainder function 
as the level-one hypercharge remainder function cannot be derived from $R_{12}[\gen{P}]$.} 
However, since $\genYfull{P}$ is the more generic Yangian generator 
we will mainly focus on this one in the subsequent discussion 
and in return shorten the level-one hypercharge discussion.

We start by rewriting the left-hand side of  \eqref{eqn:Jstar_sgauge_sgauge} 
in such a way that we can take advantage of the fact that we already know the two-point function 
of the field strength two-form, see \secref{sec:two-point-functions}. 
By plugging in the definition of $\genfull{J}^{\idxconf{c}} \sgauge$ \eqref{eqn:Jstar_sgauge_cov}, 
the left-hand side of equation \eqref{eqn:Jstar_sgauge_sgauge} becomes
\begin{align}
\genYfullnl{J}^\idxconf{d}\bigvev{  \sgauge(1) \sgauge(2) }
=f^\idxconf{d}{}_{\idxconf{b}\idxconf{c}} \, 
\bigvev{ 
\gen{J}_{\text{cov}}^\idxconf{c} X_1^\idxall{A} \, \sviel_1^{\idxall{B}} \, \sfstr^{\text{cov}}_{1 \idxall{B} \idxall{A}}\, 
\gen{J}_{\text{cov}}^\idxconf{b} X_2^\idxall{C} \, \sviel_2^{\idxall{D}} \, \sfstr^{\text{cov}}_{2 \, \idxall{D} \idxall{C}} 
} .
\end{align}
Using the interior product notation that we have introduced in \secref{sec:g-identity-new} 
the former expression can be rewritten as 
\begin{align}
\genYfullnl{J}^\idxconf{d}
\bigl\langle \, \sgauge(1) \sgauge(2) \bigr\rangle
=f^\idxconf{d}{}_{\idxconf{b}\idxconf{c}} \, i[\gen{J}^{\idxconf{c}}_1] \, i[\gen{J}^{\idxconf{b}}_2] 
\, \bigl\langle \sfstr(1) \sfstr(2) \bigr\rangle \, .
\label{eqn:Phat_doublecontraction}
\end{align}
By taking into account the decomposition 
of the field strength  $\mathcal{F}= \mathcal{F}^+ + \mathcal{F}^-$ 
we see that computing the remainder term for $\genYfullnl{P}$ 
effectively amounts to calculating the following two contraction expressions
($\idxconf{d}\sim\gen{P}$):
\begin{align}
\label{eqn:Phat_contraction_chiral}
&f^{\idxconf{d}}{}_{\idxconf{b}\idxconf{c}} 
\, i[\gen{J}^{\idxconf{c}}_1] \, i[\gen{J}^{\idxconf{b}}_2] 
\, \bigl\langle \sfstr^+(1) \sfstr^+(2) \bigr\rangle  \hspace{1.5cm} \bigl(\mbox{chiral contributions} \bigr) \, ,  
\\
&f^{\idxconf{d}}{}_{\idxconf{b}\idxconf{c}} 
\, i[\gen{J}^{\idxconf{c}}_1] \, i[\gen{J}^{\idxconf{b}}_2] 
\, \bigl\langle \sfstr^+(1) \sfstr^-(2) \bigr\rangle \hspace{1.5cm} \bigl(\mbox{mixed-chiral contributions} \bigr) \, .
\label{eqn:Phat_contraction_mchiral}
\end{align} 
Having computed these, the full result can be constructed by symmetry considerations. 

\paragraph{Level-one momentum remainder function.}

The bi-local part of the level-one momentum generator $\genYfull{P}$
can be written as combinations 
of the superconformal generators $\genfull{P}$ (bosonic translations), 
$\genfull{Q}$, $\genfull{\bar Q}$ (fermionic translations), 
$\genfull{L}$, $\genfull{\bar L}$ (rotations) and $\genfull{D}$ (dilatations)
\begin{align}
\label{eq:Phat_field}
\genYfullnl{P}{}_{,  \alpha\dot\alpha}
=
\genfull{P}{}_{,\beta\dot{\alpha}} \wedge  \genfull{L}^\beta{}_{\alpha}
+\genfull{P}{}_{,\alpha\dot{\beta}} \wedge  \genfull{\bar L}^{\dot\beta}{}_{\dot\alpha}
+2 \genfull{P}{}_{, \alpha \dot\alpha} \wedge \genfull{D}
+\genfull{Q}{}_{,\alpha a} \wedge \genfull{\bar Q}^a{}_{\dot\alpha} \, .
\end{align}
Here the action of a bi-local term $(\genfull{J}\wedge\genfull{J}')$ is
defined in analogy to \eqref{eq:newyangact} as
\begin{align}
\bigl(\genfull{J} \wedge \genfull{J}' \bigr) \mathcal{W} =
\mathcal{W}\bigl[\genfull{J}\bigl(\sgauge + \sscalar \bigr);
\genfull{J}' \bigl(\sgauge + \sscalar \bigr) \bigr] \!
-(-1)^{|\genfull{J}||\genfull{J}'|}
\mathcal{W}\bigl[\genfull{J}' \bigl(\sgauge + \sscalar \bigr);
\genfull{J}\bigl(\sgauge + \sscalar \bigr) \bigr].
\label{eqn:JwedgeJprimeW}
\end{align}
It turns out to be convenient to rewrite the expression for $\genYfullnl{P}$ 
by introducing the following modified rotation generators
\begin{align}
  \genfull{L}'{}^{\! \alpha}{}_{\beta}
&=  \genfull{L}^{\alpha}{}_{\beta} 
+ \delta_\beta^\alpha \bigl(\genfull{D} - \genfull{B}\bigr),
\\
  \genfull{\bar{L}}'{}^{\! \dot{\alpha}}{}_{\dot{\beta}}
&=  \genfull{\bar{L}}^{\dot{\alpha}}{}_{ \dot{\beta}}
+ \delta_{\dot{\beta}}^{\dot{\alpha}} \bigl(\genfull{D} + \genfull{B} \bigr).
\label{eqn:modified_rotations}
\end{align}
Using these definitions, the Yangian generator $\genYfullnl{P}$ can be rewritten as
\begin{align}
\genYfullnl{P}{}_{,  \alpha\dot\alpha}
=
\genfull{P}{}_{,\beta\dot{\alpha}}  \wedge \genfull{L}'{}^{\! \beta}{}_{\alpha}
+\genfull{P}{}_{,\alpha\dot{\beta}} \wedge \genfull{\bar{L}}'{}^{\! \dot{\beta}}{}_{\dot{\alpha}}
+\genfull{Q}{}_{,\alpha a} \wedge \genfull{\bar Q}^a{}_{\dot\alpha} \, .
\end{align}
From this equation we can now directly read off the double contraction operator 
(c.f.\ \eqref{eqn:Phat_doublecontraction}) 
that we want to use for the computation of the remainder term $R_{12}[\gen{P}]$. 
It reads
\begin{align}
f^{\idxconf{d}}{}_{\idxconf{b}\idxconf{c}} 
\, i\bigl[\gen{J}^{\idxconf{c}}_1 \bigr] \, i\bigl[\gen{J}^{\idxconf{b}}_2 \bigr]
=&+i \bigl[\gen{P}_{1, \beta \dot{\alpha}} \bigr] 
\, i \bigl[\gen{L}'_2{}^{\! \beta}{}_{\alpha} \bigr]
+i \bigl[\gen{P}_{1, \alpha \dot{\beta}} \bigr] \, 
i \bigl[\gen{\bar{L}}'_2{}^{\! \dot\beta}{}_{\dot\alpha} \bigr] 
+i \bigl[\gen{Q}_{1, \alpha a} \bigr] \, 
i \bigl[\gen{\bar{Q}}^{a}_{2 \;  \dot\alpha} \bigr] 
\nonumber \\
&-  i \bigl[\gen{L}'_1{}^{\! \beta}{}_{\alpha} \bigr] \, 
i \bigl[\gen{P}_{2, \beta \dot{\alpha}} \bigr]
- i \bigl[\gen{\bar{L}}'_1{}^{ \! \dot\beta}{}_{\dot\alpha} \bigr] \, 
i \bigl[\gen{P}_{2, \alpha \dot{\beta}} \bigr] 
 +i \bigl[\gen{\bar{Q}}^{a}_{1 \;  \dot\alpha} \bigr]  
\, i \bigl[\gen{Q}_{2, \alpha a} \bigr] \, ,
\label{eqn:Phat_contraction_expl}
\end{align}
where here and in the following the index $\idxconf{d}$ corresponds to 
the generator $\gen{P}_{\alpha\dot\alpha}$.

We start with the computation of the remainder term $R_{12}[\gen{P}]$
by investigating the chiral contributions, see \eqref{eqn:Phat_contraction_chiral}. 
The relevant correlation function was computed in \secref{sec:two-point-functions} and is given by
\begin{align}
  \bigvev{ \mathcal{F}^+(1) \mathcal{F}^+(2) } 
&= -\frac {2 g^2}{\pi^2}
  \epsilon^{\alpha \beta} \epsilon^{\gamma \delta} \Xi^{a b c d}(1,2)
   \, \der_1 \bigl(x_{12}^{-+,-1} \bar{\theta}_{12} \bigr)_{\alpha c}
  \, \der_1 \bigl(x_{12}^{-+,-1} \bar{\theta}_{12} \bigr)_{\beta d}
\nonumber\\&\qquad\cdot
  \der_2 \bigl(x_{12}^{+-,-1} \bar{\theta}_{12} \bigr)_{\gamma a}
  \, \der_2 \bigl(x_{12}^{+-,-1} \bar{\theta}_{12} \bigr)_{\delta b} \, .
\end{align}
By symmetry it is sufficient to compute the action of the double contraction operator \eqref{eqn:Phat_contraction_expl} on 
$W_{\gamma c, \delta d}=
\der_1 (x_{12}^{-+,-1} \bar{\theta}_{12})_{\gamma c}\, \der_2 (x_{12}^{+-,-1} \bar{\theta}_{12})_{\delta d}$. 
Note that the contraction operators $i[\ldots]$ effectively just replace (modulo a sign) 
the corresponding exterior derivative $\der$ by the differential operator specified in the square brackets. 
For example, we have
\begin{align}
i \bigl[\gen{P}_{1, \beta \dot{\alpha}} \bigr] \, i \bigl[\gen{L}^{\prime \beta}_{2 \; \, \alpha} \bigr] 
\, W_{\gamma c, \delta d}&= - \gen{P}_{1, \beta \dot{\alpha}} 
\bigl(x_{12}^{-+,-1} \bar{\theta}_{12} \bigr)_{\gamma c} \,  \gen{L}^{\prime \beta}_{2 \; \, \alpha} 
\bigl(x_{12}^{+-,-1} \bar{\theta}_{12} \bigr)_{\delta d} 
\nonumber \\
&=4 \bigl(x_{12}^{-+,-1}\bigr)_{\gamma \dot\alpha} 
 \bigl(x_{12}^{+-,-1} x_2^{-} x_{12}^{-+,-1} \bar{\theta}_{12}  \bigr)_{\delta c} 
\bigl(x_{12}^{+-,-1} \bar{\theta}_{12} \bigr)_{\alpha d}.
\end{align}
Note the extra minus sign in the first line coming from commuting the contraction operator 
$i \bigl[\gen{L}^{\prime \beta}_{2 \; \, \alpha} \bigr]$ past the $\der_1$. 
Carrying out the remaining contractions and adding all the contributions 
up shows that the chiral contributions vanish
\begin{align}
f^{\idxconf{d}}{}_{\idxconf{b}\idxconf{c}} \, i[\gen{J}^{\idxconf{c}}_1] \, i[\gen{J}^{\idxconf{b}}_2] \, 
\bigvev{ \sfstr^+(1) \sfstr^+(2) } =0 \, .
\label{eqn:Phat_F+F+}
\end{align}

Let us now focus on the mixed-chiral contributions. The relevant two-point
function was also computed in \secref{sec:two-point-functions} and is given by
\begin{equation}
  \bigvev{ \mathcal{F}^+(1) \mathcal{F}^-(2)} =
  -\frac {g^2}{4 \pi^2} \der_1 \bigl(x_{12}^{-+,-1} \der_2 x_{12}^{-+} \epsilon \bigr)_{\beta \gamma}
  \der_1 \bigl(\epsilon x_{12}^{-+,-1} \der_2 x_{12}^{-+} \bigr)^{\beta \gamma}.
\end{equation}
Applying the double contraction expression of \eqref{eqn:Phat_contraction_expl} to this correlator yields
\begin{align}
\label{eqn:double_contr_mchiral1}
i[\gen{J}^{\idxconf{c}}_1] \, i[\gen{J}^{\idxconf{b}}_2] \, \bigvev{\sfstr^+(1) \sfstr^-(2)}
=\mathord{}&\frac{g^2}{2 \pi^2} \, \epsilon_{\lambda \gamma} \, \epsilon^{\beta \rho} 
\Bigl\{\gen{J}^{\idxconf{c}}_1 \bigl( x_{12}^{-+,-1} \gen{J}^{\idxconf{b}}_2  \, x_{12}^{-+}\bigr)_{\beta}{}^{\lambda} \, 
 \der_1 \bigl( x_{12}^{-+,-1} \der_2 \, x_{12}^{-+} \bigr)_{\rho}{}^{\gamma}  
 \\
&\quad  -(-1)^{|\gen{J}^\idxconf{c}_1| |\gen{J}^\idxconf{b}_2|} \,  
\der_1 \bigl( x_{12}^{-+,-1}  \gen{J}^{\idxconf{b}}_2 \, x_{12}^{-+}\bigr)_{\beta}{}^{\lambda} 
 \gen{J}^{\idxconf{c}}_1 \bigl( x_{12}^{-+,-1}  \der_2 \, x_{12}^{-+}\bigr)_{\rho}{}^{\gamma} \Bigr\}  \, , \nonumber
\end{align}
where we have left out the structure constant for brevity. Some useful formulas for computing the former expression are:
\begin{align}
  \gen{L}'_1{}^{\! \alpha}{}_{\beta}
(x_{12}^{-+})^{\dot{\gamma} \gamma} &=
 -2 \delta_\beta^\gamma (x_1^-)^{\dot{\gamma} \alpha},
\\
  \gen{L}'_2{}^{ \! \alpha}{}_{\beta}
(x_{12}^{-+})^{\dot{\gamma} \gamma} &=
 2 \delta_\beta^\gamma (x_1^- - x_{12}^{-+})^{\dot{\gamma} \alpha},
\\
  \gen{\bar{L}}'_1{}^{\! \dot{\alpha}}{}_{\dot{\beta}}
(x_{12}^{-+})^{\dot{\gamma} \gamma} &=
 -2 \delta_{\dot{\beta}}^{\dot{\gamma}} (x_2^+ + x_{12}^{-+})^{\dot{\alpha} \gamma},
\\
  \gen{\bar{L}}'_2{}^{ \! \dot{\alpha}}{}_{\dot{\beta}}
(x_{12}^{-+})^{\dot{\gamma} \gamma} &=
2 \delta_{\dot{\beta}}^{\dot{\gamma}} (x_2^+)^{\dot{\alpha} \gamma}.
\end{align}
The expression \eqref{eqn:double_contr_mchiral1} obviously 
consists of two different terms: one term where both generators 
act on the same piece of the mixed-chiral two-point function 
and another term where the two generators act on different pieces. 
First we compute the term where both generators act on a single term. 
After some
computation we find that 
\begin{align}
f^{\idxconf{d}}{}_{\idxconf{b}\idxconf{c}} \, \gen{J}^{\idxconf{c}}_1 
\bigl( x_{12}^{-+,-1} \gen{J}^{\idxconf{b}}_2  \, x_{12}^{-+}\bigr)_{\beta}{}^{\lambda}
=8 \delta_\alpha^\lambda \bigl(x_{12}^{-+,-1} \bigr)_{\beta \dot{\alpha}}
- 4 \delta_\beta^\lambda \bigl(x_{12}^{-+,-1}\bigr)_{\alpha \dot{\alpha}} .
\end{align}
For the second term of \eqref{eqn:double_contr_mchiral1} 
\begin{align}
T_{\beta \rho}^{\lambda \gamma}
=f^{\idxconf{d}}{}_{\idxconf{b}\idxconf{c}}(-1)^{|\gen{J}^\idxconf{c}_1| |\gen{J}^\idxconf{b}_2|}
 \,  \der_1 \bigl( x_{12}^{-+,-1}  \gen{J}^{\idxconf{b}}_2 \, x_{12}^{-+}\bigr)_{\beta}{}^{\lambda} 
 \gen{J}^{\idxconf{c}}_1 \bigl( x_{12}^{-+,-1}  \der_2 \, x_{12}^{-+}\bigr)_{\rho}{}^{\gamma} \, ,
\end{align}
we obtain
\begin{align}
T_{\beta \rho}^{\lambda \gamma}=-4  \bigsbrk{\delta^\lambda_\alpha \bigl(x_{12}^{-+,-1} \bigr)_{\rho \dot\alpha}
\der_1 \bigl( x_{12}^{-+,-1}  \der_2 \, x_{12}^{-+}\bigr)_{\beta}{}^{\gamma} 
+ \delta^\lambda_\rho \der_1 \bigl(x_{12}^{-+,-1} \bigr)_{\beta \dot\alpha} 
\bigl( x_{12}^{-+,-1}  \der_2 \, x_{12}^{-+}\bigr)_{\alpha}{}^{\gamma} }.
\end{align}
Plugging these two expressions back into equation \eqref{eqn:double_contr_mchiral1} 
and using the identity $\epsilon_{\lambda \gamma} \epsilon^{\beta \rho} 
= \delta_\lambda^\beta \delta_\gamma^\rho -\delta_\gamma^\beta \delta_\lambda^\rho$ 
yields
\begin{equation}
 f^{\idxconf{d}}{}_{\idxconf{b}\idxconf{c}} \, i[\gen{J}^{\idxconf{c}}_1] \, i[\gen{J}^{\idxconf{b}}_2] \, 
\bigl\langle \sfstr^+(1) \sfstr^-(2) \bigr\rangle =
  \frac {2g^2}{\pi^2} \, \der_1  \der_2 \, \bigl(x_{12}^{-+,-1} \bigr)_{\alpha \dot{\alpha}} \, ,
  \label{eqn:PhatF+F-}
\end{equation}
which is a total derivative with respect to points $1$ and $2$.

Given the results \eqref{eqn:Phat_F+F+} and \eqref{eqn:PhatF+F-} 
the full result for the action of $\genYfullnl{P}$ on the correlator 
of two gauge fields can now be constructed by symmetry considerations. 
For this, note that \eqref{eqn:PhatF+F-} implies
\begin{align}
 f^{\idxconf{d}}{}_{\idxconf{b}\idxconf{c}} \, i[\gen{J}^{\idxconf{c}}_2] \, i[\gen{J}^{\idxconf{b}}_1] 
\, \bigl\langle \sfstr^+(2) \sfstr^-(1) \bigr\rangle =
  \frac {2g^2}{\pi^2} \, \der_2  \der_1 \, \bigl(x_{21}^{-+,-1} \bigr)_{\alpha \dot{\alpha}} \, .
\end{align}
Now, using the property that the two exterior derivatives anti-commute 
as well as the fact that the double contraction expression on the left-hand side 
of the above equation is symmetric under exchange of points $1$ and $2$, 
we find the following final result for the level-one momentum remainder term
\[
R_{12}[\gen{P}_{\alpha\dot\alpha}]=R_{12}^{-+}[\gen{P}_{\alpha\dot\alpha}]-R_{21}^{-+}[\gen{P}_{\alpha\dot\alpha}]\, ,
\label{eqn:remainder_general_form}
\]
where
\[
R_{12}^{-+}[\gen{P}_{\alpha\dot\alpha}]=  \frac {2g^2}{\pi^2} \bigl(x_{12}^{-+,-1} \bigr)_{\alpha \dot{\alpha}} .
\label{eqn:remainder_Phat}
\]
A common feature of all the remainder functions is that they can be cast into the form \eqref{eqn:remainder_general_form}. 
For this reason we will from now on omit the second piece of \eqref{eqn:remainder_general_form} 
and only work with the building blocks $R_{12}^{-+}[\gen{J}^\idxconf{d}]$.

\paragraph{Level-one hypercharge remainder function.}

In this subsection we compute the remainder function $R^{\idxconf{d}}_{12}$ 
for the Yangian bonus symmetry generator $\genYfull{B}$. 
The bi-local part of this generator takes the following form
\begin{align}
\genYfullnl{B}=\sfrac{1}{4} \bigl( \genfull{Q}{}_{, \alpha a} \wedge \genfull{S}^{a \alpha} 
-  \genfull{\bar{Q}}^{a}{}_{\dot{\alpha}} \wedge \genfull{\bar{S}}^{\dot{\alpha}}{}_{a} \bigr) 
\label{eqn:Bhat_field}
\end{align}
and acts on the Wilson loop operator as defined in equation \eqref{eqn:JwedgeJprimeW}. 
From the last equation we can again directly read off the relevant double contraction operator. It reads 
\begin{align}
f^{\idxconf{d}}{}_{\idxconf{b}\idxconf{c}} 
\, i\bigl[\gen{J}^{\idxconf{c}}_1 \bigr] \, i\bigl[\gen{J}^{\idxconf{b}}_2 \bigr]
=\frac{1}{4} \Bigl(& i \bigl[\gen{Q}_{1, \alpha a} \bigr] 
\, i \bigl[\gen{S}_2^{a \alpha} \bigr] \! - i \bigl[\gen{\bar{Q}}^{a}_{1 \;  \dot\alpha} \bigr] \, 
i \bigl[\gen{\bar{S}}_{2}^{\dot{\alpha}}{}_a \bigr] \nonumber \\
&+ i \bigl[\gen{S}_1^{a \alpha} \bigr] \, i \bigl[\gen{Q}_{2, \alpha a} \bigr] \! 
- i \bigl[\gen{\bar{S}}_{1}^{\dot{\alpha}}{}_a \bigr] \, i \bigl[\gen{\bar{Q}}^{a}_{2 \;  \dot\alpha} \bigr] \Bigr) \, ,
\label{eqn:Bhat_contractionexp}
\end{align}
where here and throughout this subsection $\idxconf{d} \sim \gen{B}$. 
As in the case of the level-one momentum remainder function 
we will now apply this double contraction operator to the chiral 
and mixed-chiral part of the field strength two-point function. 

Again the computations splits up into the chiral and mixed-chiral contributions.
In order to determine the chiral contributions to the level-one hypercharge remainder 
function it is again sufficient to compute the action of the contraction operator \eqref{eqn:Bhat_contractionexp}
 on the form $\der_1 (x_{12}^{-+,-1} \bar{\theta}_{12})_{\gamma c}\, \der_2 (x_{12}^{+-,-1} \bar{\theta}_{12})_{\delta d}$. 
Using the representation of the generators in terms of differential operators (see \secref{sec:supconfalgebra})
 one shows that this action vanishes and thus
\begin{align}
f^{\idxconf{d}}{}_{\idxconf{b}\idxconf{c}}  \, i[\gen{J}^{\idxconf{c}}_1] \, 
i[\gen{J}^{\idxconf{b}}_2] \, \vev{\sfstr^+(1) \sfstr^+(2)}=0 \, .
\end{align}

Turning to the mixed-chiral contributions we recall the formula for the action 
of a general double contraction operator on the mixed-chiral two-point function
of \eqn{eqn:double_contr_mchiral1}:
\begin{align}
\label{eqn:double_contr_mchiral2}
i[\gen{J}^{\idxconf{c}}_1] \, i[\gen{J}^{\idxconf{b}}_2] \, \bigvev{\sfstr^+(1) \sfstr^-(2)}
=\mathord{}&\frac{g^2}{2 \pi^2} \, \epsilon_{\lambda \gamma} \, \epsilon^{\beta \rho} 
\Bigl\{\gen{J}^{\idxconf{c}}_1 \bigl( x_{12}^{-+,-1} \gen{J}^{\idxconf{b}}_2  \, x_{12}^{-+}\bigr)_{\beta}{}^{\lambda} \, 
 \der_1 \bigl( x_{12}^{-+,-1} \der_2 \, x_{12}^{-+} \bigr)_{\rho}{}^{\gamma}  
 \\
&\quad  -(-1)^{|\gen{J}^\idxconf{c}_1| |\gen{J}^\idxconf{b}_2|} \,  
\der_1 \bigl( x_{12}^{-+,-1}  \gen{J}^{\idxconf{b}}_2 \, x_{12}^{-+}\bigr)_{\beta}{}^{\lambda} 
 \gen{J}^{\idxconf{c}}_1 \bigl( x_{12}^{-+,-1}  \der_2 \, x_{12}^{-+}\bigr)_{\rho}{}^{\gamma} \Bigr\}  . \nonumber
\end{align}
As before, we will compute the two different terms of the former expression individually.
 For the contribution where both generators act on a single term we find
\begin{align}
f^{\idxconf{d}}{}_{\idxconf{b}\idxconf{c}} \, \gen{J}^{\idxconf{c}}_1 
\bigl( x_{12}^{-+,-1} \gen{J}^{\idxconf{b}}_2  \, x_{12}^{-+}\bigr)_{\beta}{}^{\lambda}=
4  \delta^\lambda_\beta \Bigl\{\tr \bigl(x_{12}^{-+,-1} \bar{\theta}_1 \theta_2 \bigr) -1 \Bigr\} 
- 8  \bigl(x_{12}^{-+,-1} \bar{\theta}_1 \theta_2 \bigr)_{\beta}{}^{\lambda} \, .
\end{align}
For the second contribution, we obtain
\begin{align}
& f^{\idxconf{d}}{}_{\idxconf{b}\idxconf{c}} \, \der_1 \bigl( x_{12}^{-+,-1} 
 \gen{J}^{\idxconf{b}}_2 \, x_{12}^{-+}\bigr)_{\beta}{}^{\lambda}  
\gen{J}^{\idxconf{c}}_1 \bigl( x_{12}^{-+,-1}  \der_2 \, x_{12}^{-+}\bigr)_{\rho}{}^{\gamma} 
 \nonumber \\
&=-4  \bigl(x_{12}^{-+,-1} \bar{\theta}_1 \theta_2 \bigr)_{\rho}{}^{\lambda} \, 
\der_1 \bigl(x_{12}^{-+,-1} \der_2 \, x_{12}^{-+} \bigr)_{\beta}{}^{\gamma} 
\nonumber \\&\qquad
+4  \delta^{\lambda}_{\rho}  \der_1 \bigl(x_{12}^{-+,-1} \bar{\theta}_1 \bigr)_{\beta a} 
\Bigl\{\der_2 \theta_2^{a \gamma} - \bigl(\theta_2 \, x_{12}^{-+,-1}  \der_2 \, x_{12}^{-+} \bigr)^{a \gamma} \Bigr\} \, .
\end{align}
Combining both expressions and adding the correct prefactor yields
\begin{align}
f^{\idxconf{d}}{}_{\idxconf{b}\idxconf{c}}  \, i[\gen{J}^{\idxconf{c}}_1] \, 
i[\gen{J}^{\idxconf{b}}_2] \, \vev{\sfstr^+(1) \sfstr^-(2)}
= - \frac{2g^2}{\pi^2} \der_1 \der_2 \Bigl( \ln \bigl( \bigl( x_{12}^{-+} \bigr)^2 \bigr)
 + \tr \bigl( x_{12}^{-+,-1} \bar{\theta}_1 \theta_2 \bigr) \Bigr) \, .
\end{align}
From this result we can now read off the form of the essential building block $R_{12}^{-+}[\gen{B}]$
 of the level-one hypercharge remainder function, see \eqref{eqn:remainder_general_form}. It reads
\begin{align}
R_{12}^{-+}[\gen{B}]= - \frac{2g^2}{\pi^2} \Bigl( \ln \bigl( \bigl( x_{12}^{-+} \bigr)^2 \bigr)
+ \tr \bigl( x_{12}^{-+,-1} \bar{\theta}_1 \theta_2 \bigr) \Bigr)\, .
\label{eq:R-for-B}
\end{align}

\paragraph{Other level-one remainder functions.}

Having computed the remainder functions $R_{12}[\gen{P}]$ and $R_{12}[\gen{B}]$
 we can now determine all the other remainder functions 
by applying the appropriate superconformal transformations to one of them. 
This can be seen as follows. Under the action of a superconformal generator
$\gen{J}^{\idxconf{a}}=\gen{J}^{\idxconf{a}} X^{\idxall{A}} \partial_{\idxall{A}}
 = \gen{J}^{\idxconf{a}}_{\text{cov}} X^{\idxall{A}} \sdel_{\idxall{A}}$ 
the one-form $\genfull{J}^{\idxconf{c}} \sgauge$ transforms by the Lie derivative. 
A short computation shows that  $\gen{J}^{\idxconf{a}}$ acts on $\genfull{J}^{\idxconf{c}} \sgauge$
 by transforming the field as well as the conformal index
\begin{align}
\gen{J}^{\idxconf{a}} \genfull{J}^{\idxconf{c}} \sgauge
=- \genfield{J}^{\idxconf{a}} \genfull{J}^{\idxconf{c}} \sgauge + f^{\idxconf{a} \idxconf{c}}{}_{\idxconf{e}} \, 
\genfull{J}^{\idxconf{e}} \sgauge \, .
\end{align}
Using this identity one easily shows that the following equation holds true:
\begin{align}
\bigl(\gen{J}_{1}^{\idxconf{a}} + \gen{J}_{2}^{\idxconf{a}} \bigr) 
f^{\idxconf{d}}{}_{\idxconf{b}\idxconf{c}} \bigvev{ \genfull{J}^{\idxconf{c}} \sgauge(1) \, \genfull{J}^{\idxconf{b}} \sgauge(2) }
&=
 \bigl[f^{\idxconf{d}}{}_{\idxconf{b}\idxconf{c}} f^{\idxconf{a} \idxconf{c}}{}_{\idxconf{e}} 
+ (-1)^{|\idxconf{a}| |\idxconf{e}|} f^{\idxconf{d}}{}_{\idxconf{c}\idxconf{e}} 
f^{\idxconf{a} \idxconf{c}}{}_{\idxconf{b}} \bigr] 
\bigvev{ \genfull{J}^{\idxconf{e}} \sgauge(1) \, \genfull{J}^{\idxconf{b}} \sgauge(2) }
 \nonumber \\&\qquad
- f^{\idxconf{d}}{}_{\idxconf{b} \idxconf{c}} \bigvev{\bigl(\genfield{J}_{1}^{\idxconf{a}} + \genfield{J}_{2}^{\idxconf{a}} \bigr)
 \bigl( \genfull{J}^{\idxconf{c}} \sgauge(1) \, \genfull{J}^{\idxconf{b}} \sgauge(2)\bigr)} \, .
\label{eqn:J_on_JstarbiAA}
\end{align}
An important thing to note now is that the expression 
in the second line vanishes. 
This is because the term in brackets represents a conformal
variation of an object which is invariant under linearized 
gauge transformations. Thus, using the supersymmetric Jacobi 
identity \eqref{eqn:susy_jacobi}, we find
\begin{align}
\bigl(\gen{J}_{1}^{\idxconf{a}} + \gen{J}_{2}^{\idxconf{a}} \bigr) 
f^{\idxconf{d}}{}_{\idxconf{b}\idxconf{c}} \bigvev{ \genfull{J}^{\idxconf{c}} \sgauge(1) \, \genfull{J}^{\idxconf{b}} \sgauge(2) }
= f^{\idxconf{a} \idxconf{d}}{}_{\idxconf{c}} f^{\idxconf{c}}{}_{\idxconf{b}\idxconf{e}} 
 \bigvev{ \genfull{J}^{\idxconf{e}} \sgauge(1) \, \genfull{J}^{\idxconf{b}} \sgauge(2) } \, .
\label{eq:transformedremainder}
\end{align}
By combining this expression%
\footnote{Note that equation \eqref{eq:transformedremainder} 
can also be used to confirm the result of \secref{sec:sym_gauge_propagator}, 
namely, that all the level-one generators $\genYfullnl{J}^\idxconf{d}$ 
exclusively produce a double derivative term. 
For example, choosing $\idxconf{d} \sim \gen{P}$ and $\idxconf{a} \sim \gen{S}$ 
in \eqref{eq:transformedremainder} and taking into account the result of \secref{sec:RemFunct}
as well as the fact that $\gen{J}_{i}^{\idxconf{a}} \der_i= \der_i  \gen{J}_{i}^{\idxconf{a}}$, 
we learn that $\genYfullnl{\bar{Q}}$ produces a double derivative term as well. 
By continuing this analysis we can thus confirm that \eqref{eq:propsym} indeed holds true.
}
with equation \eqref{eq:propsym}, we obtain
\begin{align}
\bigl[ \gen{J}_1^\idxconf{a} +\gen{J}_2^\idxconf{a}, \der_1 \der_2 R_{12}^\idxconf{d} \bigr \}=
f^{\idxconf{a} \idxconf{d}}{}_\idxconf{c} \der_1 \der_2 R_{12}^\idxconf{c}\, .
\label{eqn:localtermsalgebra-weak}
\end{align}
In fact, it turns out that if one excludes the level-one hypercharge remainder function 
the $R_{12}^\idxconf{d}$ satisfy the adjoint transformation law on their own:
\begin{align}
\bigl[ \gen{J}_1^\idxconf{a} +\gen{J}_2^\idxconf{a}, R_{12}^\idxconf{d} \bigr \}=
f^{\idxconf{a} \idxconf{d}}{}_\idxconf{c} R_{12}^\idxconf{c} \, .
\label{eqn:localtermsalgebra-strong}
\end{align}
Using these equations one can now iteratively compute all the yet undetermined remainder functions. 
One finds:
\begin{subequations}\label{eq:R-list}
\begin{align}
  R_{12}^{-+}[\gen{\bar{L}}'] &= \frac {2 g^2}{\pi^2} x_2^+ x_{12}^{-+,-1} - \frac {2 g^2}{\pi^2} \ln(x_{12}^{-+})^2,\\
  R_{12}^{-+}[\gen{\bar{S}}] &= \frac {4 g^2}{\pi^2} x_2^+ x_{12}^{-+,-1} \bar{\theta}_1,\\
  R_{12}^{-+}[\gen{K}] &=- \frac {2 g^2}{\pi^2} x_2^+ x_{12}^{-+,-1} x_1^-,\\
  R_{12}^{-+}[\gen{\bar{Q}}] &= -\frac {4 g^2}{\pi^2} \theta_2 x_{12}^{-+,-1},\\
  R_{12}^{-+}[\gen{R}'] &= -\frac {8 g^2}{\pi^2} \theta_2 x_{12}^{-+,-1} \bar{\theta}_1 + \frac {2 g^2}{\pi^2} \ln(x_{12}^{-+})^2,\\
  R_{12}^{-+}[\gen{S}] &= \frac {4 g^2}{\pi^2} \theta_2 x_{12}^{-+,-1} x_1^-,\\
  R_{12}^{-+}[\gen{Q}] &= - \frac {4 g^2}{\pi^2} x_{12}^{-+,-1} \bar{\theta}_1,\\
  R_{12}^{-+}[\gen{L}'] &= \frac {2 g^2}{\pi^2} x_{12}^{-+,-1} x_1^- + \frac {2 g^2}{\pi^2} \ln(x_{12}^{-+})^2.
\end{align}
\end{subequations}

However, instead of carrying out this analysis in detail we shall now present a
more powerful formalism to determine all remainder functions at once. 
For this we first need to introduce some notation and define
\begin{align}
  \mathcal{X}_{\idxall{A} \idxall{B}} &=
  \begin{pmatrix}
    x^+ \\ 2 \theta \\ 1
  \end{pmatrix} \epsilon
  \begin{pmatrix}
    x^{+, \trans} & 2 \theta^\trans & 1
  \end{pmatrix} =
  \begin{pmatrix}
    x^+ \epsilon x^{+,\trans} & 2 x^+ \epsilon \theta^\trans & x^+ \epsilon\\
    2 \theta \epsilon x^{+,\trans} & 4 \theta \epsilon \theta^\trans & 2 \theta \epsilon\\
    \epsilon x^{+,\trans} & 2 \epsilon \theta^\trans & \epsilon
  \end{pmatrix},
\\
  \overline{\mathcal{X}}^{\idxall{A} \idxall{B}} &=
  \begin{pmatrix}
    -1\\ -2 \bar{\theta}^\trans \\ x^{-,\trans}
  \end{pmatrix} \epsilon
  \begin{pmatrix}
    -1 & -2 \bar{\theta} & x^-
  \end{pmatrix} =
  \begin{pmatrix}
    \epsilon & 2 \epsilon \bar{\theta} & -\epsilon x^-\\
    2 \bar{\theta}^\trans \epsilon & 4 \bar{\theta}^\trans \epsilon \bar{\theta} & -2 \bar{\theta}^\trans \epsilon x^-\\
    -x^{-,\trans} \epsilon & -2 x^{-,\trans} \epsilon \bar{\theta} & x^{-,\trans} \epsilon x^-
  \end{pmatrix}.
\end{align}
It is not too hard to show (see \appref{sec:sconf-gen})
 that these quantities transform homogeneously under superconformal transformations (up to a rescaling).  
They are graded anti-symmetric 
and also satisfy 
\[
(\overline{\mathcal{X}} \mathcal{X})^{\idxall{A}}{}_{\idxall{C}} 
\equiv \overline{\mathcal{X}}^{\idxall{A} \idxall{B}} \mathcal{X}_{\idxall{B} \idxall{C}} = 0
\qquad\text{and}\qquad  
\tr (\overline{\mathcal{X}}_1 \mathcal{X}_2) \equiv \overline{\mathcal{X}}_1{}^{\idxall{A} \idxall{B}} 
\mathcal{X}_{2, \idxall{B} \idxall{A}} = -2 (x_{12}^{-+})^2.
\]
In our conventions the North-West--South-East contraction is superconformal invariant.  
To make the South-West--North-East contraction superconformal invariant 
we need to introduce an extra sign, \((-1)^{|\idxall{A}|} T_{\idxall{A}} U^{\idxall{A}}\) 
(see also \namedref{footnote}{fn:super-contraction}).  
For the same reason, we use the notations \(\tr U \equiv U^{\idxall{A}}{}_{\idxall{A}}\) 
if \(U\) has the first index upper and the second index lower 
and \(\str V \equiv (-1)^{|\idxall{A}|} V_{\idxall{A}}{}^{\idxall{A}}\) 
if \(V\) has the first index lower and the second index upper.

Using the notations above we define
\begin{align}
  \label{eq:R-J}
  R_{12}^{-+}[\gen{J}] &\equiv \frac {2 g^2}{\pi^2}
\left[\frac {\tr (\overline{\mathcal{X}}_1 \genmatr{J} \mathcal{X}_2)}{\tr (\overline{\mathcal{X}}_1 \mathcal{X}_2)} 
+ \frac 1 2 \ln \bigl(\tr (-\tfrac 1 2 \overline{\mathcal{X}}_1 \mathcal{X}_2)\bigr)\; \str (\genmatr{J})\right].
\end{align}
Here \(\genmatr{J}\) is the \((2\vert 4\vert 2) \times (2\vert 4\vert 2)\) 
representation of the generator \(\gen{J}\).  
This representation is worked out in \appref{sec:sconf-gen}.  The answer is
\[
  \label{eq:generic-alg-elem}
  a \cdot \genmatr{P} + \chi \cdot \genmatr{Q} + \bar{\chi} \cdot \genmatr{\bar{Q}} 
+ \omega' \cdot \genmatr{L}' + \bar{\omega}' \cdot \genmatr{\bar{L}}' + r' \cdot \genmatr{R}' 
+ \rho \cdot \genmatr{S} + \bar{\rho} \cdot \genmatr{\bar{S}} + b \cdot \genmatr{K} =
    \begin{pmatrix}
    -2 \bar{\omega}' & 2 \bar{\chi} & -2 a\\
    2 \bar{\rho} & -2 r' & -2 \chi\\
    -2 b & 2 \rho & 2 \omega'      
    \end{pmatrix},
\]
where \(\genmatr{R}' = \genmatr{R} - \genmatr{B}\). See \appref{sec:sconf-gen} for more details.

This quantity transforms appropriately under superconformal transformations:
\([\gen{J}_1^\alpha + \gen{J}_2^\alpha, R_{12}[\gen{J}^\beta]\rbrace = f^{\alpha \beta}{}_\gamma R_{12}[\gen{J}^\gamma]\).
The extra rescaling which was present in the \(\mathcal{X}\) and \(\overline{\mathcal{X}}\) transformations
cancels here between the numerator and the denominator 
while the logarithm produces an additive constant which cancels between \(R_{12}^{-+}\) and \(R_{21}^{-+}\).

The quantity \(R_{12}^{-+}[\gen{J}]\) defined above can be rewritten as
\begin{equation}
  \label{eq:R-J-transf}
  R_{12}^{-+}[\gen{J}] = -\frac {g^2}{\pi^2}
  \tr \lrsbrk{ x_{21}^{+-,-1}
  \begin{pmatrix}
    -1 & -2 \bar{\theta}_1 & x_1^-
  \end{pmatrix}
  \genmatr{J}
  \begin{pmatrix}
    x_2^+ \\ 2 \theta_2 \\ 1
  \end{pmatrix}}
  + \frac {g^2}{\pi^2} \ln (x_{21}^{+-})^2\; \str(\genmatr{J}).
\end{equation}
The \(2 \vert 4 \vert 2\) row and column matrices above are
equal to the twistorial quantities \(\mathcal{Z}\) and \(\overline{\mathcal{Z}}\) 
defined in \appref{sec:sconf-gen}.

Using eq.~\eqref{eq:R-J-transf} we easily compute \(R[a \cdot \gen{P}] = -(2 g^2/\pi^2) \tr(a x_{21}^{+-,-1})\), 
which reproduces the result 
\(R_{12}^{-+}[\gen{P}] = -(2 g^2/\pi^2) x_{21}^{+-,-1}\).  
Eq.~\eqref{eq:R-J} also reproduces the expression for \(R_{12}^{-+}[\gen{B}]\)
 in eq.~\eqref{eq:R-for-B}, up to kinematics independent terms 
which cancel between \(R_{12}^{-+}[\gen{B}]\) and \(R_{21}^{-+}[\gen{B}]\).
For the generic element in eq.~\eqref{eq:generic-alg-elem} we have
\begin{align}
  R_{12}^{-+}[a \cdot \gen{P} + \chi \cdot \gen{Q} + \ldots]& =
  -\frac {g^2}{\pi^2} \Bigl[
  2 \tr(x_{21}^{+-,-1} \bar{\omega}' x_2^+)
  -4 \tr(x_{21}^{+-,-1} \bar{\theta}_1 \bar{\rho} x_2^+)
\nonumber\\&\qquad\quad
  -2 \tr(x_{21}^{+-,-1} x_1^- b x_2^+)
  -4 \tr(x_{21}^{+-,-1} \bar{\chi} \theta_2)
\nonumber\\&\qquad\quad
  +8 \tr(x_{21}^{+-,-1} \bar{\theta}_1 r' \theta_2)
  +4 \tr(x_{21}^{+-,-1} x_1^- \rho \theta_2)
\nonumber\\&\qquad\quad
  +2 \tr(x_{21}^{+-,-1} a)
  +4 \tr(x_{21}^{+-,-1} \bar{\theta}_1 \chi)
  +2 \tr(x_{21}^{+-,-1} x_1^- \omega')\Bigr]
\nonumber\\&\qquad
  + \frac {g^2}{\pi^2} \ln (x_{21}^{+-})^2 \lrsbrk{
  \tr(-2 \bar{\omega}') + \tr(2 r') + \tr(2 \omega')},
\end{align}
which reproduces all the results in eqs.~\eqref{eq:R-list}.
This concludes our discussion of the level-one remainder functions.

\subsection{Regularization}

\begin{figure}
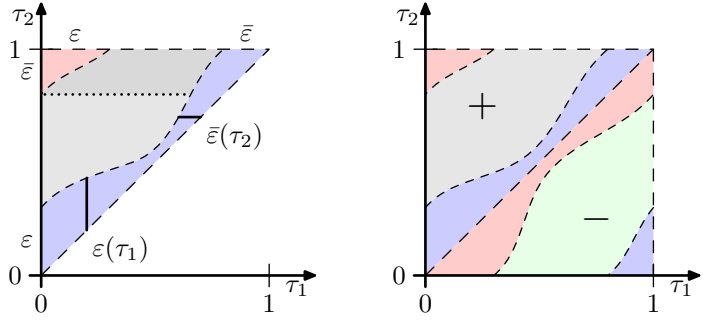
\centering
\includegraphicsbox{FigDomainReg.mps}
\qquad
\includegraphicsbox{FigDomain2Reg.mps}
\caption{Integration domain in point splitting regularization.
The left figure shows the shaded integration domain for two ordered
insertions into the Wilson loop. The colored stripe and corner
are removed by point splitting. The dotted line depicts how
we split up the domain into two convenient integration regions. 
The right figure illustrates an odd feature of the corner
when considering a Yangian level-one generator
which acts with opposite signs on the two big triangles. 
Here the upper corner is in the periodic continuation of the stripe 
in the lower triangle. 
Thus in some sense regularization has the opposite effect 
on the removed stripe and triangle.}
\label{fig:domain}
\end{figure}

The above results on the concrete form of the remainder $R$ 
show that the argument in \secref{sec:symwil} requires 
proper regularization and potentially renormalization as well.
The point is that $R$ has a linear UV-divergence
which leads to a quadratic divergence for the term
$\der_2 R_{12}|_{1=2}$ in \eqref{eq:naivewilsoninvariance}.

\paragraph{Divergence.}

Let us therefore restart the above calculation in point splitting regularization
where $|\tau_1-\tau_2|>\varepsilon(\tau_1)=\bar\varepsilon(\tau_2)$.%
\footnote{We will not make further assumptions on the nature of $\varepsilon$.
The simplest choice would be to let the coordinate $\tau$ be proportional
to the arc length of the curve and $\varepsilon=\bar\varepsilon$ 
be a constant cut-off parameter. 
However, this choice can obscure some features of renormalization,
and we therefore prefer to work with a fairly general scheme
where $\dot x^2$ and $\varepsilon,\bar\varepsilon$ are not assumed to be constant.}
The full integration domain $0<\tau_1<\tau_2<1$ needs to be adjusted not only 
where $\tau_1\approx \tau_2$, 
but also at $\tau_1\approx 0$, $\tau_2\approx 1$ 
where both points approach each other due to
periodicity of the Wilson loop, see \figref{fig:domain}.
We thus obtain for the regularized integral of 
$\der_1 \der_2 R_{12}$
\begin{align}
\label{eq:yangremainder}
\int_{\varepsilon}
\der_1 \der_2 R_{12}
&=
\int_{\varepsilon}^{1-\bar\varepsilon} \der\tau_2\int_0^{\tau_2-\bar\varepsilon}\der\tau_1\,
\partial_1 \partial_2 R_{12}
+\int_{1-\bar\varepsilon}^{1} \der\tau_2 \int_{\tau_2-1+\varepsilon}^{\tau_2-\bar\varepsilon} \der\tau_1\,
\partial_1 \partial_2 R_{12}
\nln&=
\int_{\varepsilon}^{1-\bar\varepsilon} \der\tau\, (\partial_2 R)(\tau-\bar\varepsilon,\tau)
-\int_{\varepsilon}^{1-\bar\varepsilon} \der\tau\, (\partial_2 R)(0,\tau)
\nln &\quad
+\int_{1-\bar\varepsilon}^{1} \der\tau\, (\partial_2 R)(\tau-\bar\varepsilon,\tau)
-\int_{1-\bar\varepsilon}^{1} \der\tau\, (\partial_2 R)(\tau-1+\varepsilon,\tau)
\nln&=
\int_{0}^{1} \der\tau\, (\partial_2 R)(\tau-\bar\varepsilon,\tau)
+R(0,\varepsilon)-R(0,-\bar\varepsilon)
\nln &\quad
-\int_{0}^{\varepsilon} \der\tau\, (\partial_2 R)(\tau-\bar\varepsilon,\tau)
-\int_{-\bar\varepsilon}^{0} \der\tau\, (\partial_2 R)(\tau+\varepsilon,\tau)
.
\end{align}
Again, the bi-local terms have been integrated to a 
local term and some terms located at the boundary. 
Note that the boundary terms do contribute non-trivially to the final result;
\figref{fig:domain} may be helpful in understanding this 
somewhat counter-intuitive feature: 
The local term subtracts a particular (divergent) contribution from the integral.
The boundary term does the same, however, it resides in the periodic continuation
of a region where the Yangian level-one generator (effectively) acts
with the opposite sign.
Hence the natural extension of the local term to the boundary 
does not actually absorb the boundary term, but rather reinforces it
by a factor of 2.

\paragraph{Regularization.}

Let us now expand the contributing terms in the UV-limit $\varepsilon\to 0$.%
\footnote{Here and in the following it is permissible to consider
$\varepsilon$ as a function of $\tau$ and still treat it as an expansion parameter,
i.e.\ it approaches the constant function zero.}
We find
\begin{align}
R(\tau,\tau+\varepsilon)&=
\frac{1}{\varepsilon}R_{-1}(\tau)+\half \dot R_{-1}(\tau)
+R_1(\tau)\varepsilon+\order\brk{\varepsilon^2},
\nln
(\partial_2 R)(\tau,\tau+\varepsilon)
&=-\frac{1}{\varepsilon^2}R_{-1}(\tau)+R_1(\tau)+\order\brk{\varepsilon}.
\end{align}
Note that the constant term $\half \dot R_{-1}(\tau)$ 
in the $\varepsilon$-expansion of $R(\tau,\tau+\varepsilon)$ follows from
the antisymmetry $R(\tau_1, \tau_2)=-R(\tau_2, \tau_1)$.
Inserting these expansions into the above Yangian action one obtains
after some careful calculation
\begin{align}
\label{eq:yangremainder2}
\int_{\varepsilon}
\der_1 \der_2 R_{12}
&=
\int_{0}^{1} \der\tau \, \left ( - \frac{1+\dot\varepsilon}{\varepsilon^2}\,R_{-1}(\tau) + R_1(\tau)
\right )
+\frac{4+2\dot\varepsilon}{\varepsilon}\,R_{-1}(0)
+\order\brk{\varepsilon}.
\end{align}
Note that for a non-constant function $\varepsilon$
we have to rely on the relationship $\varepsilon(\tau_1)=\bar\varepsilon(\tau_2)$ 
at the cut-off.
This implies the two equivalent relations
\[
\varepsilon(\tau)=\bar\varepsilon(\tau+\varepsilon(\tau)),
\qquad
\bar\varepsilon(\tau)=\varepsilon(\tau-\bar\varepsilon(\tau)),
\]
which can be solved perturbatively as 
\[
\bar\varepsilon=\varepsilon-\varepsilon\dot\varepsilon
+\varepsilon\dot\varepsilon^2+\half \varepsilon^2\ddot\varepsilon+\order(\varepsilon^4).
\]

Gladly, all the terms in the bi-local remainder \eqref{eq:yangremainder2}
can be absorbed by an appropriate regularization of the Yangian 
action on Wilson loops.
As discussed in \secref{sec:localrenorm},
the integral terms can be absorbed by redefining the local action 
of the Yangian on the gauge field as follows:
\[
(\genYfull{J} \sgauge)_\varepsilon = 
- \frac{N}{2}\der \tau  (\partial_2R)(\tau-\bar\varepsilon,\tau) .
\]
Note that this action maps an adjoint field to a plain number. 
Ordinarily one would expect that the gauge structure 
of the fields is preserved by the action of a symmetry.
Nevertheless, here we are in the planar limit
where such a rule makes sense (just as the bi-local insertion
of the level-one generators).
The above local action can be realized on the components
of the gauge field as follows%
\footnote{Other regularization rules are conceivable
which involve the fermionic components of the gauge field
or which make use of total derivative terms.}
\[
(\genYfull{J} \sgauge_\mu)_\varepsilon = 
-\frac{N}{2} \frac{p_\mu}{p^2}\, (\partial_2R)(\tau-\bar\varepsilon,\tau) .
\]
For the remaining boundary term in \eqref{eq:yangremainder2} 
we have to adjust the definition 
of the Yangian action on the Wilson loop \eqref{genYJfirsttry} 
by a boundary term
\[
(\genYfull{J}^\idxconf{c} \swilson)_\varepsilon =
\swilson\bigsbrk{(\genYfull{J}^\idxconf{c} \sgauge)_\varepsilon}
+f^\idxconf{c}{}_{\idxconf{b}\idxconf{a}}
\swilson\bigsbrk{\genfull{J}^\idxconf{a} \sgauge; \genfull{J}^\idxconf{b} \sgauge}_\varepsilon
-N\bigbrk{R^\idxconf{c}(0,\varepsilon)-R^\idxconf{c}(0,-\bar\varepsilon)}\swilson.
\]
Such boundary terms are natural 
because the boundary is also involved in the regularization
of the Wilson loop.

Altogether Yangian symmetry of Wilson loops
requires a $1/\varepsilon^2$ divergent local term, 
a finite local term as well as a $1/\varepsilon$ divergent boundary term as spelled out
in \eqref{eq:yangremainder2}.
The crucial point in favor of Yangian symmetry 
is that no bi-local and no bulk-boundary counterterms are needed.

\paragraph{Level-one momentum.}

For concreteness, we continue with the level-one momentum $\genYfull{P}^\mu$
for which $R_{12}[\gen{P}^\mu]$ 
is given by \eqref{eqn:remainder_general_form} and \eqref{eqn:remainder_Phat}.
The expansion of this function near coincident points takes the form
\begin{align}
  R_{12}[\gen{P}] &= \frac{2g^2}{\pi^2}\bigg[-\frac{2}{\varepsilon} p^{-1} + p^{-1} \dot{p} p^{-1}  
\nonumber\\&\qquad
+ \varepsilon p^{-1} \Bigl(\sfrac{1}{3} \ddot{p} + \sfrac{2}{3} \ddot{\bar{\theta}} \dot{\theta} 
- \sfrac{2}{3} \dot{\bar{\theta}} \ddot{\theta} - \sfrac{1}{2} \dot{p} p^{-1} \dot{p} 
- 8 \dot{\bar{\theta}} \dot{\theta} p^{-1} \dot{\bar{\theta}} \dot{\theta}\Bigr) p^{-1} + \order(\varepsilon^2)
\biggr],
\end{align}
from which the relevant functions $R_{-1}$ and $R_1$ can be read off easily.

A curious observation concerns reparametrization of 
the Wilson loop: 
The function $R_{-1}[\gen{P}]$ is covariant under a reparametrization 
$\tau\mapsto \sigma(\tau)$
\[
R_{-1}[\gen{P}]\mapsto \dot\sigma\,R_{-1}[\gen{P}].
\]
Together with the transformation $\der\tau \mapsto \der\tau \,\dot\sigma$ 
of the measure $\der \tau$ 
and the transformation $\varepsilon \mapsto \dot\sigma\,\varepsilon$
of the cut-off $\varepsilon$,
the leading contribution $\der\tau\,R_{-1}[\gen{P}]/\varepsilon^2$
to the integrand is invariant.
Conversely, the function $R_{1}[\gen{P}]$ is not covariant,
\[
R_{1}[\gen{P}]\mapsto \frac{1}{\dot\sigma}\,
\bigbrk{ R_{1}[\gen{P}] +\sfrac{1}{6} S(\sigma) R_{-1}[\gen{P}] },
\]
but rather transforms by the addition of a Schwarzian derivative
\[
 S(\sigma) := \frac {\dddot\sigma}{\dot\sigma} - \frac{3}{2}\lrbrk{\frac {\ddot\sigma}{\dot\sigma}}^2
\]
multiplying the divergence term $R_{-1}$.
Given the singularity structure of the function $R_{12}$, 
the appearance of the Schwarzian derivative is natural.
Due to this singularity we must consider 
the transformation of $\varepsilon$ at higher orders
in $\varepsilon$.
The exact transformation of $\varepsilon$ reads
\[
\varepsilon(\tau)\mapsto
 \sigma(\tau +\varepsilon(\tau))-\sigma(\tau)
= \varepsilon\dot\sigma
+\half\varepsilon^2\ddot\sigma
+\sfrac{1}{6}\varepsilon^3\dddot\sigma
+\order(\varepsilon^4).
\]
The relevant combination of divergence terms happens
to transform with a Schwarzian derivative as well
\[
\frac{1+\dot\varepsilon}{\varepsilon^2}
\mapsto
\frac{1}{\dot\sigma^2}
\lrbrk{
\frac{1+\dot\varepsilon}{\varepsilon^2}
+\sfrac{1}{6}
S(\sigma)
+\order(\varepsilon)
}.
\]
The extra term multiplying $R_{-1}$ cancels nicely with the transformation of $R_1$,
and thus the local contributions are properly reparametrization-invariant.
Moreover,
\[
\frac{1+\half\dot\varepsilon}{\varepsilon}
\mapsto
\frac{1}{\dot\sigma}
\lrbrk{
\frac{1+\half\dot\varepsilon}{\varepsilon}
},
\]
which implies that the boundary term also stays invariant 
(but its position is actually shifted).

\section{Summary and Conclusions}
\label{sec:conclusions}

This work completes the symmetry investigations of
the smooth Maldacena--Wilson loops in $\superN=4$ non-chiral superspace started in \cite{Beisert:2015jxa}, 
where we proved global superconformal and local kappa-symmetry 
of the loop along with establishing the finite result for the
vacuum expectation value at first perturbative order for general paths. 
Here we have investigated the Yangian symmetry of Wilson loops in detail and proved the
existence of this hidden symmetry at leading perturbative order. Let us briefly summarize 
the key results of this work.

The level-one generators of the $Y[\alg{psu}(2,2|4)]$ Yangian acting on a Wilson line
are represented by a bi-local and a local piece. In order to define the bi-local
piece in a gauge-covariant manner, we introduced
gauge-covariant conformal generators $\genfull{J}^{\idxconf{a}}$. These act 
compositely on the gauge fields through a conformal transformation 
plus a compensating gauge transformation. While the difference to the standard conformal 
action $\genfield{J}^{\idxconf{a}}$
disappears on the vacuum expectation value of a Wilson loop, it does make a 
crucial difference for the bi-local part of the Yangian level-one generator
$\genYfull{J}^{\idxconf{a}}$ where it implements gauge covariance. The local piece of the
Yangian level-one generators on the other hand arises from the need to cancel the UV-divergences
induced by the action of the bi-local piece in the coincidence limit. We worked out the
Yangian invariance at the leading perturbative order in detail in the presence of a 
point splitting regulator. The form of the local terms was explicitly determined.
Moreover, we demonstrated the closure of the level-one algebra 
and the compatibility of Yangian symmetry with kappa-symmetry. 
We did not verify the Serre relations for our realization of the Yangian symmetry.
However, these relations merely serve to make the algebra smaller. 
If for some reason they should not hold, the symmetry algebra of the
smooth Wilson loops would not be a Yangian, but it would in fact be much bigger.
Clarifying this is left to future work.

In fact, we did observe that there is an additional 
Yangian symmetry, namely the level-one hypercharge, 
which belongs to $Y[\alg{u}(2,2|4)]$ rather than $Y[\alg{psu}(2,2|4)]$.
This symmetry has been observed before in similar situations \cite{Beisert:2011pn,Munkler:2015xqa},
and the derivations are analogous to the other level-one generators
with minor additional complications.

While we strove to keep the discussion as
generic as possible, importantly we saw that the consistency of the
Yangian symmetry crucially depended on two properties of the underlying gauge theory:
Firstly, the vanishing of the dual Coxeter number of the underlying superconformal symmetry algebra,
secondly the existence of a technical G-identity related to the contraction of the 
field strength two-form with conformal vector fields. 
Both are fulfilled by $\superN=4$ super Yang--Mills theory
and tightly constrain the models to which integrability can possibly apply.

Finally, we would like to stress that at the leading perturbative order the bi-local
action of our conformal generators on a correlation function takes the form
\[\label{eq:yangprop}
\bigvev{
f^{\idxconf{c}}{}_{\idxconf{b}\idxconf{a}}\,
\genfull{J}^{\idxconf{a}}\sgauge_1(\tau_{1})\,
\genfull{J}^{\idxconf{b}}\sgauge_2(\tau_{2})
}_{(1)}
\sim
\der_1\der_2 R^\idxconf{c}_{12}\, ,
\]
not necessarily to be inserted into a Wilson loop. Here $R_{12}$ is the remainder
function giving rise to the local piece of the level-one Yangian $\genYfull{J}^{\idxconf{a}}$.
It is interesting to compare this to the conformal symmetry:
\[
\bigvev{
\genfull{J}^{\idxconf{a}}\sgauge_1(\tau_{1})\,\sgauge_2(\tau_{2})
+
\sgauge_1(\tau_{1})\,\genfull{J}^{\idxconf{a}}\sgauge_2(\tau_{2})
}_{(1)}
\sim
\der_1 Q^\idxconf{a}_{12}
+\der_2 Q^\idxconf{a}_{21}
.
\]
While the level-zero transformations provide only one
total derivative, both derivatives in the level-one action are required
for Yangian invariance of the Wilson loop. The relation \eqref{eq:yangprop} might be of future use.

A useful point to mention at the end of our study concerns UV-finiteness:
We focused on the special case of kappa-symmetric Maldacena--Wilson loops
in order to avoid UV-regularization which might obscure the symmetries. 
Our derivations and results, however, turn out to be independent of this assumption. 
Whether or not the Wilson loop is kappa-symmetric, 
the bi-local insertion of the Yangian level-one generators 
are UV-divergent and require regularization. 
Renormalization of the Yangian action removes the undesired terms
and makes the Yangian become a symmetry.
Thus our result in fact applies to the larger class of
generic smooth Wilson loops \eqref{eq:ourloop}
whose couplings $q^i$ to the scalars are unconstrained.

\medskip

In this paper we only discussed a single Wilson loop. 
What can we say about Yangian symmetry of the expectation value 
of a product of two Wilson loops, $\langle \tr W_1\, \tr W_2\rangle$? 
At leading order in $1/N$ the expectation value
factorizes into the leading orders of the expectation values 
of the individual Wilson loops. There is nothing new to be learned here.
The first sub-leading terms in the $1/N$ expansion are either connected 
or disconnected. 
The disconnected terms are non-planar, 
and Yangian symmetry is most likely spoiled. 
In any case, this result again follows from a single Wilson loop.
The connected term has the large-$N$ topology of an annulus.
As such it is analogous to the planar two-point function 
of single trace operators for which integrability made a big difference.
This connected term is still expected to be constrained by the center of the Yangian, 
as explained in ref.~\cite{Beisert:2010jq}. This point deserves further study.

The general smooth super Wilson loop depends on an infinite number of bosonic and fermionic variables. 
In order to employ the uncovered hidden symmetries towards finding the exact expression
for the vacuum expectation value of such operators, it is wise to look at special types of contours
depending on a finite number of variables while preserving the smoothness of the curve.
In \cite{Beisert:2015jxa} we proposed spline-shaped Wilson loops composed of piecewise-quadratic
contours joined in a way ensuring UV-finiteness at the first perturbative order.
It would be equally interesting to study situations where 
UV-singularities arise, namely corners, light-like points 
or segments as well as self-intersections. These may or may not
lead to anomalies or violations of Yangian symmetry. 
It would be interesting to study such simple super Wilson loops
and particular situations thereby putting the uncovered integrability 
to test and to work.

\paragraph{Acknowledgments.}

We thank James Drummond, Nadav Drukker and Hagen M\"unkler for important discussions. 
The work of NB is partially supported
by grant no.\ 200021-137616 from the Swiss National Science Foundation,
through the SNSF NCCR SwissMAP
and by grant no.\ 615203 from the European Research Council under the FP7.
DM gratefully acknowledges the support of the DFG funded Graduate School GK 1504.
JP thanks the Pauli Center for Theoretical Studies, Z\"urich and the Institute for 
Theoretical Physics at ETH Z\"urich for hospitality and support in the framework of a visiting
professorship.
No non-supersymmetric gauge theories were harmed in this investigation.

\appendix

\section{Conventions and Notations}
\label{sec:conventions}

We use the following notations
  \begin{itemize}
    \item 4d vector indices: \(\mu\), \(\nu,\rho,\ldots=0,\ldots,3\)
    \item  4d left spinor indices: \(\alpha\), \(\beta,\gamma,\ldots=1,2\)
    \item 4d right spinor indices: \(\dot{\alpha}\), \(\dot{\beta},\dot\gamma,\ldots=\dot 1,\dot 2\)
    \item 6d vector indices: \(i\), \(j,k,\ldots=1,\ldots,6\)
    \item 6d spinor indices: \(a\), \(b,c,\ldots=1,\ldots,4\)
    \item 4d $\superN=4$ non-chiral superspace: $\{x^{\mu},\theta^{a\alpha},\bar\theta^{\dot\alpha}{}_{a}\}$
    \item 4d $\superN=4$ super connections and scalar superfield: $\mathcal{A}^{\text{cov}}_{\mu}$,
$\mathcal{A}^{\text{cov}}_{\alpha a}$,  $\mathcal{A}^{\text{cov} \, a}{}_{\dot\alpha}$ and  $\Phi_{i}$ or $\Phi_{ab}$
    \item 4d super-momentum: $p^{\mu}=\dot x^{\mu}+ \theta \sigma^{\mu}\dot{\bar\theta} - \dot\theta\sigma^{\mu}\bar\theta$
    \item Scalar coupling: $q^{i}=n^{i}\sqrt{p^{\mu}p_{\mu}}$ with $q^{i}q^{i}=p^{\mu}p_{\mu}$ and $n^{i}n^{i}=1$
    \item Gauge coupling constant: $g$
    \item $N$ is the rank of the $\grp{U}(N)$ gauge group.
  \end{itemize}

We use the mostly minus signature.  The four-dimensional Pauli
matrices are given by \(\sigma_{\mu} = (\mathbf{1}, \vec{\sigma})\)
and \(\bar{\sigma}_{\mu} = (\mathbf{1}, -\vec{\sigma})\), where
\[
  \sigma_{1} = \begin{pmatrix}0&1\\1&0\end{pmatrix}, \qquad
  \sigma_{2} = \begin{pmatrix}0&-i\\i&0\end{pmatrix}, \qquad
  \sigma_{3} = \begin{pmatrix}1&0\\0&-1\end{pmatrix}.
\]
They satisfy the usual algebra
\begin{equation}
  \sigma_{\mu} \bar{\sigma}_{\nu} + \sigma_{\nu} \bar{\sigma}_{\mu} =
  2 \eta_{\mu \nu}, \qquad
  \bar{\sigma}_{\mu} \sigma_{\nu} + \bar{\sigma}_{\nu} \sigma_{\mu} =
  2 \eta_{\mu \nu}.
\end{equation}

Here is a list of notations we use for the generators of the superconformal algebra and its Yangian:
  \begin{itemize}
  \item \(\gen{J}\) an abstract generator of the superconformal algebra,
  \item \(\genmatr{J}\) the \((2\vert 4\vert 2) \times (2\vert 4\vert 2)\) matrix representation,
  \item \(\genpath{J}\) the representation of \(\gen{J}\) on the paths,
  \item \(\genYpath{J}\) the level-one Yangian generator corresponding to \(\genpath{J}\),
  \item \(\genfield{J}\) a representation of \(\gen{J}\) on the fields,
  \item \(\genYfield{J}\) the level-one Yangian generator corresponding to \(\genfield{J}\),
  \item \(\genfull{J}\) a different representation of \(\gen{J}\) on the fields.  
This maps fields to gauge-covariant fields.  
It differs from the other representation \(\genfield{J}\) by a gauge transformation.
  \item \(\genYfull{J}\) the level-one Yangian corresponding to \(\genfull{J}\),
  \item \(\genYfullnl{J}\) the bi-local part of \(\genYfull{J}\).
  \end{itemize}

\section{Superconformal Generators}
\label{sec:sconf-gen}

Let us define
\[
  \mathcal{Z} =
  \begin{pmatrix}
    x^+ \\ 2 \theta \\ 1
  \end{pmatrix}, \qquad
  \overline{\mathcal{Z}} =
  \begin{pmatrix}
    -1 & -2 \bar{\theta} & x^-
  \end{pmatrix}.
\]
We have \(\overline{\mathcal{Z}} \mathcal{Z} = 0\).
A superconformal transformation determined by a \((2\vert 4\vert 2) \times (2\vert 4\vert 2)\) 
supermatrix \(u\), acts on the \(\mathcal{Z}\) and \(\overline{\mathcal{Z}}\) as follows
\[
  \label{eq:Z-transf}
  \delta \mathcal{Z} = u \mathcal{Z} + \mathcal{Z} \lambda, \qquad
  \delta \overline{\mathcal{Z}} = -\overline{\mathcal{Z}} u + \overline{\lambda}\overline{\mathcal{Z}},
\]
where \(\lambda\), \(\overline{\lambda}\) are \(2 \times 2\) matrices.
Due to these compensating transformations parameterized by \(\lambda\), \(\overline{\lambda}\), 
the action of the central charge in \(\alg{u}(2, 2 \vert 4)\) is not well defined.  
Notice also that the transformations in eq.~\eqref{eq:Z-transf} 
preserve the constraint \(\overline{\mathcal{Z}} \cdot \mathcal{Z} = 0\).

A generic element of the super-algebra is represented by
the matrix
\[
  a \cdot \genmatr{P} + \chi \cdot \genmatr{Q} 
+ \bar{\chi} \cdot \genmatr{\bar{Q}} 
+ \omega' \cdot \genmatr{L}' 
+ \bar{\omega}' \cdot \genmatr{\bar{L}}' 
+ r' \cdot \genmatr{R}' 
+ \rho \cdot \genmatr{S} 
+ \bar{\rho} \cdot \genmatr{\bar{S}} 
+ b \cdot \genmatr{K} 
=
    \begin{pmatrix}
    -2 \bar{\omega}' & 2 \bar{\chi} & -2 a\\
    2 \bar{\rho} & -2 r' & -2 \chi \\
    -2 b & 2 \rho & 2 \omega'      
    \end{pmatrix}.
\]
The action of the supermatrix \(u\) yields transformations 
for the superspace coordinates \((x, \theta, \bar{\theta})\), 
which can be expressed as differential operators.  
This implies the following representation as differential operators
\begin{subequations}
\begin{align}
  a \cdot \genmatr{P} &= - a \cdot \partial_x,
\\
  \chi \cdot \genmatr{Q} &= - \chi \cdot \partial_\theta 
- (\bar{\theta} \chi) \cdot \partial_x,
\\
  \bar{\chi} \cdot \genmatr{\bar{Q}} &= (\bar{\chi} \theta) \cdot \partial_x 
- \bar{\chi} \cdot \partial_{\bar{\theta}},
\\
  \bar{\omega}' \cdot \genmatr{\bar{L}}' &= -(\bar{\omega}' x) \cdot \partial_x 
- 2 (\bar{\omega}' \bar{\theta}) \cdot \partial_{\bar{\theta}},
\\
  \omega' \cdot \genmatr{L}' &= - (x \omega') \cdot \partial_x 
- 2 (\theta \omega') \cdot \partial_\theta,
\\
  r' \cdot \genmatr{R}' &= -2 (r' \theta) \cdot \partial_\theta + 2 (\bar{\theta} r') \cdot \partial_{\bar{\theta}},\\
  \rho \cdot \genmatr{S} &= -(x^+ \rho \theta) \cdot \partial_x 
- 4 (\theta \rho \theta) \cdot \partial_\theta + (x^- \rho) \cdot \partial_{\bar{\theta}},
\\
  \bar{\rho} \cdot \genmatr{\bar{S}} &= (\bar{\theta} \bar{\rho} x^-) 
\cdot \partial_x + 4 (\bar{\theta} \bar{\rho} \bar{\theta}) \cdot \partial_{\bar{\theta}} 
+ (\bar{\rho} x^+) \cdot \partial_\theta,
\\
  b \cdot \genmatr{K} &= \sfrac 1 2 (x^+ b x^+ + x^- b x^-) \cdot \partial_x 
+ 2 (x^- b \bar{\theta}) \cdot \partial_{\bar{\theta}} + 2 (\theta b x^+) \cdot \partial_\theta,
\end{align}
\end{subequations}
where we used the notations 
\(a \cdot \partial_x = a^{\dot{\alpha} \alpha} \partial_{\alpha \dot{\alpha}}\), 
\(\chi \cdot \partial_\theta = \chi^{a \alpha} \partial_{\alpha a}\), 
\(\bar{\chi} \cdot \partial_{\bar{\theta}} = \bar{\chi}^{\dot{\alpha}}{}_a \bar{\partial}^a{}_{\dot{\alpha}}\), 
etc.
We emphasize that these generators only reproduce the action 
of the superconformal group on the coordinates \((x, \theta, \bar{\theta})\); 
the action on the coordinates \(q\) is not contained in these expressions.

\bibliographystyle{nb}
\bibliography{Superwilson10d}

\end{document}